\def\combineddocument{}
\documentclass[a4paper,fleqn]{cas-sc}

\usepackage[authoryear]{natbib}
\usepackage{xcolor}

\ifdefined\combineddocument
\else
\usepackage{xr-hyper}
\fi

\usepackage{longtable}

\usepackage{tikz}
\usetikzlibrary{positioning,arrows.meta, matrix, arrows, calc}

\usepackage{caption}
\usepackage{subcaption}

\usepackage{float}
\def\tsc#1{\csdef{#1}{\textsc{\lowercase{#1}}\xspace}}
\tsc{ES}
\tsc{GAF}
\begin{document}
\let\WriteBookmarks\relax
\def\floatpagepagefraction{1}
\def\textpagefraction{.001}

% Allow figures to appear in-text
\renewcommand{\topfraction}{0.9}       % max fraction of page for floats at top
\renewcommand{\bottomfraction}{0.9}    % max fraction of page for floats at bottom
\renewcommand{\textfraction}{0.1}      % min fraction of page for text
\renewcommand{\floatpagefraction}{0.7} % min fraction of float page that should have floats
\setcounter{topnumber}{4}              % max floats at top of page
\setcounter{bottomnumber}{4}           % max floats at bottom of page
\setcounter{totalnumber}{8}            % max floats per page

\shorttitle{Non-stationary Spatial Modeling Using Fractional SPDEs}    

\shortauthors{E. Svee and G.-A. Fuglstad}

\title [mode = title]{Non-stationary Spatial Modeling Using Fractional SPDEs}  


% \tnotetext[1]{Title footnote text} 


% \author[1]{Elling Svee}%[<options>]
%
% % % Corresponding author indication
% % \cormark[1]
% %
% % % Footnote of the first author
% % \fnmark[1]
%
% % Email id of the first author
% \ead{elling.svee@ntnu.no}
%
% % URL of the first author
% \ead[url]{https://ellingsvee.github.io/}
%
% % Credit authorship
% % eg: \credit{Conceptualization of this study, Methodology, Software}
% % \credit{}
%
% \author[2]{Geir-Arne Fuglstad}%[]
%
% % Address/affiliation
% \affiliation[1, 2]{organization={Norwegian University of Science and Technology},
%             addressline={Høgskoleringen 1}, 
%             city={Trondheim},
% %          citysep={}, % Uncomment if no comma needed between city and postcode
%             postcode={7034}, 
%             state={Trøndelag},
%             country={Norway}}
%
%
% % Footnote of the second author
% % \fnmark[2]
%
% % Email id of the second author
% \ead{geir-arne.fuglstad@ntnu.no}
%
% % URL of the second author
% \ead[url]{https://www.ntnu.no/ansatte/geir-arne.fuglstad}

% === Authors ===
\author[1]{Elling Svee}[
  orcid=0009-0008-4225-964X]
\ead{elling.svee@ntnu.no}
\ead[url]{https://ellingsvee.github.io/}

\author[1]{Geir-Arne Fuglstad}[
  orcid=0000-0003-4995-2152]
\ead{geir-arne.fuglstad@ntnu.no}
\ead[url]{https://www.ntnu.no/ansatte/geir-arne.fuglstad}

% === Affiliation ===
\affiliation[1]{organization={Norwegian University of Science and Technology},
             addressline={Alfred Getz’ vei 1},
            % city={Trondheim},
          postcode={7034},
           state={Trondheim},
            country={Norway}}

% \affiliation[2]{organization={Norwegian University of Science and Technology},
%             addressline={Høgskoleringen 1}, 
%             city={Trondheim},
% %          citysep={}, % Uncomment if no comma needed between city and postcode
%             postcode={7034}, 
%             state={Trøndelag},
%             country={Norway}}

% % Corresponding author text
% \cortext[1]{Corresponding author}
%
% % Footnote text


\begin{abstract}
  We construct a Gaussian random field (GRF) that combines fractional smoothness with spatially varying anisotropy. The GRF is defined through a stochastic partial differential equation (SPDE), where the range, marginal variance, and anisotropy vary spatially according to a spectral parametrization of the SPDE coefficients. Priors are constructed to reduce overfitting in this flexible covariance model, and we estimate parameters using a gradient-based optimization approach based on automatic differentiation. In a simulation study, we investigate how many observations are required to reliably estimate fractional smoothness and non-stationarity, and find that one realization containing 500 observations or more is needed in the scenario considered. We also find that the proposed penalization prevents overfitting across varying numbers of observation locations. Two case studies demonstrate that the relative importance of fractional smoothness and non-stationarity is application dependent. Non-stationarity improves predictions in an application to ocean salinity, whereas fractional smoothness improves predictions in an application to precipitation. Predictive ability is assessed using mean squared error and the continuous ranked probability score. In addition to prediction, the proposed approach can be used as a tool to explore the presence of fractional smoothness and non-stationarity.
\end{abstract}


% % Research highlights
% \begin{highlights}
% \item 
% \item 
% \item 
% \end{highlights}

% Keywords
% \begin{keywords}
% \sep\sep\sep 
% % Spatial non-stationary \sep Fractional smoothness \sep Stochastic partial differential equations \sep Gaussian Markov random fields \sep
% \end{keywords}
\begin{keywords}
Spatially varying anisotropy\sep Fractional smoothness\sep SPDE approach \sep Gaussian Markov random fields\sep Ocean salinity\sep Precipitation
\end{keywords}

\maketitle

% Main text
\section{Introduction}\label{seq:Introduction}
Gaussian random fields (GRFs) are a common tool for spatial modeling in fields such as environmental sciences and ecology \citep{MR2848400, banerjee2003hierarchical}. Isotropic covariance models are widely used, and the Matérn family of covariance functions \citep{stein1999interpolation} is an appealing choice because it has three interpretable parameters that can be controlled separately: 1) marginal variance, 2) spatial range, and 3) smoothness. However, in many applications, isotropy and stationarity are unrealistic assumptions. This has motivated work on more flexible covariance structures \citep{sampson2010constructions,schmidt2020flexible} and on combining them with approaches that enable efficient inference for spatial models \citep{heaton2019case}. In this paper, we extend the popular \textit{SPDE approach} \citep{lindgren2011explicit, 2022_SPDE_10_years} to simultaneously incorporate both spatially varying anisotropy and fractional smoothness, and use the extended model to explore their relative importance for two datasets.

Several approaches have been proposed to model non-stationary spatial processes. One prominent class of methods is the deformation approach, where the spatial domain is transformed through a bijective mapping such that the process appears stationary in the transformed space \citep{sampson1992nonparametric, damian2001bayesian, schmidt2003bayesian}. Another strategy relies on convolution-based constructions, which represent the spatial process as a smoothed integral of a latent process, allowing the covariance structure to vary across space \citep{higdon1998process, paciorek2006spatial}. Covariate-driven models offer an alternative, incorporating spatially varying covariates directly into the covariance function to capture heterogeneity \citep{neto2014accounting, risser2015regression}. Notably, several of these approaches have associated software implementations, such as the local covariance models provided by \citet{risser2017local, risser2020bayesian}. However, while the methods give considerable flexibility, they use dense covariance matrices which often come with a substantial computational cost.

The SPDE approach proposed by \citet{lindgren2011explicit} provides an efficient alternative by approximating non-stationary GRFs using sparse precision matrices. \citet{hu2015spatialmodellingtemperaturehumidity} applied the approach to model annual precipitation in southern Norway, while \citet{fuglstad2015exploring} incorporated spatially varying anisotropy. \citet{fuglstad2015doesnonstationaryspatialdata} demonstrated that non-stationary models can improve predictive performance, but also emphasized the importance of carefully controlling model flexibility to avoid overfitting. Another important extension is numerical methods to handle fractional spatial operators \citep{bolin2018weak,bolin2020numerical}. In particular, the rational approximation for fractional SPDEs introduced by \citet{bolin2020rational}. This enables GRFs with arbitrary smoothness parameters, overcoming the limitation of the original SPDE formulation being restricted to a limited set of smoothness values. The example in \citet[Section~7]{bolin2020rational} combined non-stationary range and variance with a fractional smoothness parameter, but assumed isotropy. Other computational alternatives to speed up computations with fractional non-stationary SPDE models also exist, such as the matrix-free approach \citep{pereira_geostatistics_2023} and Krylov subspace methods \citep{antil_efficient_2022}. However, they rely on approximations of the covariance matrix and the likelihood, and are not discussed further.

This paper extends existing work on non-stationary fractional SPDE models by incorporating non-stationarity in the anisotropy. The result is a more flexible class of models capable of capturing a wider range of spatial processes. A key challenge in non-stationary SPDE-based modeling is practical inference \citep{2022_SPDE_10_years}, and it is well established that penalizing the flexibility of non-stationary components through appropriate priors is essential \citep{fuglstad2019constructing}. We therefore develop priors for both stationary and non-stationary parameters. To obtain gradients during parameter estimation, we differentiate the log-likelihood using \textit{automatic differentiation} (AD). Compared to finite-difference approaches, this technique scales well with the number of parameters, making it beneficial for non-stationary models. There is evidence to suggest that AD can be helpful for optimizing the smoothness parameter with a Matérn covariance structure \citep{geoga2023fitting}, but it is also known that marginal variance, range, and smoothness are not consistently estimable under infill asymptotics \citep{zhang2004inconsistent}. Thus a key focus in this paper is to assess how well the flexible model works in practice in terms of computational robustness.

The proposed model is assessed and compared to simpler alternatives through a simulation study. We investigate the relative importance of non-integer smoothness and non-stationarity, and how this depends on prior specification and data availability. Our primary focus is on predictive performance, measured using the \textit{root mean square error} (RMSE) and the \textit{continuous ranked probability score} (CRPS). The parameter estimates are evaluated through their bias. We also investigate the importance of non-stationarity versus fractional smoothness in two case studies: 1) emulation of a numerical model describing the ocean salinity in Trondheimsfjorden \citep{sintef1987, SLAGSTAD20051}, and 2) analysis of data from a climate reanalysis model describing the average summer precipitation over the conterminous U.S. \citep{genton2015cross}. These are both data-rich cases in which we emulate the output of numerical models, and in which non-stationarity or fractional smoothness is expected to improve performance.

In Section \ref{sec:Specifying and discretizing the covariance structure} we present the proposed SPDE model, detailing its discretization and how we approximate the fractional smoothness. Section \ref{sec:Parametrization and prior for covariance structure} specifies our choice of parameterization and priors for the stationary and non-stationary parameters. Then in Section \ref{sec:Spatial regression model}, we outline the hierarchical modeling framework used for inference, and explain how we apply AD to compute gradients during parameter estimation. The performance of the model is evaluated through a simulation study in Section \ref{sec:Simulation study} and case studies in Section \ref{sec:Real-world application}. Lastly, the paper concludes with Section \ref{seq:Discussion}, summarizing our main results and outlining potential directions for future research. A Python implementation of the model can be found at \href{https://github.com/ellingsvee/FracNonStatSPDE}{\texttt{https://github.com/ellingsvee/FracNonStatSPDE}}.

\section{Specifying and discretizing the covariance structure}\label{sec:Specifying and discretizing the covariance structure}

\subsection{Covariance model}\label{sub:Exact specification}
Inspired by \citet{bolin2020rational,bolin2024covariance}, we consider a GRF $u(\cdot)$ on the domain $\mathcal{D}\subset\mathbb{R}^2$ given by an open, bounded and convex polygon. Its covariance structure is described through the SPDE
\begin{equation}\label{eq:SPDE_model_general_thesis_model}
  \begin{cases}
    \left(\kappa(\boldsymbol{s}) - \nabla\cdot \mathbf{H}(\boldsymbol{s})\nabla \right)^{\beta} u(\boldsymbol{s}) = \tau(\boldsymbol{s})\mathcal{W}(\boldsymbol{s}), & \quad\boldsymbol{s}\in\mathcal{D},         \\
    \left(\mathbf{H}(\boldsymbol{s})\nabla u(\boldsymbol{s})\right)\cdot \boldsymbol{n}(\boldsymbol{s}) = 0,                                                                             & \quad\boldsymbol{s}\in\partial\mathcal{D},
  \end{cases}
\end{equation}
where $\mathcal{W}(\cdot)$ is Gaussian white noise, $\boldsymbol{n}(\cdot)$ is the outwards normal vector, $\kappa(\cdot)$ and $\tau(\cdot)$ are positive functions, and $\mathbf{H}(\cdot)$ is a spatially varying positive-definite $2\times 2$ matrix. For the SPDE to give a GRF with finite variance, we need $\beta > 1/2$. In this paper, we choose parametrizations of $\kappa(\cdot)$, $\tau(\cdot)$ and $\mathbf{H}(\cdot)$ that ensure that these functions are smooth on $\overline{\mathcal{D}}=\mathcal{D}\cup\partial\mathcal{D}$, $\kappa(\cdot) \geq \kappa_{1} > 0$, and that the smallest eigenvalue of $\mathbf{H}(\cdot)$ is greater than some $\lambda_{1}>0$. These conditions are stricter than necessary conditions for a solution to exist, and interested readers can see milder conditions on the coefficients in \citet{bolin2020rational}.

For stationary coefficients $\kappa(\cdot)\equiv \kappa_0$, $\mathbf{H}(\cdot)\equiv \mathbf{H}_0$ and $\tau(\cdot)=\tau_0$, the stationary solution of Equation \eqref{eq:SPDE_model_general_thesis_model} on $\mathcal{D}=\mathbb{R}^2$ has a Matérn covariance function with geometric anisotropy
\begin{equation}\label{eq:Matern Covariance Function}
        c(\boldsymbol{h}) = \frac{\sigma_{0}^{2}}{2^{\nu-1}\Gamma(\nu)}\left(\kappa_0\left\|\mathbf{H}_0^{-1/2}\boldsymbol{h}\right\|\right)^{\nu}K_{\nu}\left(\kappa_0\left\|\mathbf{H}_0^{-1/2}\boldsymbol{h}\right\|\right),
  \quad \boldsymbol{h} \in \mathbb{R}^2.
\end{equation}
Here, $\boldsymbol{h}$ is a displacement vector, $\nu = 2\beta-1$ is the smoothness, $K_\nu(\cdot)$ is the modified Bessel function of the second kind with order $\nu$, and 
the marginal standard deviation is
\[
    \sigma_{0} = \sqrt{\frac{\Gamma(2\beta-1)}{4\pi\Gamma(2\beta)}}\frac{\tau_0}{\kappa_0^{2\beta-1}\left|\mathbf{H}_0\right|^{1/4}}.
\]
This is shown in \citet{lindgren2011explicit} for $\mathbf{H}_0=\mathbf{I}$, and the extension to general $\mathbf{H}_0>0$ can be shown by a linear change of coordinates in Equation \eqref{eq:SPDE_model_general_thesis_model} starting with the case with $\mathbf{H}_0=\mathbf{I}$; see, e.g., \citet{hildeman2021deformed}.  The change of coordinates induces an anisotropic Laplacian and a scaling factor for the Gaussian white noise. The former changes the distance measure in Equation \eqref{eq:Matern Covariance Function}, and the latter gives rise to the factor $|\mathbf{H}_0|^{-1/4}$ in the expression for $\sigma_0$. However, for solutions of Equation \eqref{eq:SPDE_model_general_thesis_model}, Equation \eqref{eq:Matern Covariance Function} is only an approximation due to the boundary conditions imposed on the bounded domain. Empirically, the approximation becomes better when computing the covariance between a pair of locations further away from the boundary, and \citet{lindgren2011explicit} suggest extending the domain outside the area of interest by two times the correlation range.

When $\kappa(\cdot)$ and $\mathbf{H}(\cdot)$ are non-stationary, they affect both marginal variance and correlation structure. We therefore choose to introduce the positive and smooth function $\sigma(\cdot)$, and let 
\begin{equation}\label{eq:tau_RHS_term}
  \tau(\boldsymbol{s}) = \sigma(\boldsymbol{s})\sqrt{4\pi\frac{\Gamma(2\beta)}{\Gamma(2\beta - 1)}} \kappa(\boldsymbol{s})^{2\beta - 1} \left|\mathbf{H(\boldsymbol{s})}\right|^{1/4}, \quad \boldsymbol{s}\in\mathcal{D}.
\end{equation}
The aim is to reduce the dependence of marginal variance on $\kappa(\cdot)$ and $\mathbf{H}(\cdot)$, and to introduce a separate function $\sigma(\cdot)$ that controls marginal standard deviation. This is not exact for spatially varying parameters, but it would decouple correlation and marginal variance exactly (up to boundary conditions) for the case of constant coefficients.

In \citet{fuglstad2015exploring}, the spatially varying positive definite matrix $\mathbf{H}(\cdot)$ is described through a baseline isotropic component and a vector field describing the direction and strength of extra dependence. We instead use the parameterization recently proposed by \citet{llamazareselias2024parameterization},
\begin{equation}\label{eq:DiffusionMatrixParam}
  \mathbf{H} = \cosh\left(\|\boldsymbol{v}\|\right)\mathbf{I} + \frac{\sinh\left(\|\boldsymbol{v}\|\right)}{\|\boldsymbol{v}\|}
  \begin{bmatrix}
    v_{x} & v_{y}  \\
    v_{y} & -v_{x}
  \end{bmatrix},
\end{equation}
where $\boldsymbol{v}$ is a spatially varying vector field. For ease of notation, we have suppressed the dependence of location $\boldsymbol{s}$. The benefit of this parameterization is that it is a smooth bijection between vectors $\boldsymbol{v}\in\mathbb{R}^2$ and positive definite matrices $\mathbf{H}$ with determinant $1$. This means that it removes the identifiability issue of $\boldsymbol{v}$ and $-\boldsymbol{v}$, resulting in the same $\mathbf{H}$ without needing constraints on the parameter space. This should make the parameter space more well behaved, and \citet{llamazareselias2024parameterization} show how the parameterization can be leveraged to construct a penalized complexity prior \citep{simpson2015penalisingmodelcomponentcomplexity} for anisotropy. However, their work is currently limited to stationary models, and we take a different approach for penalizing model complexity in this paper.

Let correlation range be defined as the distance where correlation is approximately $0.13$. Denote the eigenpairs of $\mathbf{H}$ at a fixed location as $(\lambda_{\max},\boldsymbol{v}_{\max})$ and $(\lambda_{\min},\boldsymbol{v}_{\min})$, where $\lambda_{\max}\geq \lambda_{\min}$. Then the largest correlation $\rho_{\max} = \sqrt{8\nu\lambda_{\max}}/\kappa$ is in the direction $\boldsymbol{v}_{\max}$, and smallest in the direction $\boldsymbol{v}_{\min}$ and is given by $\rho_{\min} = \sqrt{8\nu\lambda_{\min}}/\kappa$. By \citet[]{llamazareselias2024parameterization}, the eigenvalues of $\mathbf{H}$ are
\begin{equation*}
    \lambda_{\max}=\exp\{ \|\boldsymbol{v}\| \} \quad \text{and} \quad \lambda_{\min}=\exp\{ -\|\boldsymbol{v}\| \}
.\end{equation*}
This gives rise to interpretable parameters: 1) geometric average of minimum and maximum ranges $\rho = \sqrt{\rho_{\max}\rho_{\min}} = \sqrt{8\nu}/\kappa$, 2) ratio between maximum and minimum ranges $a = \rho_{\max}/\rho_{\min} = \exp(||\boldsymbol{v}||)$, and 3) angle $\psi$ between direction with max correlation and the $x$-axis, defined as $\left(\cos 2\psi, \sin 2\psi \right) = \boldsymbol{v} / \|\boldsymbol{v}\|$ with $\psi \in (-\pi/2, \pi/2]$. Figure~\ref{fig:IsoCorrelationCurve} illustrates an iso-correlation curve, showing the relation between $\boldsymbol{v}$, the angle $\psi$ and the correlation ranges. 
\begin{figure}[pos=H]
  \centering
  \begin{tikzpicture}[scale=1.0]
    \def\a{2.0}
    \def\b{\a * 0.5}
    \def\veclength{\a * 0.5}
    \def\angle{35}
    \def\vecangle{2*\angle}
    \def\const{1.0}

    \draw[thick, rotate=\angle] (0,0) ellipse (\a cm and \b cm);

    \draw[->] (-2.5 * \const, 0) -- (2.5 * \const, 0) node[anchor=north west, yshift=2mm] {$x$};
    \draw[->] (0, -1.5 * \const) -- (0, 1.5 * \const) node[anchor=south east, xshift=2mm] {$y$};

    \draw[->, line width=0.3mm] (0,0) -- ({\veclength*cos(\vecangle)},{\veclength*sin(\vecangle)})
      node[pos=1, anchor=south west, xshift=-1.5mm, yshift=-0.5mm] {\footnotesize $\boldsymbol{v}$};

    \draw[-, line width=0.3mm] (0,0) -- ({\a*cos(\angle)},{\a*sin(\angle)})
      node[pos=1, anchor=south west, xshift=-0.5mm, yshift=-1.0mm] {\footnotesize $\rho_{\max} = \frac{\sqrt{8 \nu \lambda_{\max}}}{\kappa}$};

    \draw[thick,-] (\b,0) arc[start angle=0, end angle=\angle, radius=\b];
    \node at (\b * 1.1, \a * 0.18) {\footnotesize $\psi$}; 

    \draw[dashed,-] (\b*cos{\angle}, \b*sin{\angle}) arc[start angle=\angle, end angle=2*\angle, radius=\b];

  \end{tikzpicture}
% }
  \caption{Relation between the parameterized vector $\boldsymbol{v}$ and the iso-correlation curve at a fixed location. $\lambda_{\max}$ denotes the maximum eigenvalue of $\mathbf{H}$, and the angle $\psi$ of the anisotropy is half of the angle of $\boldsymbol{v}$.}
  \label{fig:IsoCorrelationCurve}
\end{figure}

In summary, we have a Whittle-Matérn covariance structure, which reduces to approximately Matérn with geometric anisotropy for constant coefficients $\kappa(\cdot)\equiv \kappa_0$,  $\mathbf{H}(\cdot) \equiv \mathbf{H}_0$, and $\sigma(\cdot)\equiv \sigma_0$. Isotropy is achieved by setting $\mathbf{H}_0 = \mathbf{I}$. In the stationary case, one can see that the resulting covariance function is overparametrized since $\tau_0$ is superfluous. Due to the parameterization of $\mathbf{H}(\cdot)$ in Equation \eqref{eq:DiffusionMatrixParam}, the number of parameters required to describe $\mathbf{H}_0$ reduces from three to one, meaning $\tau_0$ is necessary. For the non-stationary GRF, we assume latent models where $\log\kappa(\cdot)$, $\log\sigma(\cdot)$, $v_x(\cdot)$ and $v_y(\cdot)$ are independent GRFs. Log-transformations are used here to ensure that $\kappa(\cdot)$ and $\sigma(\cdot)$ remain positive. The choice of latent model or prior is discussed in Sections \ref{sec:Parametrization and prior for covariance structure} and \ref{sec:Spatial regression model}. The associated interpretable parameters, $\rho(\cdot)$, $a(\cdot)$ and $\psi(\cdot)$, are also spatially varying, but their interpretation does not directly correspond to that of their stationary counterparts when the covariance structure is not stationary.

\subsection{Discretization for efficient computations}\label{sub:Discretization}
\subsubsection{Non-fractional SPDE}
We follow the approach in \citet{lindgren2011explicit}, and consider a triangular mesh of the convex polygon $\mathcal{D}$ with $m$ vertices. We define the piecewise linear basis functions $\left\{\varphi_{i}\right\}_{i=1}^{m}$. These equal one on their corresponding vertex and zero on all others, together forming a compactly supported and finite-dimensional function space. The solution of the SPDE in Equation \eqref{eq:SPDE_model_general_thesis_model} is approximated by the GRF
\begin{equation}\label{eq:FEMApproximation}
      \tilde{u}(\boldsymbol{s}) = \sum_{i=1}^{m}w_{i}\varphi_{i}(\boldsymbol{s}), \quad\boldsymbol{s}\in\mathcal{D}
,\end{equation}
where $\boldsymbol{w}=\left(w_{1}, \ldots, w_{m}\right)^{\mathrm{T}}\sim \mathcal{N}\left(\boldsymbol{0},\mathbf{Q}^{-1}\right)$. We approximate the precision matrix $\mathbf{Q}$ using a \textit{finite element method} (FEM).

The triangulation $\mathcal{T}$ is a set of triangle elements $T\subset\mathbb{R}^2$. For $T\in\mathcal{T}$, let $\langle f,g \rangle_T = \int_T f(\boldsymbol{s})g(\boldsymbol{s})\mathrm{d}\boldsymbol{s}$ and let $\boldsymbol{s}_T$ denote the centroid of $T$. Considering first $\beta = 1$ and following the steps in \citet{lindgren2011explicit}, the FEM gives rise to a finite dimensional $m\times m$ linear system
\begin{equation}\label{eq:FEMMatrixEquality}
    \mathbf{L}\boldsymbol{w} \overset{\mathrm{def}}{=} \left( \mathbf{C}_{\kappa^{2}} + \mathbf{G} \right)\boldsymbol{w} 
    \overset{\mathrm{d}}{=}
    \mathbf{C}_{\tau^{2}}^{1/2}\boldsymbol{z},
    \quad \boldsymbol{z} \sim \mathcal{N} \left( \boldsymbol{0}, \mathbf{I} \right)
,\end{equation}
where $\overset{\mathrm{d}}{=}$ denotes equality in distribution. The mass-matrices are computed using mass-lumping
\begin{equation*}
    \left( \mathbf{C}_{f} \right)_{ij} = \begin{cases}
        \hfil \sum_{T \in \mathcal{T}} f( \boldsymbol{s}_{T} )  \langle 1,\varphi_{i}\rangle_{T} \approx \int_\mathcal{D} f(\boldsymbol{s})\varphi_i(\boldsymbol{s})\mathrm{d}\boldsymbol{s} , & i = j, \\
        \hfil 0, & i \neq j,
    \end{cases}
    \quad \text{for} \quad f(\cdot) \in \left\{ \kappa^{2}(\cdot), \tau^{2}(\cdot) \right\},\end{equation*}
and we let $\mathbf{C}$ denote the matrix with $f(\cdot) \equiv 1$. The stiffness matrix $\mathbf{G}$ is computed as
\begin{equation*}
    \mathbf{G}_{ij} = \sum_{T \in \mathcal{T}_{h}}  \langle\nabla\varphi_{i},\mathbf{H}\left(\boldsymbol{s}_{T}\right)\nabla\varphi_{j}\rangle_{T}\approx \int_\mathcal{D} \nabla \varphi_i(\boldsymbol{s})^\mathrm{T}\boldsymbol{H}(\boldsymbol{s})\nabla\varphi_j(\boldsymbol{s})\mathrm{d}\boldsymbol{s} 
    .\end{equation*}
Solving Equation \eqref{eq:FEMMatrixEquality} for $\boldsymbol{w}$ we obtain $\mathbf{Q}_{1} = \mathbf{L}^{\mathrm{T}}\mathbf{C}_{\tau^{2}}^{-1}\mathbf{L}$. Following the iterative approach from \cite[Appendix 4]{lindgren2011explicit}, we can also determine the precisions for higher integer values of $\beta$, giving
\begin{equation}\label{eq:SPDE_non_frac_precision_matrices}
    \begin{cases}
        \mathbf{Q}_{1} = \mathbf{L}^{\mathrm{T}}\mathbf{C}_{\tau^{2}}^{-1}\mathbf{L},&\quad\beta=1, \\
        \mathbf{Q}_{\beta} = \mathbf{L}^{\mathrm{T}}\mathbf{C}^{-1}\mathbf{Q}_{\beta-1}\mathbf{C}^{-1}\mathbf{L},&\quad\beta=2, 3, \ldots \\
    \end{cases}.
\end{equation}
The resulting precision matrices are sparse, but include more and more non-zero terms for increasing values of $\beta$.

\subsubsection{Fractional SPDE}\label{ssub:RationalApproximation}

The solution for non-integer $\beta$ is approximated using the rational SPDE approach. We refer to \citet{bolin2020rational} for the rigorous details, and instead focus on presenting the main ideas. Equation \eqref{eq:FEMMatrixEquality} suggests that the discrete version of the operator $L = \left(\kappa(\cdot)-\nabla\cdot\mathbf{H}(\cdot)\nabla\right)$ is $\mathbf{C}^{-1}\mathbf{L}$ for the coefficients of the FEM basis. This means that in Equation \eqref{eq:FEMMatrixEquality}, $\mathbf{C}^{-1}\mathbf{L}$ would map between the coefficients of the solution, $\boldsymbol{w}$, and the coefficients $\mathbf{C}^{-1}\mathbf{C}_{\tau^{2}}^{1/2}\boldsymbol{z}$ of the right-hand side. \citet{bolin2020rational} argue that one should  work with the operator $L^{-1}$, defined in the sense that $f = L^{-1}g$ solves $Lf=g$ under the no-flow boundary conditions. The inverse operator behaves better than the original operator since the eigenvalues are bounded from above and belong to some interval $J = \left(0, \lambda_0^{-1}\right]$, where $\lambda_0>0$. Consequently, the problem reduces to finding a computationally efficient approximation of $\left(\mathbf{C}^{-1}\mathbf{L}\right)^{-\beta}$ for $\beta > 1/2$.

Inspired by \citet{bolin2020rational}, we rescale $\mathbf{L}$ as $\mathbf{K} = \kappa_{\min}^{-2}\mathbf{L}$, where $\kappa_{\min} = \inf_{\boldsymbol{s} \in \mathcal{D}} \kappa(\boldsymbol{s})$. This will guarantee that the eigenvalues of $(\mathbf{C}^{-1}\mathbf{K})^{-1}$ are contained in $J = \left(\delta, 1\right]$ for a sufficiently small $\delta >0$. This follows from Equation \eqref{eq:FEMMatrixEquality} since $\mathbf{C}^{-1}\mathbf{K}=\mathbf{I}+\kappa_{\min}^{-2}\mathbf{C}^{-1}(\mathbf{C}_{\kappa^2-\kappa_{\min}^2}+\mathbf{G})$, where the last term has non-zero eigenvalues and the first term makes eigenvalues greater than or equal to 1.

Consider the real function $r(x) = x^\beta$ for $x\in (\delta,1]$, and let $k_\beta = \max\{1, \lfloor \beta\rfloor\}$. We choose an approximation of the form $\tilde{r}(x)=P_\mathrm{R}(x)P_\mathrm{L}(x)^{-1}$ for $x \in (\delta,1]$, where $P_\mathrm{R}(x) = \sum_{i = 0}^k c_i x^i$, $P_\mathrm{L}(x)=x^{k_\beta}\sum_{i=0}^{k+1} b_i x^i$, and $k$ is the order of the rational approximation. It is called a rational approximation because both the denominator and the numerator are polynomials. The coefficients $b_0, \ldots, b_{k+1}$ and $c_0,\ldots, c_{k}$ are chosen according to the Clenshaw–Lord Chebyshev–Padé algorithm \citep{baker1996pade}. We can then express
\[
    r\left(x^{-1}\right)  \approx \tilde{r}\left(x^{-1}\right)= \left(\sum_{i=0}^kc_i x^{k-i}\right)\left(x^{k_\beta-1}\sum_{i = 0}^{k+1}b_ix^{k+1-i}\right)^{-1}, \quad x^{-1}\in (\delta,1].
\]
This approximation is constructed for a real function, but can be extended to an approximation for positive definite matrices. As in \cite{bolin2020rational}, we do not have a theoretical result for the required value of $\delta$ to contain all eigenvalues, but choose $\delta = 10^{-(5+k)/2}$. 

Evaluating $P_{\mathrm{R}}$ and $P_{\mathrm{L}}$ for  $\mathbf{C}^{-1}\mathbf{K}$ gives $\left(\mathbf{C}^{-1}\mathbf{K}\right)^{-\beta} \approx \mathbf{P}_\mathrm{R}\mathbf{P}_\mathrm{L}^{-1}$, where $\mathbf{P}_\mathrm{L} = (\mathbf{C}^{-1}\mathbf{K})^{k_\beta-1}\sum_{i = 0}^{k+1}b_i(\mathbf{C}^{-1}\mathbf{K})^{k+1-i}$ and $\mathbf{P}_\mathrm{R} = \sum_{i = 0}^{k}c_i(\mathbf{C}^{-1}\mathbf{K})^{k-i}$. Combined with $\mathbf{C}^{-1}\mathbf{L} = \kappa_{\min}^{2}\mathbf{C}^{-1}\mathbf{K}$, this yields $\left(\mathbf{C}^{-1}\mathbf{L}\right)^{-\beta} \approx \kappa_{\min}^{-2\beta}\mathbf{P}_\mathrm{R}\mathbf{P}_\mathrm{L}^{-1}$. Defining $\tilde{\tau}^{2}(\cdot) = \kappa_{\min}^{-4\beta}\tau^{2}(\cdot)$, so that $\mathbf{C}_{\tilde{\tau}^{2}}^{1/2} = \kappa_{\min}^{-2\beta}\mathbf{C}_{\tau^{2}}^{1/2}$, we may then write the approximate solution for the FEM basis coefficients as
\[
  \boldsymbol{w} \overset{d}{=} (\mathbf{C}^{-1}\mathbf{L})^{-\beta}\mathbf{C}^{-1}\mathbf{C}_{\tau^{2}}^{1 / 2} \boldsymbol{z} \approx \mathbf{P}_\mathrm{R}\mathbf{P}_\mathrm{L}^{-1} \mathbf{C}^{-1}\mathbf{C}_{\tilde{\tau}^{2}}^{1 / 2} \boldsymbol{z}, \quad \boldsymbol{z}\sim \mathcal{N}\left(\boldsymbol{0}, \mathbf{I}\right)
.\]
This can be more compactly  written as $\boldsymbol{w} = \mathbf{P}_\mathrm{R}\widetilde{\boldsymbol{w}}$ and $\widetilde{\boldsymbol{w}}\sim\mathcal{N}(\boldsymbol{0}, \widetilde{\mathbf{Q}}^{-1})$, where $\widetilde{\mathbf{Q}} = \mathbf{P}_\mathrm{L}^{\mathrm{T}}\mathbf{C}\mathbf{C}_{\tilde{\tau}^{2}}^{-1}\mathbf{C}\mathbf{P}_\mathrm{L}
$. To maintain a consistent notation across fractional and non-fractional smoothnesses, let $\mathbf{P}_{\mathrm{R}} = \mathbf{I}$, $\mathbf{P}_\mathrm{L} = (\mathbf{C}^{-1}\mathbf{L})^\beta$, and $\tilde{\tau}^{2}(\cdot) = \tau^{2}(\cdot)$ for $\beta\in\mathbb{N}$.
 
\section{Parametrization and prior for covariance structure}\label{sec:Parametrization and prior for covariance structure}

Our GRF distribution depends on the smoothness $\nu$ and the four non-stationary parameters $\log\kappa(\cdot)$, $\log\sigma(\cdot)$, $v_{x}(\cdot)$ and $v_{y}(\cdot)$. We choose to parameterize the functions on a rectangular domain $\tilde{\mathcal{D}}=(A_1, A_2)\times(B_1, B_2)\subset\mathbb{R}^2$, where the rectangular domain contains the original domain used to solve Equation \eqref{eq:SPDE_model_general_thesis_model}, i.e., $\mathcal{D}\subset\tilde{\mathcal{D}}$. Each non-stationary parameter is decomposed into a stationary part and a non-stationary part described by a finite basis. For example, $\log\kappa:\tilde{\mathcal{D}}\to \mathbb{R}$ is split into a stationary contribution $\log\kappa_0$ and a spatially-varying $\log\kappa_{\mathrm{NS}}:\tilde{\mathcal{D}}\to \mathbb{R}$, giving $\log\kappa(\boldsymbol{s}) = \log\kappa_0 + \log\kappa_{\mathrm{NS}}(\boldsymbol{s})$ for $\boldsymbol{s}\in\tilde{\mathcal{D}}$. We impose the constraint $\int_{\tilde{\mathcal{D}}}\log\kappa_{\mathrm{NS}}(\boldsymbol{s})\,\mathrm{d}\boldsymbol{s} = 0$, so that the non-stationary contribution is orthogonal to the constant functions in $L^{2}(\tilde{\mathcal{D}})$ and the splitting between $\log\kappa_0$ and $\log\kappa_{\mathrm{NS}}(\cdot)$ is unique. The zero-frequency mode is therefore handled entirely by the stationary prior on $\log\kappa_0$, while the non-stationary prior introduced in Section \ref{ssub:Parametrization and prior for non-stationarity} only acts on the orthogonal complement.

\subsection{Parametrization and prior for stationarity}\label{ssub:Parametrization and prior for stationarity}

Instead of placing priors directly on $\kappa_{0}$, $v_{x, 0}$ and $v_{y, 0}$, we define them through the stationary parts $\rho_{0}$, $a_{0}$ and $\psi_{0}$ of the interpretable parameters. The prior for the angle $\psi_{0}$ is set to the uniform distribution $\psi_{0} \sim \mathrm{Unif}\left( -\pi / 2, \pi / 2 \right)$, meaning we do not favor some directions more than others. For range and marginal standard deviation, we follow \citet{fuglstad2019constructing}, and specify a joint PC prior for $\left( \rho_{0}, \sigma_{0} \right)$ as
\begin{equation}\label{eq:PCPriorJointRangeAndMargVar}
  \pi\left(\rho, \sigma\right) = \lambda_{\rho}\lambda_{\sigma} \rho^{-2} \exp\left\{-\lambda_{\rho}\rho^{-1} - \lambda_{\sigma}\sigma\right\},\quad\rho, \sigma>0, \quad \lambda_{\rho}, \lambda_{\sigma}> 0,
\end{equation}
where $\lambda_\rho$ and $\lambda_\sigma$ are hyperparameters selected so that $C_\rho$ is the prior median for range and $C_\sigma$ is the prior median for marginal standard deviation. 

For selecting $\pi(a)$, we view the anisotropic model as a flexible extension of an isotropic base model. Let $\boldsymbol{v} \sim \mathcal{N}\left( \boldsymbol{0}, \sigma_{\boldsymbol{v}}^{2}\mathbf{I} \right)$, with the hyperparameter $\sigma_{\boldsymbol{v}}>0$ controlling the marginal standard deviation. As the variance approaches zero, the anisotropy approaches $a=1$. Since $\left(v_{x}^{2} + v_{y}^{2}\right)/\sigma_{\boldsymbol{v}}^{2}$ is chi-square distributed with $2$ degrees of freedom, the CDF of $a = \exp(\|\boldsymbol{v}\|)$ is
\begin{equation}\label{eq:CDFAnisotropyPrior}
    \mathrm{P}\left(a \leq k\right) 
        = \mathrm{P} \left( \frac{v_{x}^{2} + v_{y}^{2}}{\sigma_{\boldsymbol{v}}^{2}} \leq \left( \frac{\log k }{\sigma_{\boldsymbol{v}}} \right)^{2} \right)
        = 1- \exp\left\{-\frac{1}{2}\left(\frac{\log k}{\sigma_{\boldsymbol{v}}}\right)^{2}\right\}
,\end{equation}
which we differentiate to obtain
\begin{equation*}
    \pi(a) = \sigma_{\boldsymbol{v}}^{-2}\frac{\log a}{a}\exp\left\{-\frac{1}{2}\left(\frac{\log a}{\sigma_{\boldsymbol{v}}}\right)^{2}\right\},
    \quad a \geq 1.
\end{equation*}
Instead of having to specify $\sigma_{\boldsymbol{v}}$ directly, we control the prior with a hyperparameter $C_a$ chosen so that $\mathrm{P}\left(a > C_{a}\right) = 0.05$.

Lastly, inspired by the rSPDE package \citep{rSPDEpackage}, $\pi(\nu)$ is set to a beta distribution constructed on the user-specified interval $\left(0, \nu_{\max}\right)$. The prior density becomes
\begin{equation}\label{eq:PriorDensidySmootness}
  \pi(\nu) = \frac{1}{\nu_{\max} B(p, q)}\left(\frac{\nu}{\nu_{\max}}\right)^{p-1}\left(1-\frac{\nu}{\nu_{\max}}\right)^{q-1},
  \quad 0 < \nu < \nu_{\max}, \quad p,q>0
,\end{equation}
where $B(p, q)$ denotes the beta function. To simplify prior specification by avoiding having to select $p$ and $q$ directly, we reparameterize the beta distribution using its mean value $C_{\nu}$ and the length $C_{\nu, \mathrm{HPD}}$ of the $95\%$ \textit{highest posterior density} (HPD) interval. This length is defined as the width of the interval $\left\{ \nu : \pi(\nu) \geq k \right\}$ where 
\begin{equation*}
  \int_{\left\{ \nu : \pi(\nu) \geq k \right\}} \pi(\nu)\mathrm{d}\nu = 95\%
.\end{equation*}
A shorter interval implies a more concentrated and informative prior, while a longer interval results in a less informative one.

\subsection{Parametrization and prior for non-stationarity }\label{ssub:Parametrization and prior for non-stationarity} 

The parameterization of the non-stationarity is the same for all four functions, so we use $\log \kappa_{\mathrm{NS}}(\cdot)$ for demonstration. We follow \citet{fuglstad2015doesnonstationaryspatialdata} and impose a GRF on $\log\kappa_{\mathrm{NS}}(\cdot)$ through an SPDE. First, let $f_{kl}(\cdot)$ be the eigenfunctions of $-\Delta$ on $\tilde{\mathcal{D}}$ with zero Neumann boundary conditions,
\begin{equation}\label{eq:NonStatParamBasisFunc}
    f_{kl}(x, y) = \frac{1}{C_{kl}\sqrt{AB}}\cos\left(\frac{k\pi\left(x-A_{1}\right)}{A}\right)\cos\left(\frac{l\pi\left(y-B_{1}\right)}{B}\right), \quad \boldsymbol{s}=(x,y)^\mathrm{T} \in \tilde{\mathcal{D}}, \quad (k,l)\in \mathbb{N}_0^2,\end{equation}
where $A=A_{2}-A_{1}$, $B=B_{2}-B_{1}$. $C_{kl} = 2$ if $k$ and $l$ are both non-zero, $C_{kl} = \sqrt{2}$ if exactly one of $k$ and $l$ is non-zero, and $C_{kl} = 1$ if $k = l = 0$.

We choose a spline-like prior that does not introduce a separate range hyperparameter, and lets us control the variability through a single penalty parameter $\tau_{\kappa}>0$,
\begin{equation}\label{eq:param_penalty_SPDE}
  \begin{cases}
    -\Delta \log\kappa_{\mathrm{NS}}(\boldsymbol{s})=\mathcal{W}_{\kappa}(\boldsymbol{s})/\sqrt{\tau_{\kappa}}, & \quad\boldsymbol{s}\in\tilde{\mathcal{D}},         \\
    \nabla \log\kappa_{\mathrm{NS}}(\boldsymbol{s})\cdot \boldsymbol{n}(\boldsymbol{s}) = 0,                      & \quad\boldsymbol{s}\in\partial\tilde{\mathcal{D}},
  \end{cases}
\end{equation}
where $\mathcal{W}_\kappa(\cdot) = \sum_{(k,l)\in \mathbb{N}_0^2\setminus\{(0,0)\}}z_{kl}f_{kl}(\cdot)$ with
$z_{kl} \overset{\text{iid}}{\sim} \mathcal{N}(0,1)$. I.e., $\mathcal{W}_\kappa(\cdot)$ is Gaussian white noise without the $(0,0)$ frequency. This is a slight abuse of notation since the series cannot be evaluated pointwise and $\mathcal{W}_\kappa$ is a generalised GRF. To have a unique solution, we solve with
the condition $\int_{\tilde{\mathcal{D}}}\log\kappa_{\mathrm{NS}}(\boldsymbol{s})\mathrm{d}\boldsymbol{s} = 0$, which is equivalent to excluding the $(0,0)$ frequency also in the solution.

Let $E = (\{0,1,\ldots,M\}\times\{0,1,\ldots, N\})\setminus\{(0,0)\}$, then the left-hand side is approximated by
\begin{equation}\label{eq:fourier_basis}
  \log\kappa_{\mathrm{NS}}(\boldsymbol{s}) \approx \sum_{(k,l)\in E}\alpha_{kl}f_{kl}(\boldsymbol{s}),\quad \boldsymbol{s} \in \tilde{\mathcal{D}}.
\end{equation}
Let $\boldsymbol{\alpha}_{\kappa}=\left(\alpha_{\kappa, kl}\right)_{(k,l)\in E}^\mathrm{T}$ be the vector of coefficients, then the discretized SPDE is
\[
  -\Delta\left(\sum_{(k,l)\in E}\alpha_{\kappa, kl}f_{kl}(\boldsymbol{s})\right) = \frac{1}{\sqrt{\tau_{\kappa}}} \sum_{(k,l)\in E}z_{kl}f_{kl}(\boldsymbol{s}), \quad\boldsymbol{s}\in\tilde{\mathcal{D}}.\]
Due to orthogonality of the eigenfunctions, we can equate the coefficients of each $f_{kl}$ and solve for each coefficient separately, giving
\[
  \alpha_{\kappa, kl}\left[\left(\frac{\pi k}{A}\right)^{2}+\left(\frac{\pi l}{B}\right)^{2}\right]= \frac{z_{kl}}{\sqrt{\tau_{\kappa}}}, \quad \quad (k,l)\in E.
\]
Thus, the coefficient vector $\boldsymbol{\alpha}_{\kappa}$ contains $(M+1)(N+1) - 1$ elements, and has distribution $\boldsymbol{\alpha}_{\kappa} \sim \mathcal{N}\left(\boldsymbol{0}, \tau_{\kappa}^{-1}\mathbf{Q}_{\mathrm{NS}}^{-1}\right)$. $\mathbf{Q}_{\mathrm{NS}}$ is a diagonal precision matrix with $Q_{\mathrm{NS},i,i} = \left[(\pi k_i/A)^2+(\pi l_i/B)^2\right]^{2}$ for $i=1,\ldots, (M+1)(N+1)-1$, where $k_i$ and $l_i$ are the index pair corresponding to the ordering used for $\boldsymbol{\alpha}_\kappa$.

In the same way, $\log\sigma(\cdot)$, $v_x(\cdot)$ and $v_y(\cdot)$ respectively give rise to $\boldsymbol{\alpha}_\sigma$, $\boldsymbol{\alpha}_{v_x}$ and $\boldsymbol{\alpha}_{v_y}$ with the corresponding distribution having penalty parameters $\tau_\sigma$, $\tau_{v_x}$ and $\tau_{v_y}$. When specifying the priors, we have to select coefficients $\boldsymbol{\tau} = ( \tau_{\kappa}, \tau_{\sigma}, \tau_{v_{x}}, \tau_{v_{y}} )^\mathrm{T}$ which determine how much the non-stationarity is penalized. The non-stationarity can be controlled by setting a bound limiting the variability \citep[Appendix G.2]{fuglstad2019constructing}. Using the interpretable parameter $\rho(\cdot)$ with stationary component $\rho_{0}$, we choose the penalty $\tau_{\kappa}$ such that
\begin{equation}\label{eq:PenaltyProbabilityExpression}
    \mathrm{P} \left( \max_{\boldsymbol{s}\in\tilde{\mathcal{D}}} \left\lvert \log \left( \frac{\rho(\boldsymbol{s})}{\rho_0} \right) \right\rvert > \log C_{\mathrm{NS}, \rho} \right) = 0.05
.\end{equation}
Here, $C_{\mathrm{NS}, \rho}>0$ is a user-defined threshold determining the upper bound. The probability in Equation \eqref{eq:PenaltyProbabilityExpression} is approximated by Monte Carlo simulation using the known distribution of $\boldsymbol{\alpha}_{\kappa}$. We construct a regular grid $S\subset\tilde{\mathcal{D}}$ at which the ratio is evaluated, and draw $N_{\mathrm{MC}}$ independent samples $\boldsymbol{\alpha}_{\kappa}^{(1)}, \ldots, \boldsymbol{\alpha}_{\kappa}^{(N_{\mathrm{MC}})}$ from the prior $\mathcal{N}\left(\boldsymbol{0}, \tau_{\kappa}^{-1}\mathbf{Q}_{\mathrm{NS}}^{-1}\right)$. Letting $\rho^{(i)}(\cdot)$ denote the realization of $\rho(\cdot)$ obtained from $\boldsymbol{\alpha}_{\kappa}^{(i)}$, the probability is approximated as
\begin{equation*}
    \mathrm{P} \left( \max_{\boldsymbol{s}\in\tilde{\mathcal{D}}} \left\lvert \log \left( \frac{\rho(\boldsymbol{s})}{\rho_0} \right) \right\rvert > \log C_{\mathrm{NS}, \rho} \right) \approx
    \frac{1}{N_{\mathrm{MC}}}\sum_{i=1}^{N_{\mathrm{MC}}}\mathbb{1}\left( \max_{\boldsymbol{s} \in S} \left\lvert \log \left( \frac{\rho^{(i)}(\boldsymbol{s})}{\rho_0} \right) \right\rvert > \log C_{\mathrm{NS}, \rho} \right)
,\end{equation*}
where $\mathbb{1}(\cdot)$ is the indicator function, and the maximum is taken over the finite grid $S$. By repeating this process for different $\tau_{\kappa}$, we select the value that yields an estimated probability close to the desired value $0.05$. The penalties for the remaining non-stationary parameters are set through a similar procedure, where we specify thresholds limiting the ratio between the parameter and its stationary component. For $v_{x}(\cdot)$ and $v_{y}(\cdot)$, we set $\tau_{v_{x}} = \tau_{v_{y}}$ and choose this common penalty through a single $C_{\mathrm{NS}, \boldsymbol{v}}$. Since $\boldsymbol{v}(\cdot)$ is vector-valued, we instead impose the threshold on the anisotropy ratio $\log\left(a(\boldsymbol{s})/a_{0}\right)$. The penalty is then selected by a similar Monte Carlo procedure as for $\tau_{\kappa}$.

For ease of notation, we construct a vector of stationary parameters $\boldsymbol{\theta}_\mathrm{S} = (\kappa_0, \sigma_0, v_{x,0}, v_{y,0}, \nu)^\mathrm{T}$, and $\boldsymbol{\theta}_{\mathrm{NS}} = (\boldsymbol{\alpha}_{\kappa}^{\mathrm{T}}, \boldsymbol{\alpha}_{\sigma}^{\mathrm{T}}, \boldsymbol{\alpha}_{v_x}^{\mathrm{T}}, \boldsymbol{\alpha}_{v_y}^{\mathrm{T}})^{\mathrm{T}}$ with coefficients controlling the non-stationary parameters. These vectors are used during optimization for parameter estimation. Summarizing the priors, the stationary $\boldsymbol{\theta}_{\mathrm{S}}$ is controlled by hyperparameters $C_\rho$, $C_\sigma$, $C_a$, $C_\nu$, $C_{\nu, \mathrm{HPD}}$ and $\nu_{\max}$, while the hyperparameters for the non-stationarity $\boldsymbol{\theta}_{\mathrm{NS}}$ are $C_{\mathrm{NS}, \rho}$, $C_{\mathrm{NS}, \sigma}$ and $C_{\mathrm{NS}, \boldsymbol{v}}$.

\section{Spatial regression model}\label{sec:Spatial regression model}
\subsection{Hierarchical model}
Assume we have a vector of $n$ observations $\boldsymbol{y}=\left(y_{1},\ldots,y_{n}\right)^{\text{T}}$ made at locations $\boldsymbol{s}_{1}, \ldots, \boldsymbol{s}_{n} \in \mathcal{D}$, respectively. For $i=1,\ldots,n$, the observations are independently modeled as
\[
    y_i|\eta(\boldsymbol{s}_i), \sigma_\mathrm{N}^{2} \sim \mathcal{N}\left(\eta(\boldsymbol{s}_i), \sigma_\mathrm{N}^2\right)
,\]
where $\eta(\cdot)$ is a spatially varying true signal. It is modeled as the GRF
\[
    \eta(\boldsymbol{s}) = \boldsymbol{x}(\boldsymbol{s})^\mathrm{T}\boldsymbol{\beta}+u(\boldsymbol{s}),\quad \boldsymbol{s}\in\mathcal{D},
\]
where $\boldsymbol{x}(\cdot)$ is a spatially-varying vector of $p$ covariates, $\boldsymbol{\beta}$ are the $p$ coefficients, and $u(\cdot)$ is the GRF from Section \ref{sub:Exact specification}. For computations, we replace $u(\cdot)$ by the approximation $\tilde{u}(\cdot)$ described in Section \ref{sub:Discretization}. This gives the vector representation
\begin{equation*}
    \boldsymbol{y} = \mathbf{X}\boldsymbol{\beta} + \mathbf{A}\mathbf{P}_\mathrm{R}\widetilde{\boldsymbol{w}} + \boldsymbol{\epsilon},
\end{equation*}
where $\mathbf{X}$ is the $n\times p$ design matrix, $\mathbf{A}$ is the $n \times m$ matrix mapping from location to basis functions, $\widetilde{\boldsymbol{w}}\sim\mathcal{N}(0, \widetilde{\mathbf{Q}}^{-1})$ as described in Section \ref{sub:Discretization}, and $\boldsymbol{\epsilon}|\sigma_\mathrm{N}^2 \sim \mathcal{N} \left( \boldsymbol{0}, \sigma_\mathrm{N}^{2}\mathbf{I}\right)$. The covariance model is described by parameters $\boldsymbol{\theta} = \left(\boldsymbol{\theta}_\mathrm{S}^\mathrm{T}, \boldsymbol{\theta}_{\mathrm{NS}}^\mathrm{T}\right)^\mathrm{T}$ that control $\nu$, $\log\kappa(\cdot)$, $\log\sigma(\cdot)$, $v_x(\cdot)$, $v_y(\cdot)$ as discussed in Section \ref{sec:Parametrization and prior for covariance structure}, and we can write the hierarchical model 
\begin{align*}
  \boldsymbol{y}|\boldsymbol{\beta},\widetilde{\boldsymbol{w}}, \sigma_\mathrm{N}^2 \sim \mathcal{N}\left(\mathbf{X}\boldsymbol{\beta}+\mathbf{A}\mathbf{P}_\mathrm{R}\widetilde{\boldsymbol{w}}, \sigma_\mathrm{N}^2\mathbf{I}\right), \\
    \boldsymbol{\beta}\sim\mathcal{N}(\boldsymbol{0}, \mathbf{I}/\tau_\beta), \widetilde{\boldsymbol{w}}|\boldsymbol{\theta}\sim\mathcal{N}\left(\boldsymbol{0}, \widetilde{\mathbf{Q}}^{-1}(\boldsymbol{\theta})\right), \\
    \sigma_\mathrm{N}^2\sim\pi\left(\sigma_\mathrm{N}^2\right), \boldsymbol{\theta}\sim\pi\left(\boldsymbol{\theta}\right).
\end{align*}
$\pi\left(\sigma_\mathrm{N}^2\right)$ is given the PC prior from \citet{simpson2015penalisingmodelcomponentcomplexity} with the median $C_{\sigma_{\mathrm{N}}}$ as hyperparameter, and $\tau_\beta$ is a small value corresponding to a vague prior on the coefficients. Additionally we have the hyperparameters described in Sections \ref{ssub:Parametrization and prior for stationarity} and \ref{ssub:Parametrization and prior for non-stationarity} for the covariance model.

\subsection{Parameter estimation and prediction}\label{sub:Parameter estimation and prediction}

Collecting the vectors controlling the spatial and fixed effects in $\boldsymbol{z}=\left(\widetilde{\boldsymbol{w}}^{\text{T}}, \boldsymbol{\beta}^{\text{T}}\right)^{\text{T}}$, define the matrices 
\[
  \mathbf{S}=
  \begin{bmatrix}
    \mathbf{A}\mathbf{P}_{\mathrm{R}} && \mathbf{X}
  \end{bmatrix}
  \quad\text{and}\quad
  \mathbf{Q}_{\boldsymbol{z}} =
  \begin{bmatrix}
    \widetilde{\mathbf{Q}} && \mathbf{0}\\
    \mathbf{0} && \tau_{\beta}\mathbf{I}
  \end{bmatrix}.
  \]
Then we can write the hierarchical model as
\begin{align*}
  \boldsymbol{y}|\boldsymbol{z},\boldsymbol{\theta}, \sigma_\mathrm{N}^2&\sim \mathcal{N}\left(\mathbf{S}\boldsymbol{z}, \sigma_{\mathrm{N}}^{2}\mathbf{I}\right) \\
  \boldsymbol{z}|\boldsymbol{\theta}&\sim \mathcal{N}\left(\boldsymbol{0}, \mathbf{Q}_{\boldsymbol{z}}^{-1}\right)
  .
\end{align*}
Denote the posterior distribution as $\pi(\boldsymbol{\theta}|\boldsymbol{y})$, then the procedure described in \citet[Appendix C]{fuglstad2015doesnonstationaryspatialdata} gives
\begin{equation}\label{eq:HierarchicalModelLL}
  \begin{aligned}
    \log\pi\left(\boldsymbol{\theta}, \sigma_\mathrm{N}^2|\boldsymbol{y}\right) 
    \propto& \log\pi(\sigma_\mathrm{N}^2) +\log\pi\left(\boldsymbol{\theta}_{\mathrm{S}}\right) +  \log\pi \left(\boldsymbol{\theta}_{\mathrm{NS}} \right) \\
    &  + \frac{1}{2}\log|\mathbf{Q}_{\boldsymbol{z}}| - \frac{n}{2}\log\sigma_{\mathrm{N}}^{2} -\frac{1}{2}\log|\mathbf{Q}_{\mathrm{C}}| \\
    & -\frac{1}{2}\boldsymbol{\mu}_{\mathrm{C}}^{\text{T}}\mathbf{Q}_{\boldsymbol{z}}\boldsymbol{\mu}_{\mathrm{C}} - \frac{1}{2}\sigma_{\mathrm{N}}^{-2}\left(\boldsymbol{y}-\mathbf{S}\boldsymbol{\mu}_{\mathrm{C}}\right)^{\text{T}}\left(\boldsymbol{y}-\mathbf{S}\boldsymbol{\mu}_{\mathrm{C}}\right) ,
  \end{aligned}
\end{equation}
where $\mathbf{Q}_{\mathrm{C}}=\mathbf{Q}_{\boldsymbol{z}}+\mathbf{S}^{\text{T}}\mathbf{S}/\sigma_{\mathrm{N}}^{2}$ and $\boldsymbol{\mu}_{\mathrm{C}}=\mathbf{Q}_{\mathrm{C}}^{-1}\boldsymbol{S}^{\text{T}}\boldsymbol{y}/\sigma_{\mathrm{N}}^{2}$. Here $\boldsymbol{\mu}_\mathrm{C}$ and $\mathbf{Q}_\mathrm{C}$ are the mean vector and precision matrix for $\boldsymbol{z}|\boldsymbol{y}, \sigma_\mathrm{N}^2, \boldsymbol{\theta}$. The MAP $\left( \widehat{\boldsymbol{\theta}}^{\mathrm{T}}, \hat{\sigma}_{\mathrm{N}}^{2} \right)^{\mathrm{T}}$ of $\left(\boldsymbol{\theta}^\mathrm{T}, \sigma_\mathrm{N}^2\right)^\mathrm{T}$ are the parameters maximizing this log-likelihood. 

For a new set of $k$ unobserved locations $\boldsymbol{s}_{1}^*, \ldots, \boldsymbol{s}_{k}^*\in\mathcal{D}$, the posterior mean and marginal variance of $\eta(\cdot)$ can be predicted at the unobserved locations using the density of $\boldsymbol{z}|\boldsymbol{y},\widehat{\boldsymbol{\theta}}, \hat{\sigma}_\mathrm{N}^2$. Define the matrix
$\mathbf{S}_{\mathrm{P}}=
  \begin{bmatrix}
    \mathbf{A}_{\mathrm{P}}\mathbf{P}_{\mathrm{R}} && \mathbf{X}_{\mathrm{P}}
  \end{bmatrix}$,
where $\mathbf{X}_{\mathrm{P}}$ is the $k\times p$ dimensional design matrix for the prediction locations, and $\mathbf{A}_{\mathrm{P}}$ is the $k\times m$ dimensional matrix mapping the triangulation to the prediction locations. Let $\boldsymbol{\eta}^*=\left(\eta\left(\boldsymbol{s}_1^*\right),\ldots, \eta\left(\boldsymbol{s}_k^*\right)\right)^\mathrm{T}$, then 
\[
  \boldsymbol{\eta}^*|\boldsymbol{y}, \widehat{\boldsymbol{\theta}},\hat{\sigma}_\mathrm{N}^2\sim \mathcal{N}\left(\mathbf{S}_{\mathrm{P}}\boldsymbol{\mu}_{\mathrm{C}}, \mathbf{S}_{\mathrm{P}}\mathbf{Q}_{\mathrm{C}}^{-1}\mathbf{S}_{\mathrm{P}}^{\mathrm{T}}\right)
,\]
where we only need to compute the diagonal of $\mathbf{S}_{\mathrm{P}}\mathbf{Q}_{\mathrm{C}}^{-1}\mathbf{S}_{\mathrm{P}}^{\mathrm{T}}$ for marginal properties. We use this distribution when the goal is to estimate the latent $\eta(\cdot)$. For predicting a new value at a held-out location, as in cross-validation, the predictive mean is unchanged but the variance is obtained from the diagonal of $\mathbf{S}_{\mathrm{P}}\mathbf{Q}_{\mathrm{C}}^{-1}\mathbf{S}_{\mathrm{P}}^{\mathrm{T}} + \hat{\sigma}_{\mathrm{N}}^{2}\mathbf{I}$ to account for the additional independent measurement error.

\subsection{Optimization strategy}\label{sub:Optimization strategy}
\subsubsection{Gradient-based optimization with automatic differentiation}
Gradient information is required for reliable convergence when estimating the model parameters. \citet[Appendix D]{fuglstad2015doesnonstationaryspatialdata} derive analytical gradients for a related non-stationary model with $\nu = 1$, but extending these derivations to fractional models is considerably more involved due to the rational approximation. The rSPDE package \citep{rSPDEpackage} instead relies on numerical differentiation. However, as later shown in Figure~\ref{fig:benchmark_rspde_fd}, this approach does not scale well as the number of parameters increases. The scaling could be improved by parallelizing the finite-difference approach over multiple processes, but this would require inter-process communication and a more complex implementation. As an alternative, we use AD to obtain gradients of the log-likelihood. AD evaluates derivatives by systematically applying the chain rule to the sequence of operations defining the log-likelihood, a process referred to as \textit{backpropagation}. This provides gradients that are exact up to machine precision and can be computed efficiently in high-dimensional parameter spaces. See \citet{baydin_automatic_2015} for a comprehensive introduction to AD, and the Supplementary Material for the formulas needed to differentiate the log-likelihood.

When applying this strategy to the rational approximation, special care is needed for differentiating with respect to the smoothness parameter~$\nu$. The rational approximation depends on $k_{\beta} = \max\{1,\lfloor \beta \rfloor\}$ and on whether $\nu$ is integer-valued. Consequently, the computational structure of the log-likelihood changes at $\nu \in \mathbb{N}$, and the log-likelihood is not differentiable with respect to $\nu$ at these points. In practice, this is not problematic, since the optimizer is unlikely to propose integer values for $\nu$. Moreover, because $k_\beta = 1$ for all $\nu \in (0,3)$, the computational graph remains unchanged for non-integer smoothnesses throughout this interval. We therefore apply the AD approach under the assumption that $\nu \in (0,3)\setminus\{1,2\}$. The parameter $\nu$ also enters the likelihood directly through the coefficients of the rational approximation. Rather than recomputing these coefficients at each optimization step, we precompute them on a grid of $200$ values of $\nu$. For intermediate values, the coefficients are obtained using differentiable cubic spline interpolation. This reduces the computational cost and allows gradients with respect to $\nu$ to be evaluated without differentiating through the construction of the rational approximation itself.

Once gradients of the log-likelihood are available, the model parameters are optimized using a two-stage procedure. We first use the Adam algorithm \citep{kingma2017adammethodstochasticoptimization}, with the learning rate set to $0.01$ and the maximum number of iterations to $500$. The remaining parameters are kept at their default values from the Optax implementation \citep{Optax2020}. We then refine the estimates using the SciPy implementation of L-BFGS-B \citep{LBFGSB, scipy2020}. This second step is run for at most $200$ iterations. In both stages, optimization is stopped early when the relative change in log-likelihood between successive iterations is below $10^{-6}$. This two-stage strategy works well for the simulation and case studies in Sections~\ref{sec:Simulation study} and~\ref{sec:Real-world application}, although tuning may be needed for other applications.

\subsubsection{Sparse AD implementation and computational benchmarks}
A practical challenge in realizing the AD-based optimization strategy is the treatment of sparse matrix operations. In general, the derivative of a sparse matrix operation need not have the same sparsity pattern as the original operation. Although sparse AD is supported to some extent in high-level programming languages, existing tools do not provide the combination of sparse operations required for our log-likelihood \citep[Appendix B]{hill_sparser_2025}. PyTorch \citep{paszke2019pytorchimperativestylehighperformance} and JAX \citep{jax2018github} revert to dense operations during backpropagation for the operations considered here, while more general-purpose tools like SparseDiffTools.jl \citep{juliadiffcontributorsJuliaDiffSparseDiffToolsjl2024} are not designed for differentiating through sparse Cholesky factorizations.

To make the AD-based optimization computationally feasible, we implement the sparse AD operations required for our log-likelihood in the custom library sparsejax \citep{svee_sparsejax}. This uses JAX arrays for most operations, but relies on the JAX foreign function interface to support sparse linear algebra on different backends. On CPUs, sparse computations are performed using SciPy \citep{scipy2020} and CHOLMOD \citep{cholmod2008}, while GPU acceleration is enabled through cuSPARSE and cuDSS \citep{cusparse2009implementing,nvidia_cudss}. During backpropagation, the VJPs are evaluated only at the stored coordinates of each sparse input. This restriction is closely related to the masking strategy of \citet{nytko2023optimizedsparsematrixoperations}, and avoids forming dense gradient matrices before applying a mask. Since the sparsity pattern is fixed by the triangulation and the structure of the SPDE, entries outside the pattern are structurally zero for all parameter values, and the resulting parameter gradients remain exact. 

We evaluate the implementation by benchmarking the runtime and memory use of a single optimization iteration, consisting of one log-likelihood evaluation and one gradient evaluation. The experiments are run on a workstation with an AMD Ryzen 7 5800X CPU with $8$ physical cores and $16$ threads, $32$ GB of system memory, and an NVIDIA GeForce RTX 3090 GPU with $24$ GB of device memory. Memory use is reported as peak host memory for CPU runs and peak device memory for GPU runs. All experiments use an order-$k=2$ rational approximation, and timings are reported after JAX's just-in-time compilation.

\begin{figure}[pos=H]
  \centering
  \begin{subfigure}[b]{0.33\textwidth}
    \centering
    \includegraphics[width=\textwidth]{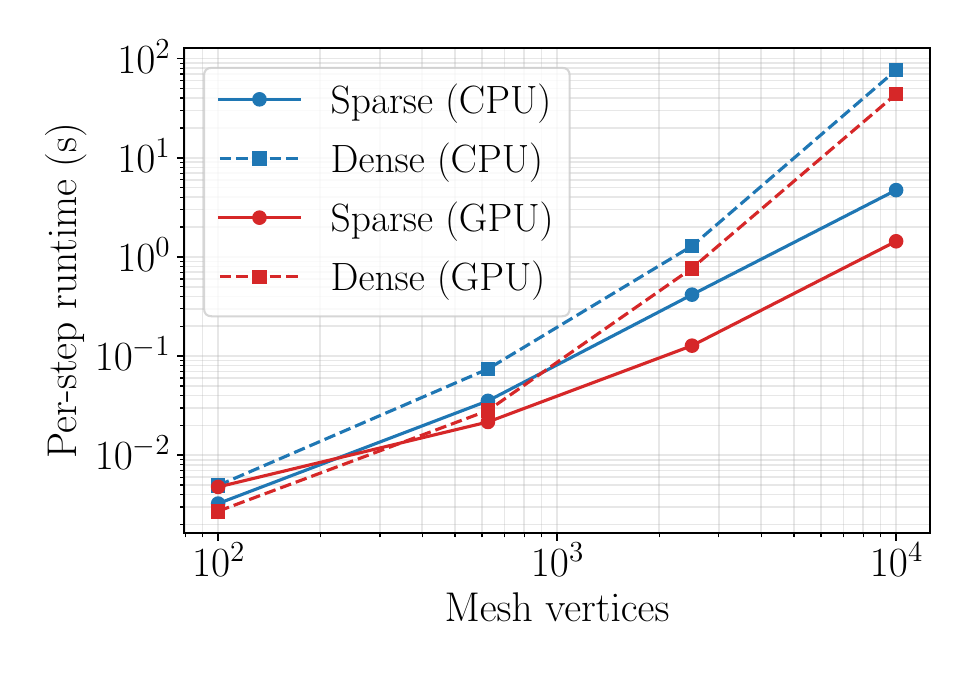}
    \caption{Runtime}
    \label{fig:benchmark_rspde_runtime_scaling}
  \end{subfigure}\hfill
  \begin{subfigure}[b]{0.33\textwidth}
    \centering
    \includegraphics[width=\textwidth]{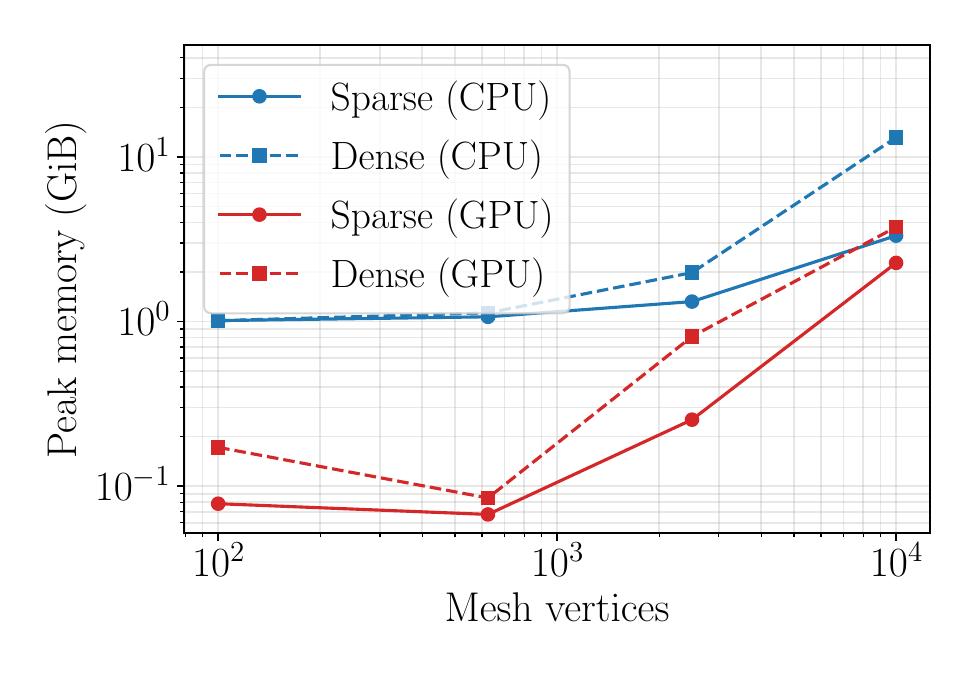}
    \caption{Peak memory use}
    \label{fig:benchmark_rspde_memory_scaling}
  \end{subfigure}\hfill
  \begin{subfigure}[b]{0.33\textwidth}
    \centering
    \includegraphics[width=\textwidth]{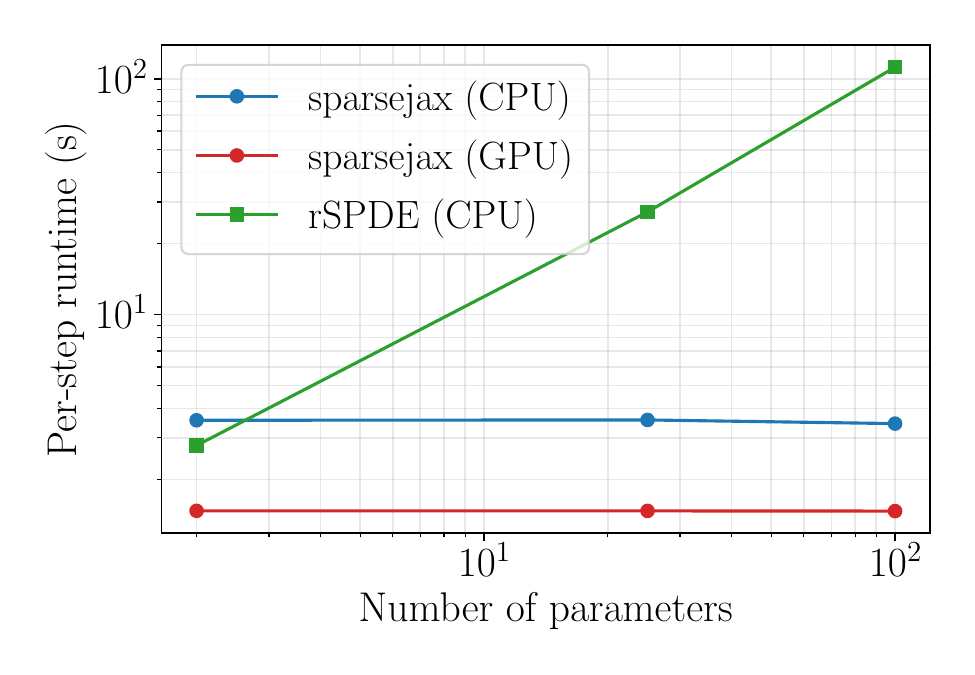}
    \caption{AD and finite-difference comparison}
    \label{fig:benchmark_rspde_fd}
  \end{subfigure}
  \caption{Runtime and peak memory use of the AD-based implementation, and runtime scaling with the number of parameters compared to a finite-difference implementation. Each measurement corresponds to a single optimization step.}
\end{figure}

Comparing the sparse implementation to doing all computations using dense matrices, Figure~\ref{fig:benchmark_rspde_runtime_scaling} shows that the runtime of the sparse implementation scales better than the dense as the problem size increases. For $10^{4}$ vertices, a single sparse optimization step is approximately $16$ times faster than a dense step on the CPU and $31$ times faster on the GPU. Figure~\ref{fig:benchmark_rspde_memory_scaling} also shows improvements in memory use on the CPU. However, on the GPU, the sparse and dense implementations have comparable memory. This is not a methodological limitation, but rather that the partial inverse algorithm has not yet been implemented for the GPU. Instead, the gradient of the log-determinant is currently computed by solving a dense linear system. GPU-accelerated partial inverse algorithms exist \citep{fattah_gpu_accelerated_2025}, and could be included to improve the scalability of the implementation.

Lastly, Figure~\ref{fig:benchmark_rspde_fd} compares the AD-based implementation with the finite-difference implementation from rSPDE. On the CPU, finite differences are slightly faster for few parameters, whereas AD scales substantially better as the number of parameters increases. This makes it more suitable for the fractional and non-stationary models considered here. Nevertheless, finite differences may be preferable in some settings. For example, the covariance-based rational approximation \citep{bolin2024covariance} in rSPDE integrates with R-INLA \citep{rue_approximate_2007}, which requires Hessian information. In this case, finite-differencing the gradient may be more attractive than differentiating the full AD-based gradient. Furthermore, Sparse AD reduces differentiation costs, but does not remove the scaling limitations associated with very large datasets or fine meshes. This is not specific to our model, but a general feature of SPDE-based approaches. In such settings, alternative approximations may be preferable. Nearest-neighbor methods approximate the Gaussian likelihood through low-dimensional conditional distributions, and have been applied to non-stationary spatial models \citep{risser2020bayesian}. Alternatively, compactly supported non-stationary covariance kernels yield sparse covariance matrices. When combined with distributed high-performance computing, they have scaled Gaussian process models to more than ten million observations \citep{risser_compactly-supported_2025,noack_gp2scale_2026}.

\section{Simulation study}\label{sec:Simulation study}

To assess the importance of non-stationarity and fractional smoothness in prediction, we study the parameter estimation and predictive performance of the proposed model. Key results are presented in this section, while additional experiments are provided in the Supplementary Material.

\subsection{Data-generating models}

\begin{figure}[pos=H]
  \centering
  \begin{subfigure}[b]{0.263\textwidth}
    \centering
    \includegraphics[width=\textwidth]{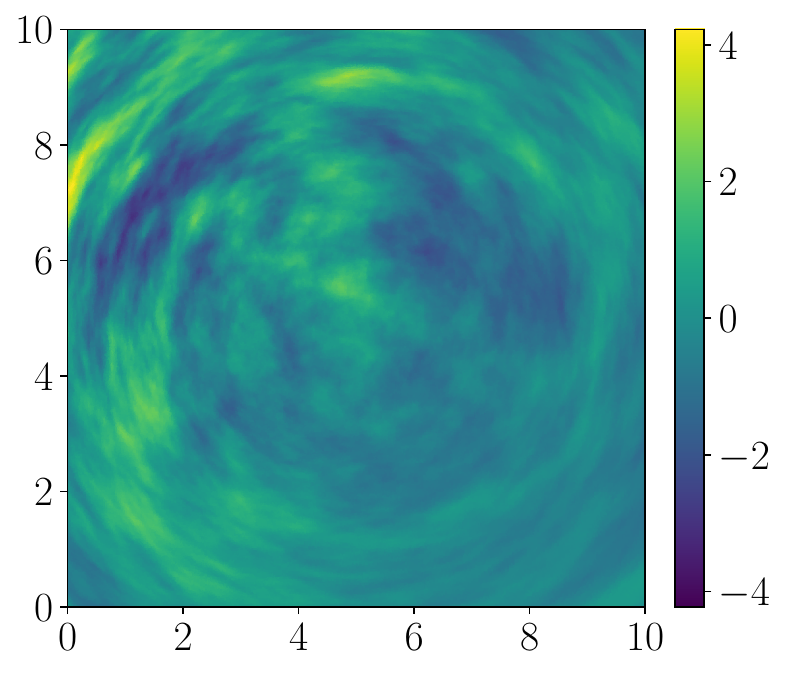}
    \caption{Realization}
    \label{fig:VizNonStatProcessRealization}
  \end{subfigure}\hfill
  \begin{subfigure}[b]{0.25\textwidth}
    \centering
    \includegraphics[width=\textwidth]{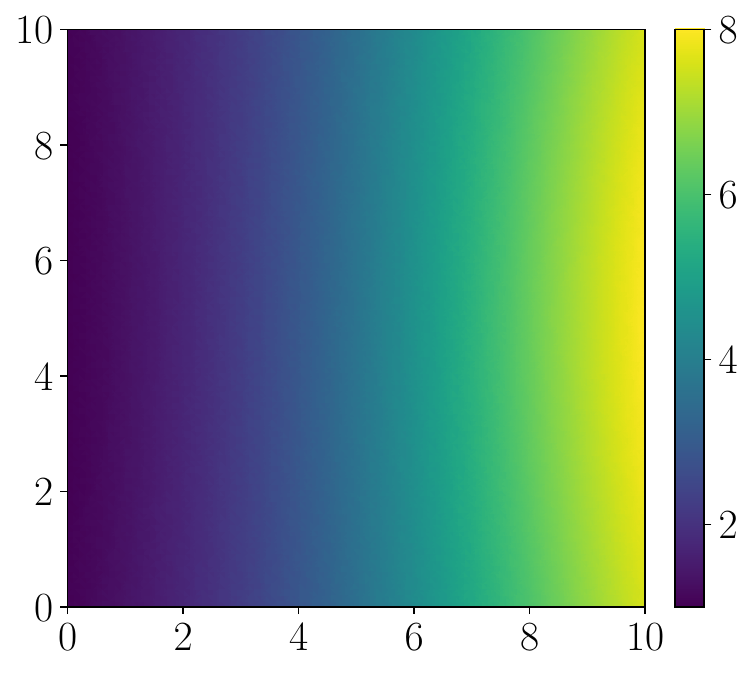}
    \caption{Correlation range}
    \label{fig:VizNonStatProcessRange}
  \end{subfigure}\hfill
  \begin{subfigure}[b]{0.26\textwidth}
    \centering
    \includegraphics[width=\textwidth]{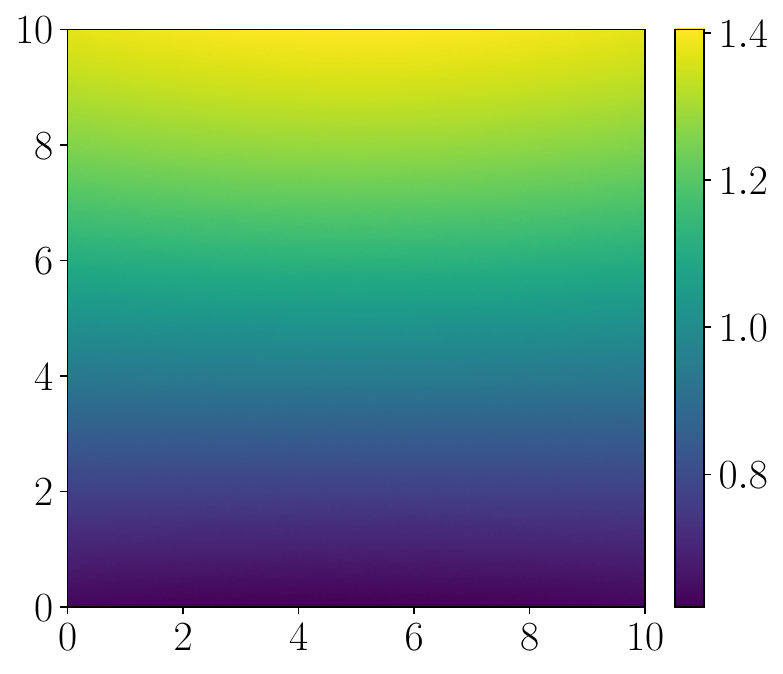}
    \caption{Marginal standard deviation}
    \label{fig:VizNonStatProcessSD}
  \end{subfigure}\hfill
  \begin{subfigure}[b]{0.225\textwidth} \centering
    \includegraphics[width=\textwidth]{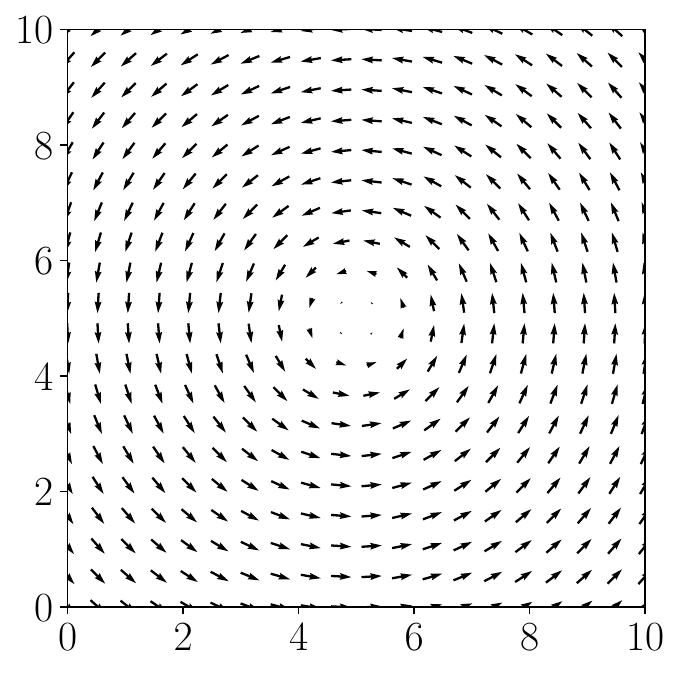}
    \caption{Anisotropy direction}
    \label{fig:VizNonStatProcessVecField}
  \end{subfigure}

  \caption{Non-stationary SPDE model with $\nu = 0.5$, used for generating evaluation data in the simulation study.}
  \label{fig:VizNonStatProcess}
\end{figure}

We use data generated from a stationary and a non-stationary model, both with the same smoothness parameter $\nu=0.5$. In the stationary model, $\kappa(\cdot)$, $\sigma(\cdot)$, $v_x(\cdot)$, and $v_y(\cdot)$ are constant over the domain. In the non-stationary model, $\kappa(\cdot)$, $\sigma(\cdot)$, $v_{x}(\cdot)$ and $v_{y}(\cdot)$ vary spatially, and are each represented using eight basis functions. The SPDE approach is used to generate data from a centered GRF on the domain $\mathcal{D} = [0,10]\times[0,10]$. We embed $\mathcal{D}$ in an extended domain obtained by adding $10$ units on each side. Constructing a finite-element discretization on the extended domain, we get triangular mesh with $13\,280$ vertices and $26\,318$ elements.  For the rational approximation, we use order $k = 2$. Figure \ref{fig:VizNonStatProcess} shows a realization from the non-stationary field. It has a correlation range that increases when moving from the left to the right of the domain, and a marginal standard deviation that is higher in the top part of the domain. The anisotropy is selected to rotate around the center of the domain.

\subsection{Candidate models}\label{sub:Simulation study Models and evaluation}

We compare four classes of candidate models. All are based on the hierarchical specification from Section \ref{sec:Spatial regression model}, but where the fixed effects are omitted. The first class is a non-fractional and stationary model (NF-S), which assumes fixed smoothness $\nu = 1$ and stationary parameters. The second class is a non-fractional and non-stationary model (NF-NS), which allows $\kappa(\cdot)$, $\sigma(\cdot)$, $v_x(\cdot)$, and $v_y(\cdot)$ to vary spatially. We also consider a fractional and stationary model (F-S), and our proposed fractional and non-stationary model (F-NS). Here the smoothness $\nu$ is estimated using observations. Each model is discretized using the same mesh as in the data-generating step, and the fractional models employ an order $k = 2$ rational approximation.

For parameter estimation, stationary parameters are initialized at random values within $\pm 50\%$ of the true value, while the coefficients governing non-stationarity are initialized to zero. The hyperparameters $C_{\rho}$, $C_{\sigma}$ and $C_{\sigma_{N}}$ for the prior medians are set to the true value, while we let $C_{a} = 4$. When estimating smoothness, the prior for $\nu$ is specified with $C_{\nu} = 1.0$, $C_{\nu, \mathrm{HPD}} = 1.8$ and $\nu_{\max} = 2.0$. For the non-stationary models, we consider either $8$ or $24$ basis functions per non-stationary parameter. To reduce the number of candidates, all parameters use the same number of basis functions. Every non-stationary parameter is also assumed to use the same hyperparameter $C_{\mathrm{NS}}$ for controlling the flexibility, where we consider $C_{\mathrm{NS}} \in \{ 2, 5, 10 \}$, corresponding to strict, moderate and small penalization. For ease of notation, we denote a model with $8$ basis functions and a $C_{\mathrm{NS}}=2$ as B$8$-C$2$, and introduce similar abbreviations for the other combinations.

\subsection{Evaluation}

For the evaluation, we generate $25$ independent datasets from each data-generating model. For each dataset $i=1, \ldots, 25$, we sample $1000$ locations $\boldsymbol{s}_{i,1}, \ldots, \boldsymbol{s}_{i,1000}$ uniformly from the interior of the spatial domain and interpolate the latent field at these locations. Observation error is added as Gaussian noise with marginal variance $\sigma_\mathrm{N}^{2} = 0.1$, yielding observations $y_{i,1}, \ldots, y_{i,1000}$. To investigate the effect of data density, we fit the candidate models using subsets of the $1000$ observed locations with size $125$, $500$ and $1000$. For each subset size, parameter bias is evaluated by averaging the parameter estimates over the $25$ simulated datasets. Predictions are made on a regular $100 \times 100$ grid covering $\mathcal{D}$. For each simulation and each subset size, the estimated parameters are used to compute the posterior distribution of the latent field at all prediction locations. Interpolating the values of the true latent field at the prediction locations, we compare them to the predicted posterior means and standard deviations by computing the average RMSE and CRPS across the $25$ simulations.

\subsection{Impact of basis functions and penalties}

We begin by examining how model performance depends on the number of basis functions and the choice of penalty parameters. To isolate these effects, we focus on the stationary F-S model and the different parameterizations of the non-stationary F-NS model. All models use a fixed smoothness parameter $\nu = 0.5$. Recall that the true non-stationary data were generated by an F-NS model with $8$ basis functions. Models using $8$ or $24$ basis functions should therefore in principle both be capable of representing the spatial variation of the data-generating model.

\begin{figure}[pos=H]
  \centering
  \includegraphics[width=0.8\textwidth]{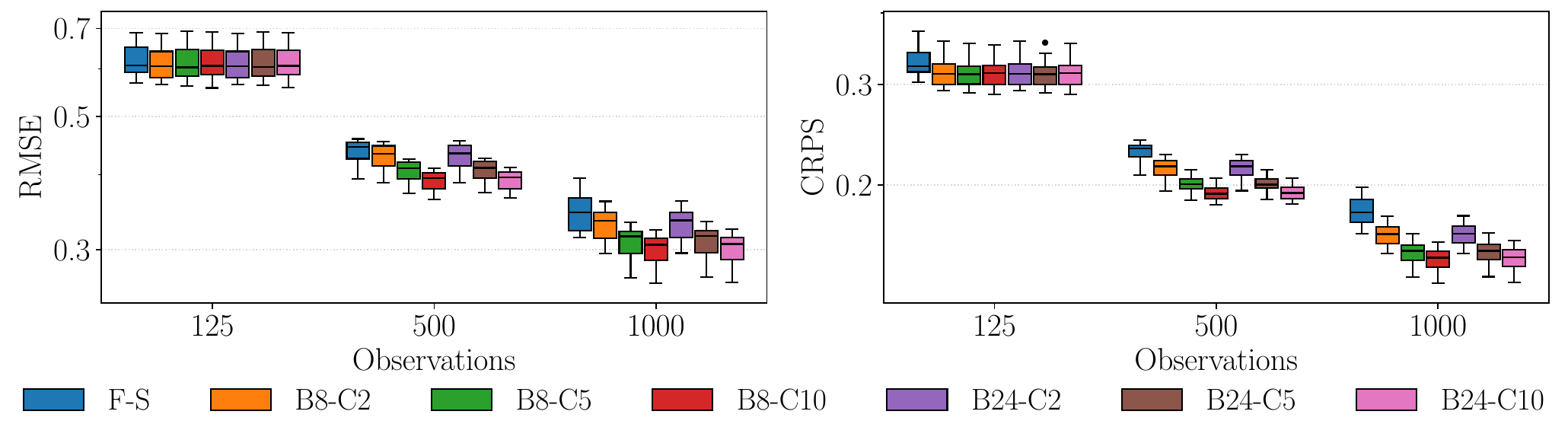}
  \caption{Log-scaled RMSE and CRPS scores for data generated by a non-stationary model with smoothness $\nu = 0.5$. Compares fractional models with different numbers of basis functions and penalty parameters for the non-stationary components, and results are plotted as functions of the number of observations.}
  \label{fig:viz_non_stat_fixed_nu_rmse_crps}
\end{figure}

Fitting the models to observations from the non-stationary data-generating model, Figure \ref{fig:viz_non_stat_fixed_nu_rmse_crps} shows RMSE and CRPS as functions of the number of observations. Using only $125$ observations, all models have a similar performance, except that the stationary model has a slightly worse CRPS. When at least $500$ observations are available, the more flexible models outperform the ones with a higher penalization. This indicates that increased flexibility allows the model to capture non-stationarity when sufficient data are available. Comparing the models with $8$ and $24$ basis functions, we find the models to perform similarly. This suggests that overspecifying the non-stationary structure by using more basis functions does not necessarily degrade performance. There does also not appear to be an issue of the flexible model overfitting when the sample size is small. However, this does not rule out overfitting in general, particularly for highly parameterized models with weak penalization.

\subsection{Estimating the smoothness}

To assess the effect of estimating the smoothness parameter, we no longer fix $\nu = 0.5$. Instead, we estimate $\nu$ using the AD approach and compare the fractional models F-S and F-NS to their non-fractional counterparts NF-S and NF-NS. For the non-stationary models F-NS and NF-NS, we use $8$ basis functions and set $C_{\mathrm{NS}} = 10$, which was found to provide the best performance when at least $500$ observations are available.

\begin{figure}[pos=H]
  \centering
  \includegraphics[width=0.7\textwidth]{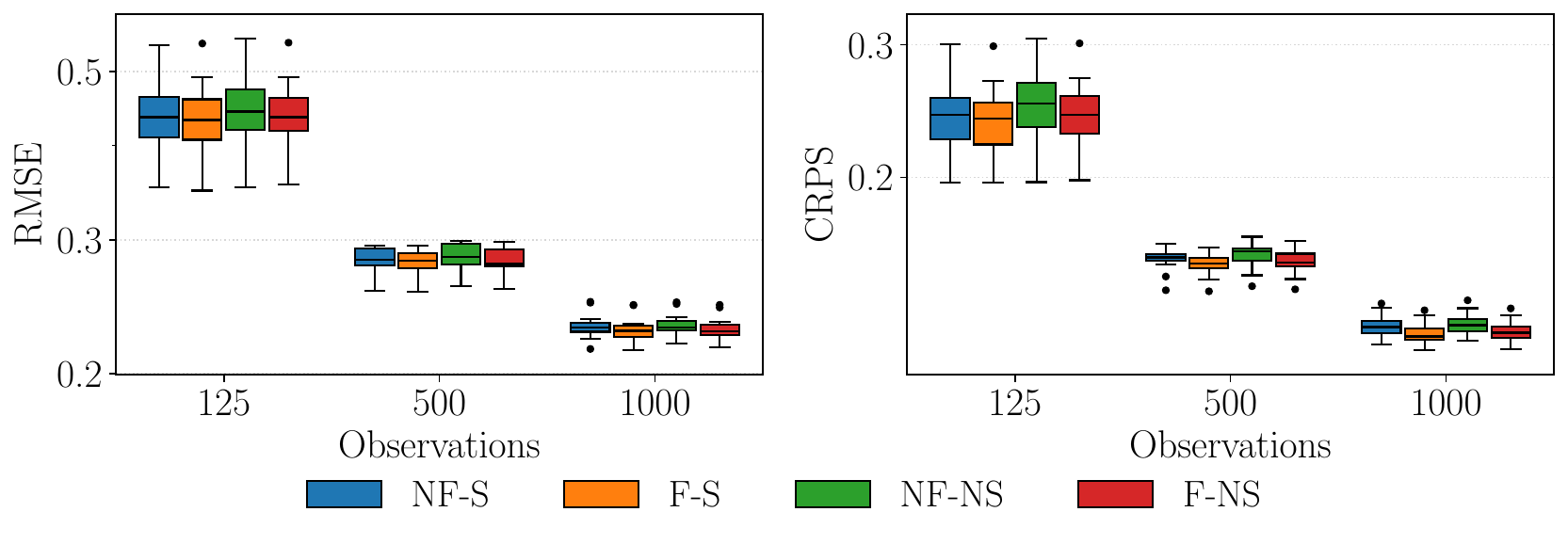}
  \caption{Log-scaled RMSE and CRPS scores for data generated by a stationary model with smoothness $\nu = 0.5$. Compares fractional and non-fractional model to simpler non-fractional and stationary alternatives, and results are plotted as functions of the number of observations.}
  \label{fig:viz_frac_stat_rmse_crps}
\end{figure}

Fitting to observations from the stationary data-generating model, Figure \ref{fig:viz_frac_stat_rmse_crps} shows predictive RMSE and CRPS as functions of the number of observations. The four model performs very similarly across all observation sizes. Except for slightly higher CRPS scores for $125$ observations, the non-stationary models does not overfit to the stationary data. Estimating the smoothness does only give a small improvement in predictive performance. Table S.1 in the Supplementary Material shows parameter estimates based on $500$ observations. Both F-S and F-NS estimate the smoothness parameter with relatively small bias. This suggests that even when smoothness is estimated accurately, the resulting gains in predictive performance are limited in this setting.

\begin{figure}[pos=H]
   \centering
   \includegraphics[width=0.7\textwidth]{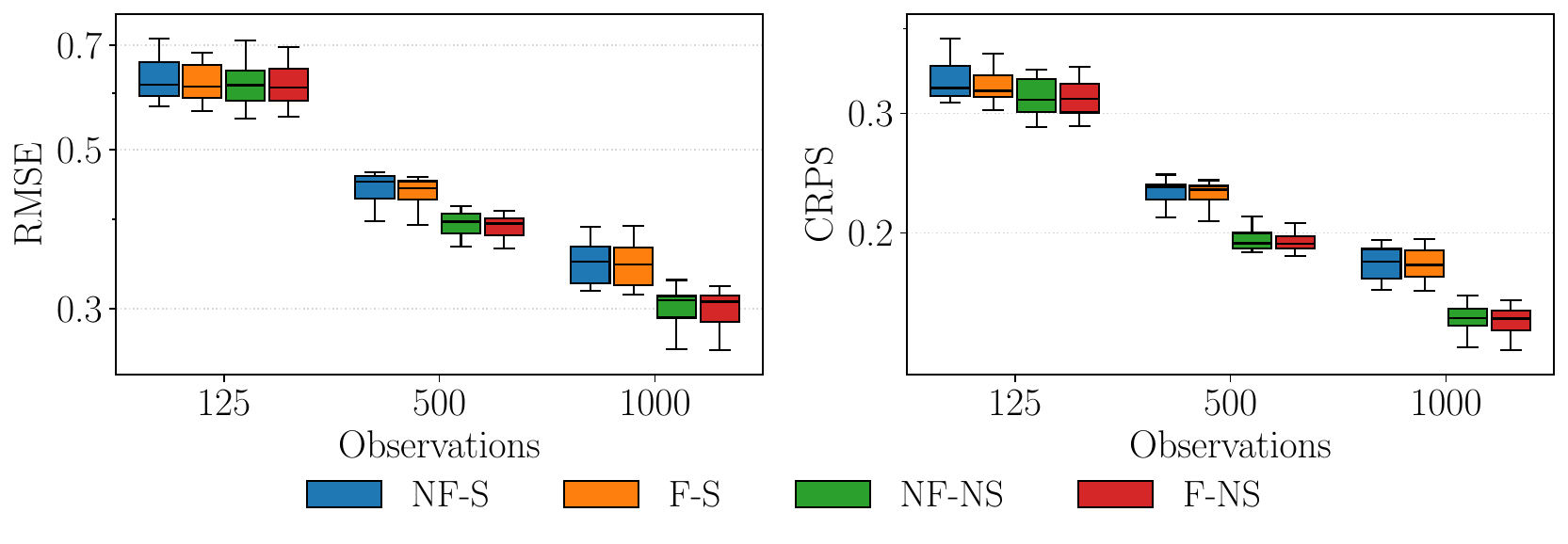}
   \caption{Log-scaled RMSE and CRPS scores for data generated by a non-stationary model with smoothness $\nu = 0.5$. Compares fractional and non-fractional model to simpler non-fractional and stationary alternatives, and results are plotted as functions of the number of observations.}
   \label{fig:viz_frac_non_stat_rmse_crps}
 \end{figure}

Next, we fit the candidate models to data generated from the non-stationary process. The parameter estimates from Table S.2 of the Supplementary Material show that the fractional models again estimate the smoothness parameter with relatively low bias. However, the F-NS estimates the smoothness to be slightly higher than the true value. Predictive RMSE and CRPS are shown in Figure \ref{fig:viz_frac_non_stat_rmse_crps}. As in the stationary case, differences between fractional and non-fractional models are modest. In contrast, the non-stationary models consistently outperform the stationary models when at least $500$ observations are available, highlighting the importance of modeling non-stationarity when sufficient data are present.

\section{Case studies}\label{sec:Real-world application}

We evaluate the added value of the spatial regression model through two case studies. 
\subsection{Ocean salinity}

\begin{figure}[pos=H]
  \centering
  \begin{subfigure}[b]{0.37\textwidth}
    \centering
    \includegraphics[width=\textwidth]{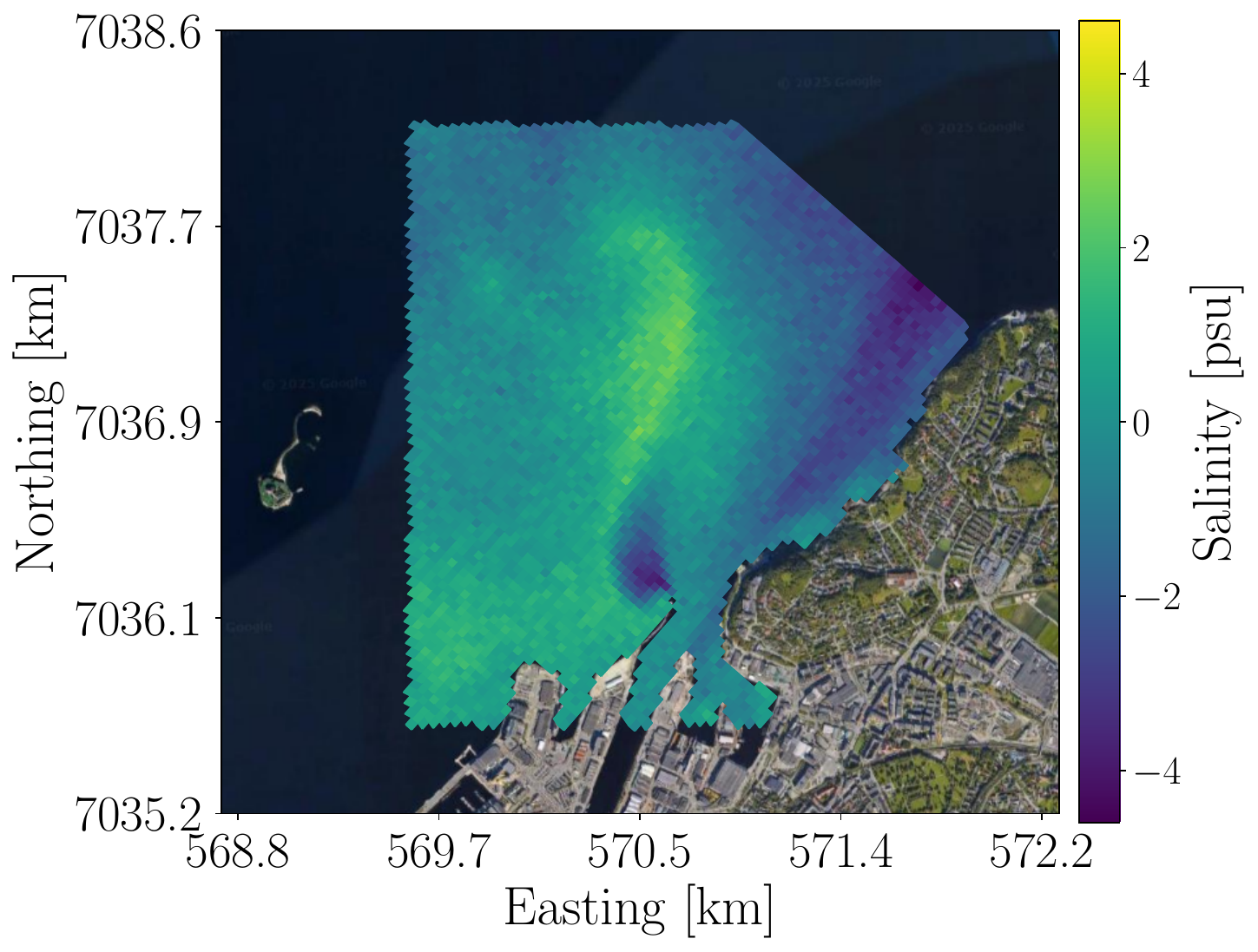}
    \caption{Centered salinity anomalies}
    \label{fig:SINMOD_SatelliteAndRealization}
  \end{subfigure}%
  \hspace{0.005\textwidth} \begin{subfigure}[b]{0.28\textwidth} \centering \includegraphics[width=\textwidth]{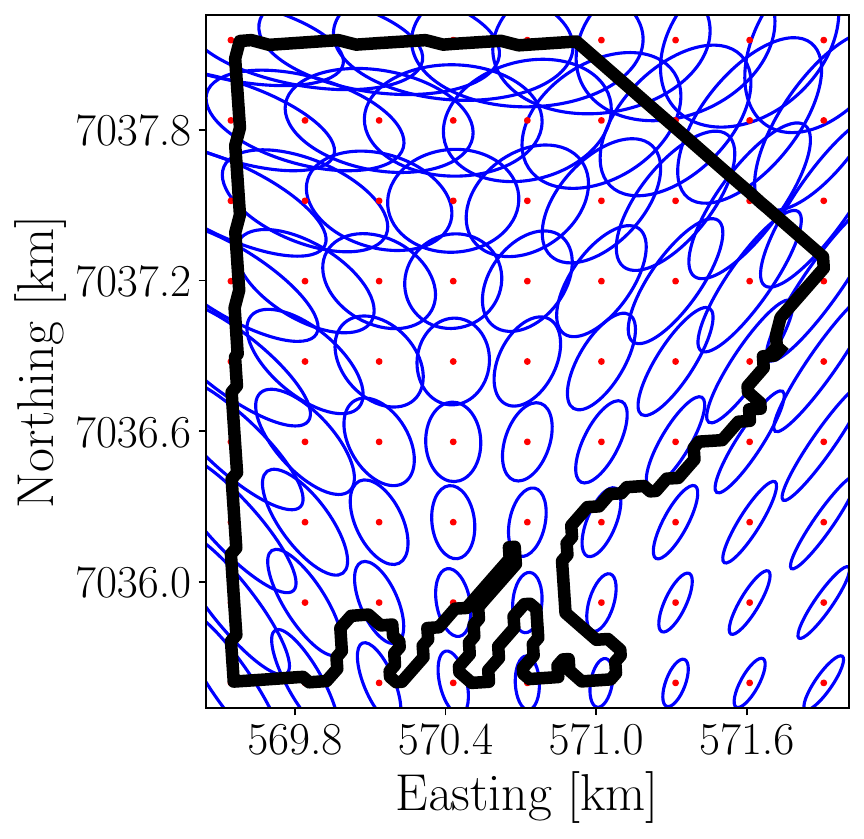}
    \caption{Contours of $0.7$ correlation}
    \label{fig:SINMODEstimCorrRange}
  \end{subfigure}%
  % \hspace{-0.050\textwidth}
  \begin{subfigure}[b]{0.315\textwidth}
    \centering
    \includegraphics[width=\textwidth]{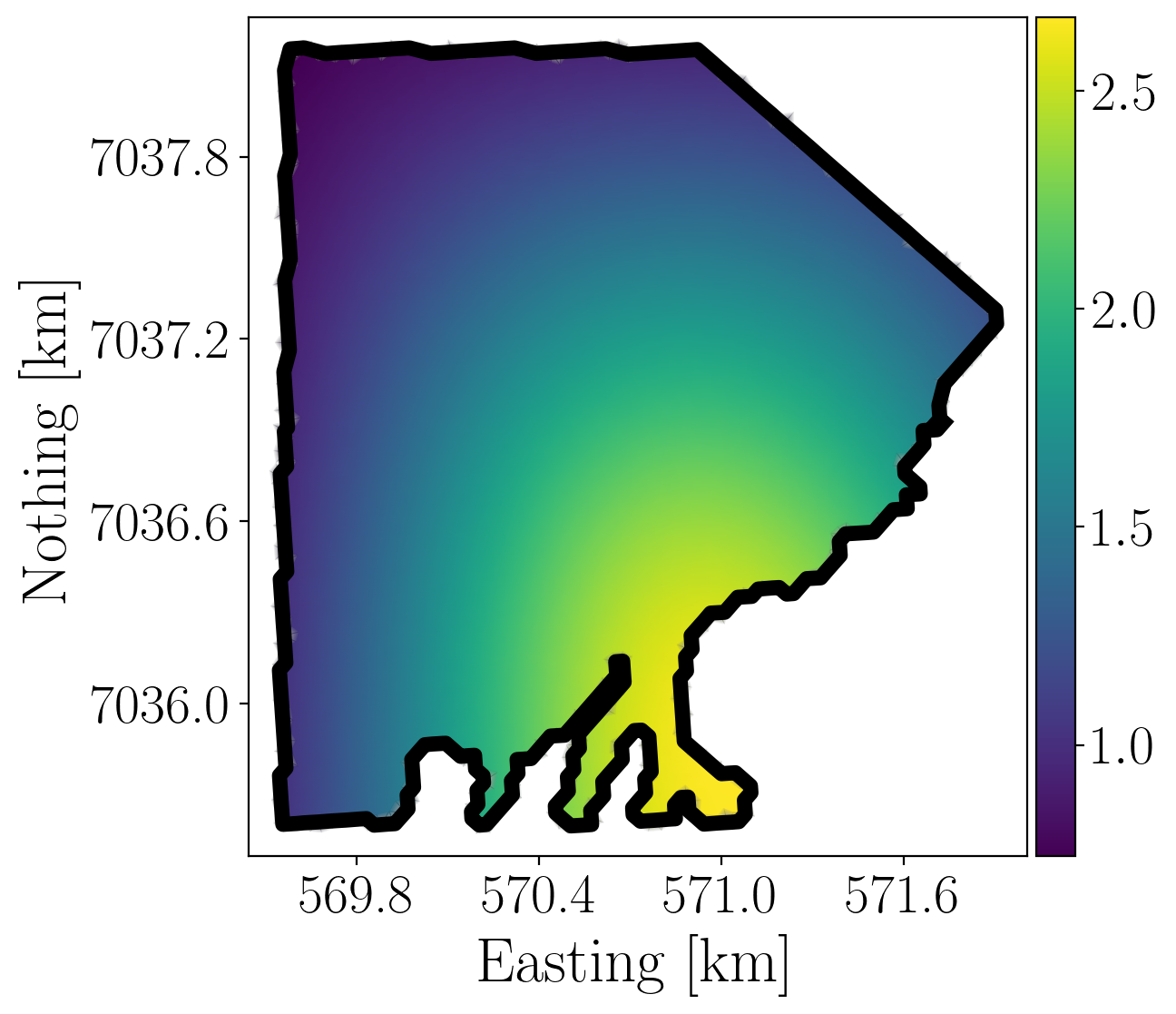}
    \caption{Marginal standard deviation}
    \label{fig:SINMODEstimStdDev}
  \end{subfigure}
  \caption{True sample from the salinity dataset, and estimated non-stationary correlation and marginal standard deviation when fitting F-NS to all available data.}
  \label{fig:SINMODEstimCorrRangeAndStdDev}
\end{figure}

The first dataset consists of salinity values in Trondheimsfjorden, and comes from the SINMOD model developed by SINTEF Ocean \citep{sintef1987, SLAGSTAD20051}. This is a numerical ocean model that simulates ocean dynamics based on atmospheric forces, freshwater input and tides \citep{berild2023spatiallyvaryinganisotropygaussian}. SINMOD reports the salinity in \textit{practical salinity units} (psu), which describe the grams of salt per kilogram of seawater \citep{webb_introduction_nodate}. The data are provided on a grid with a spatial resolution of $32 \,\mathrm{m}$, using a polar stereographic projection centered on the North Pole. For this analysis, the coordinates are transformed to the \textit{Universal Transverse Mercator} (UTM) zone $32$V projection. We examine the hourly salinity on the 2nd of May 2020, simulated at $4196$ locations and at a depth of $0.5\,\mathrm{m}$. The $24$ hourly fields are treated as independent spatial realizations, and we center them by subtracting the location-wise mean across realizations. To replicate measurement error, we add Gaussian noise with a marginal standard deviation of $\sigma_{\mathrm{N}} = 0.01\,\mathrm{psu}$. This corresponds to a signal-to-noise ratio of approximately $1\%$. Figure \ref{fig:SINMOD_SatelliteAndRealization} shows a realization of the centered salinity anomalies, with a satellite image of the region placed underneath for reference. The realization shows notable spatial variation near the mouth of the Nidelva River, where freshwater mixes with the salty seawater.

We continue using the same four model classes as in the simulation study. Based on the results of the simulation study, we found that using $8$ basis functions and $C_{\mathrm{NS}}=10$ performed well for the non-stationary models, and we therefore adopt this configuration for the analysis. To guide the choice of priors, we first perform a preliminary analysis of the full dataset using a NF-S model with diffuse priors. We estimate $\hat{\rho}_{0} = 1.2\,\mathrm{km}$ and $\hat{\sigma}_{0} = 1\,\mathrm{psu}$, and use these to set the hyperparameters $C_{\rho}$ and $C_{\sigma}$. From the added noise we set $C_{\sigma_{\mathrm{N}}} = 0.1$, and use $C_{a} = 4$ for the anisotropy. We have little prior knowledge about the smoothness of the data, as previous analyses have used a fixed $\nu = 1$ \citep{lilleborge2021bivariate, berild2023spatiallyvaryinganisotropygaussian, berild2024nonstationaryspatiotemporalmodelingusing}. Therefore, we choose the weakly informative $C_{\nu} = 1.0$, $C_{\nu, \mathrm{HPD}} = 1.8$ and $\nu_{\max} = 2.0$. Discretizing the models, the domain is first expanded by $2.4\,\mathrm{km}$ on all sides, and we use fmesher \citep{lindgrenFmesher} to construct a triangulation with $10654$ vertices and $21143$ elements.

First, we examine whether the non-stationary models can capture any non-stationary behavior that aligns with known physical processes in the domain. Fitting the F-NS model to the full dataset, Figure \ref{fig:SINMODEstimCorrRangeAndStdDev} shows the estimated iso-correlation curves and marginal standard deviation. At the bottom of the domain, in the area near where Nidelva flows into the fjord, the model estimates a low correlation range and high marginal standard deviation. This could be explained by the mixing of freshwater and saltwater, which increases variability and reduces the spatial range compared to areas further out in the fjord. Moreover, the estimated anisotropy indicates stronger correlation along the outflow direction, consistent with the movement of the water.

Next, we evaluate the predictive performance of the models, which requires specifying training and test datasets. Because the data are observed on a dense spatial grid, randomly selecting training and test locations across the domain would place many test locations close to observed ones, causing the evaluation to be dominated by short-range interpolation and observation noise. Instead, we adopt an evaluation strategy in which one realization is held out at a time, and model parameters are estimated using observations from all locations in the remaining realizations. Predictions are then made for the held-out realization, conditioning on observations at only a randomly selected subset of locations, with the number of observed locations varied between $10$ and $250$. Predictive accuracy at the unobserved locations is assessed using RMSE and CRPS. This procedure is repeated for each of the $24$ realizations, allowing us to compute empirical means and standard deviations of the scores. Although this setup is less common in practice, it arises in several real-world applications, such as the autonomous underwater vehicle studies \citet{lilleborge2021bivariate, berild2023spatiallyvaryinganisotropygaussian, berild2024nonstationaryspatiotemporalmodelingusing}.

\begin{figure}[pos=H]
  \centering
  \includegraphics[width=0.7\textwidth]{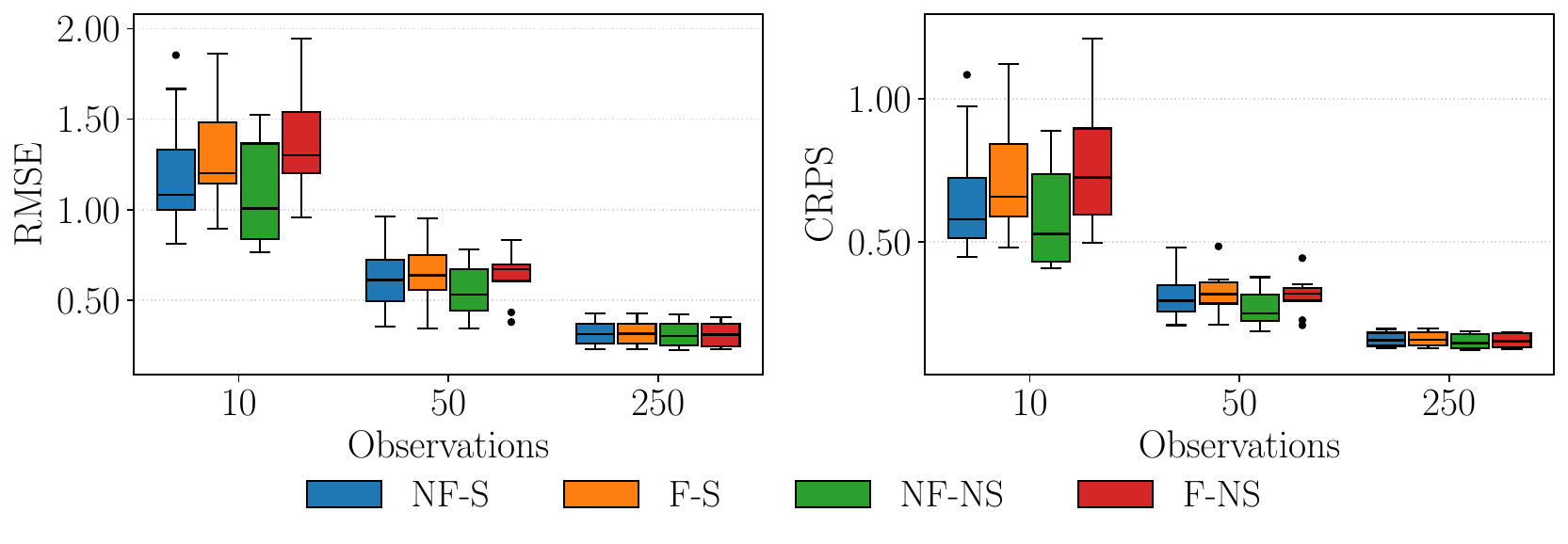}
  \caption{Log-scaled RMSE and CRPS scores for the evaluation using the salinity dataset. Results plotted as functions of the number of observations.}
  \label{fig:SINMODRMSEAndCRPSAsFunctionOfNObs}
\end{figure}

Figure \ref{fig:SINMODRMSEAndCRPSAsFunctionOfNObs} shows RMSE and CRPS as functions of the number of observations in the held-out realization. The non-stationary models consistently outperform the stationary ones. For $250$ observations, the performance gap narrows as most test locations lie close to observed ones. In contrast, fractional and non-fractional models show similar predictive accuracy. From Table S.7 in the Supplementary Material, the estimated fractional smoothness parameters are slightly below one, suggesting near non-fractional behavior.

\subsection{Precipitation data}
The second dataset contains climate reanalysis data describing average summer precipitation over the conterminous U.S. Climate reanalysis combines a climate model with observations to reconstruct past climate patterns \citep{CopernicusClimateReanalysis}. This dataset was generated by the Experimental Climate Prediction Center Regional Spectral Model as part of the North American Regional Climate Change Assessment Program. It covers a 26-year period from 1979 to 2004, and includes average precipitation values measured in centimeters at 4112 locations given in geographic coordinates across the domain. We select this dataset because prior studies have indicated the presence of a fractional smoothness. In particular, \citet{genton2015cross} found that an exponential covariance model may be appropriate in parts of the domain, while \citet{bolin2020rational} estimated a smoothness parameter of $\nu = 0.44$.

\begin{figure}[pos=H]
  \centering
  \begin{subfigure}[b]{0.50\textwidth}
    \centering
    \includegraphics[width=\textwidth]{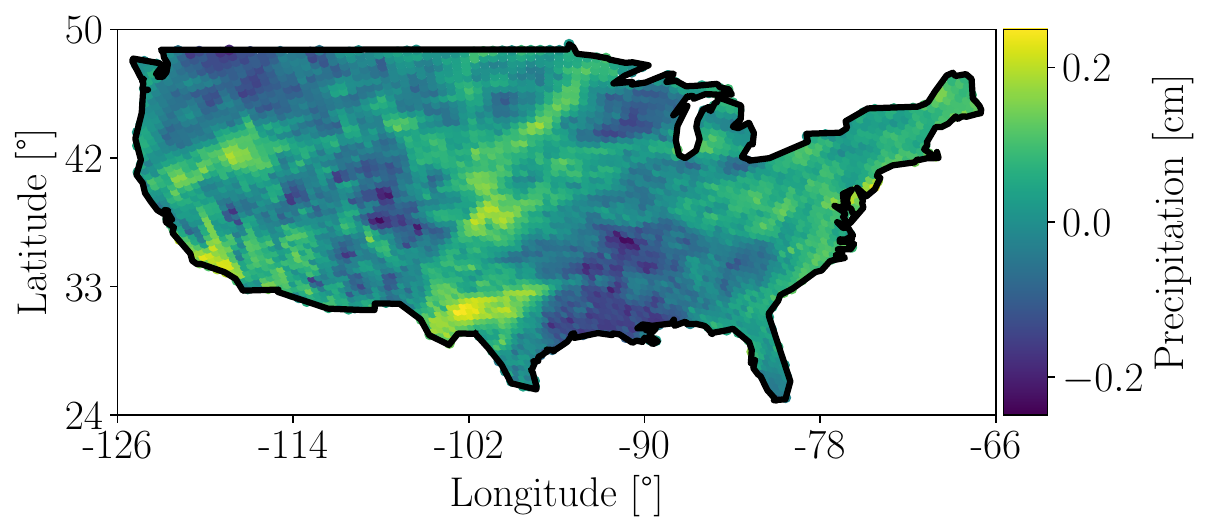}
    \caption{Standardized and centered precipitation anomalies.}
    \label{fig:PercipObs1979}
  \end{subfigure}%
  \begin{subfigure}[b]{0.48\textwidth}
    \centering
    \includegraphics[width=\textwidth]{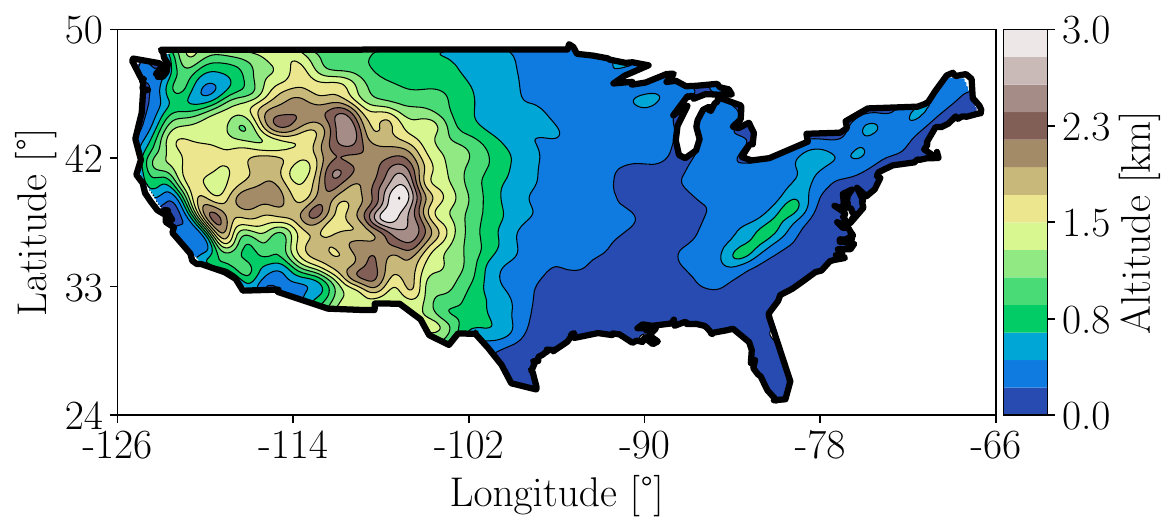}
    \caption{Altitude over the conterminous U.S.}
    \label{fig:PercipAltitude}
  \end{subfigure}
  \caption{Precipitation anomalies and altitude over the spatial domain. Data are from $1979$.}
  \label{fig:PercipObsAndAltitude}
\end{figure}

To prepare the precipitation data for spatial modeling, a cube-root transformation is applied so that a Gaussian distribution can be assumed \citep{genton2015cross}. Each realization of the transformed data is also centered by subtracting the mean over the realizations. Figure \ref{fig:PercipObs1979} shows the processed anomalies for $1979$. The spatial correlation range appears shorter in the southwest and increases toward the east. The pattern is not unique to this specific year, but appears in most realizations. A possible explanation is the differences in altitude over the domain, as shown in Figure \ref{fig:PercipAltitude} that the Rocky Mountains lead to the western region being much more mountainous.

Specifying the stationary priors, we again perform a preliminary analysis using a NF-S model. The hyperparameters $C_{\rho}$, $C_{\sigma}$, and $C_{\sigma_{\mathrm{N}}}$ are set based on the estimated values $\hat{\rho}_{0} = 5^{\circ}$, $\hat{\sigma}_{0}=0.1\,\mathrm{cm}$, and $\hat{\sigma}_{\mathrm{N}}=0.02\,\mathrm{cm}$, while $C_{a}$ is set to $4$. Choosing $\pi(\nu)$, we base the prior on previous studies and adopt a more informative specification than for the salinity data. Specifically, we let $C_{\nu} = 0.5$, $C_{\nu, \mathrm{HPD}}=0.8$, and $\nu_{\max} = 1.0$. Discretizing the model, we expand the domain by $20^{\circ}$ on each side and triangulate it to obtain $9555$ vertices and $18939$ elements. 

\begin{figure}[pos=H]
        \centering
        \begin{subfigure}[b]{0.45\textwidth}
                \centering
                \includegraphics[width=\textwidth]{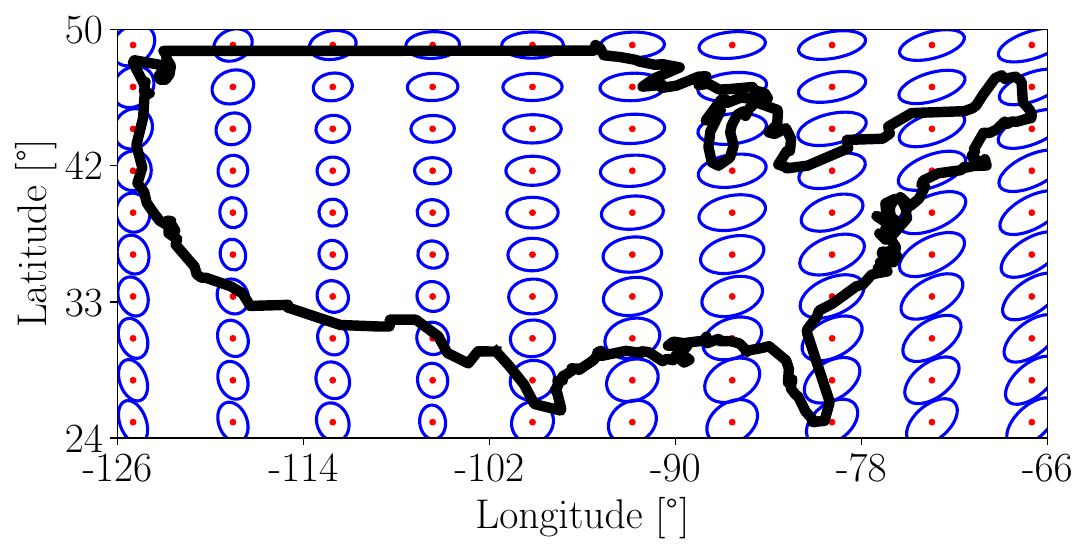}
                \caption{Contours of $0.7$ correlation}
                \label{fig:PrecipIsoCorr}
        \end{subfigure}
        \begin{subfigure}[b]{0.50\textwidth}
                \centering
                \includegraphics[width=\textwidth]{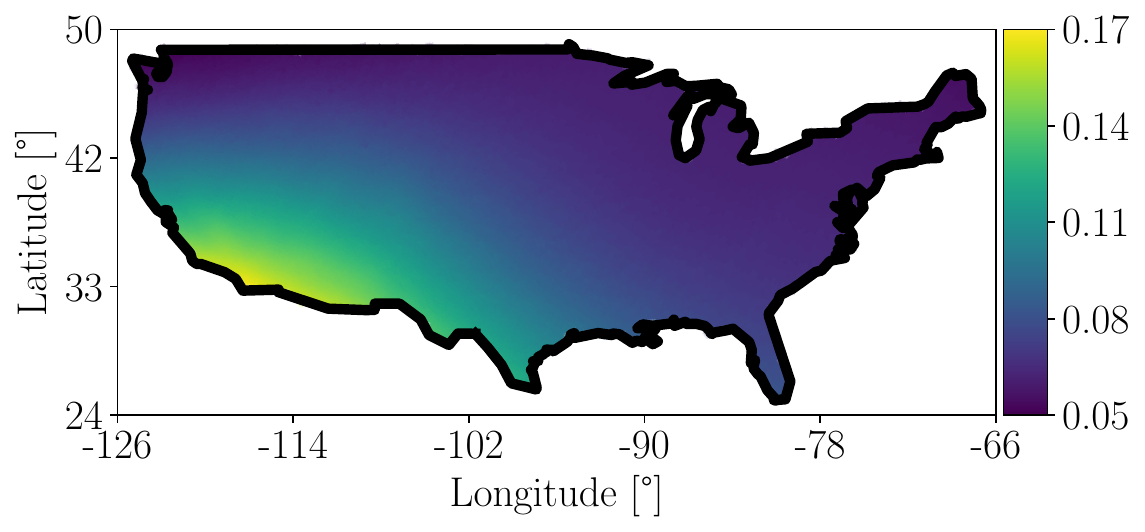}
                \caption{Marginal standard deviation}
                \label{fig:PrecipMargSD}
        \end{subfigure}%
    \caption{Estimates of non-stationary correlation and marginal standard deviation when fitting F-NS to all available precipitation data.}
    \label{fig:PrecipIsoCorrAndMargSD}
\end{figure}

As for the salinity data, we begin our analysis by fitting the F-NS to all available data. The estimated iso-correlation curves and marginal standard deviation are shown in Figure \ref{fig:PrecipIsoCorrAndMargSD}. In the western part of the domain, the model estimates a shorter correlation range. In the east, a slight non-stationary anisotropic pattern emerges, with stronger correlation along the longitudinal axis. This may be a result of the Gulf Stream affecting the eastern climate. Figure \ref{fig:PrecipMargSD} also shows a higher marginal standard deviation in the southwest. This could be due to the complex terrain of the Rocky Mountains or Sierra Nevada, which introduces localized weather phenomena. 

\begin{figure}[pos=H]
  \centering
  \includegraphics[width=0.7\textwidth]{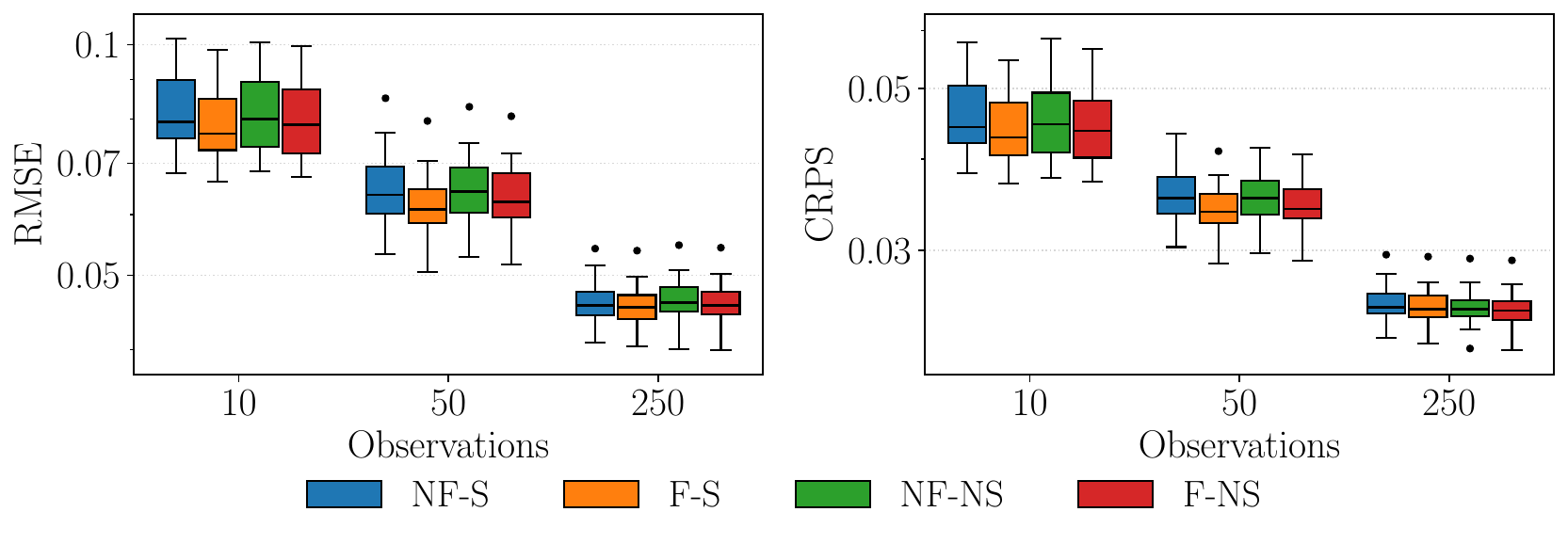}
  \caption{Log-scaled RMSE and CRPS scores for the evaluation using the precipitation dataset. Results plotted as functions of the number of observations.}
  \label{fig:PercipRMSEandCRPSAsFunctionOfNObs}
\end{figure}

Fitting the candidate models to the precipitation data by following the same procedure as for the salinity data, Figure \ref{fig:PercipRMSEandCRPSAsFunctionOfNObs} presents the RMSE and CRPS scores as functions of the number of observations. Unlike the salinity data, there is a clearer difference between the fractional and non-fractional models. Here we see the fractional models yield better predictions across all numbers of observations. Table S.8 of the Supplementary Material shows that the smoothness estimates for the fractional models are close to $0.4$, and therefore consistent with the $\nu = 0.44$ reported by \citet{bolin2020rational}. 

Another key difference from the salinity data is that the non-stationary models no longer outperform the stationary ones. This is surprising, as the differences in elevation led us to expect a difference in rainfall between the eastern and western regions. However, our findings align with those of \citet{fuglstad2015doesnonstationaryspatialdata}. When analyzing a similar dataset using a NF-NS model, they found that a stationary model performed approximately as well as the non-stationary. They argued that this was because the non-stationarity was incorporated into the wrong part of the model, and that $\sigma_{\mathrm{N}}$ should vary across the domain.

\section{Discussion}\label{seq:Discussion}
This paper extends the class of non-stationary SPDE models by combining non-stationary anisotropy with a fractional smoothness. Specifically, we combine the framework for non-stationary anisotropy developed by \citet{fuglstad2015exploring, fuglstad2015doesnonstationaryspatialdata}, the parameterization of the diffusion matrix by \citet{llamazareselias2024parameterization}, and the rational approximation from \citet{bolin2020rational}. The increased complexity comes with a higher risk of overfitting. To address this, we develop priors for the stationary and non-stationary components of the model. For the parameters specifying the stationary properties, we adopt PC-like priors, while the non-stationary terms are controlled using spline-based penalties from \citet{fuglstad2015doesnonstationaryspatialdata}. We also suggest an approach to setting these priors, and describe an interpretable approach for selecting penalties for the non-stationarity.

We further demonstrate the use of AD for parameter estimation in non-stationary fractional SPDEs. The proposed methodology provides a practical and extensible framework for gradient-based inference. By combining AD with sparse matrix operations, we compute gradients without deriving analytical expressions or relying on numerical differentiation. This is particularly valuable for non-stationary models, where the number of parameters can be large. The approach uses a masking strategy that preserves sparsity during backpropagation while still yielding exact gradients. Our implementation in sparsejax supports both CPU and GPU backends. The benchmarks show that it scales better than finite-difference approaches as the parameter dimension increases, and more favorably than dense AD for finer discretizations.

The simulation study demonstrates that the proposed model can reliably detect both fractional smoothness and non-stationarity when sufficient data are available. With at least $500$ observations from a single realization, the fractional models estimate the smoothness parameter with a relatively low bias, and the non-stationary models outperform their stationary counterparts under non-stationary data-generating processes. For the non-stationary field, weaker penalization of the non-stationary components improved performance without degrading results for smaller sample sizes. However, this conclusion may depend on the specific non-stationary pattern and observation locations, and selecting an appropriate penalization level remains an open challenge. The predictive gains from estimating fractional smoothness are modest, as non-fractional models can partially compensate by adjusting other parameters such as the correlation range.

The two case studies show that the relative importance of fractional smoothness and non-stationarity depends on the application. For the salinity data, the non-stationary models outperform the stationary models. The estimated non-stationary patterns also reflect known physical processes, such as shorter correlation ranges near the river mouth, where freshwater mixes with seawater. However, the estimated smoothness is close to one, so the fractional and non-fractional models perform similarly. For the precipitation data, the situation is reversed. The smoothness is estimated to be around 0.4, and the fractional models outperform the non-fractional models across all sample sizes. However, the non-stationary models offer no improvements over the stationary models. This is consistent with \citet{fuglstad2015doesnonstationaryspatialdata}, who for a similar dataset argued that non-stationarity may be more relevant in the measurement noise than in the spatial covariance. Together, these results show that the proposed model can help identify which type of flexibility is most relevant for a given dataset. We did not observe issues with estimating smoothness jointly with the other covariance parameters in the simulation study and the case studies.

Based on our findings, we identify three key use cases for the proposed model. First, it serves as a tool for exploratory analysis, helping practitioners determine whether added complexity in the form of non-stationarity or fractional smoothness is needed. If the analysis reveals little evidence for such features, a simpler model can be used with greater confidence. Second, the model is valuable when the goal is to extract interpretable parameters that relate to underlying physical processes. Non-stationary anisotropy can reveal patterns such as water currents or terrain effects, while the estimated smoothness characterizes the local regularity of the field. Third, the model can improve predictive performance when there is reason to expect fractional or non-stationary behavior, as demonstrated by the case studies.

For future work, there are several promising directions. A current limitation of the computational framework is that the partial-inverse algorithm is not yet supported on the GPU, leaving room for further improvements in runtime and memory efficiency. The covariance-based rational SPDE approach by \citet{bolin2024covariance} offers better numerical stability for larger values of $\nu$, and adapting this approximation to the non-stationary model could extend its applicability and robustness. Furthermore, developing more principled strategies for penalizing non-stationary components remains an open problem. Progress on these fronts would make the proposed modeling framework even more practical for applied spatial modeling.

% \clearpage %%Remove this from your manuscript

% Supplementary Material is included here only when compiling
% main_combined.tex.  The standalone main.tex and supplementary.tex files
% remain available for the journal submission workflow.
\ifdefined\combineddocument
\clearpage
\setcounter{section}{0}
\setcounter{subsection}{0}
\setcounter{subsubsection}{0}
\setcounter{figure}{0}
\setcounter{table}{0}
\setcounter{equation}{0}
\renewcommand{\thesection}{S.\arabic{section}}
\renewcommand{\thesubsection}{S.\arabic{section}.\arabic{subsection}}
\renewcommand{\thesubsubsection}{S.\arabic{section}.\arabic{subsection}.\arabic{subsubsection}}
\renewcommand{\thefigure}{S.\arabic{figure}}
\renewcommand{\thetable}{S.\arabic{table}}
\renewcommand{\theequation}{S.\arabic{equation}}
% Keep PDF anchors unique after the counters are reset for the supplement.
\renewcommand{\theHsection}{supplement.\arabic{section}}
\renewcommand{\theHsubsection}{supplement.\arabic{section}.\arabic{subsection}}
\renewcommand{\theHsubsubsection}{supplement.\arabic{section}.\arabic{subsection}.\arabic{subsubsection}}
\renewcommand{\theHfigure}{supplement.\arabic{figure}}
\renewcommand{\theHtable}{supplement.\arabic{table}}
\renewcommand{\theHequation}{supplement.\arabic{equation}}

\shorttitle{Supplementary Material: Non-stationary Spatial Modeling Using Fractional SPDEs}
\shortauthors{E. Svee and G.-A. Fuglstad}
\thispagestyle{empty}

\begin{center}
  {\LARGE\bfseries Supplementary Material}\\[0.5em]
  {\large Non-stationary Spatial Modeling Using Fractional SPDEs}\\[0.5em]
  {\normalsize Elling Svee \quad Geir-Arne Fuglstad}
\end{center}

\vspace{1em}

\section{Sparse reverse-mode automatic differentiation for the likelihood}\label{sec:SparseAD}

\subsection{Sparse adjoints for efficient backpropagation}\label{sub:SparseAdjoints}

Evaluating the log-likelihood in Equation (15) of the main paper requires two log-determinants, one linear solve, and multiple sparse matrix products. The conditional precision matrix $\mathbf{Q}_{\mathrm{C}}=\mathbf{Q}_{\boldsymbol{z}}+\mathbf{S}^{\mathrm{T}}\mathbf{S}/\sigma_{\mathrm{N}}^{2}$ has dimension $(m+p)\times(m+p)$, while the conditional mean $\boldsymbol{\mu}_{\mathrm{C}}=\mathbf{Q}_{\mathrm{C}}^{-1}\mathbf{S}^{\mathrm{T}}\boldsymbol{y}/\sigma_{\mathrm{N}}^{2}$ has length $m+p$. Consequently, differentiating the log-likelihood using standard dense AD becomes prohibitively expensive for fine meshes. To preserve the efficiency of the SPDE approach, it is therefore necessary to exploit the matrix sparsity during backpropagation.

In general, the adjoint of a sparse matrix is not necessarily sparse. However, in our case, the sparsity pattern is fixed by the discretization and does not change during optimization. Therefore, only adjoint entries associated with this structural sparsity pattern are needed. To see this, consider an operation $\mathbf{U}=f(\mathbf{M})$ contributing to the log-likelihood $\log\pi$, for which the vector-Jacobian product propagates the downstream adjoint $\overline{\mathbf{U}}=\partial\log\pi/\partial\mathbf{U}$ to $\overline{\mathbf{M}}=\partial\log\pi/\partial\mathbf{M}$ \citep{giles_collected_2008}. Let $\mathbf{M}=\mathbf{M}(\boldsymbol{\theta})$ be sparse with fixed structural pattern $\mathcal{I}_{\mathbf{M}}=\{(i_{1},j_{1}),\ldots,(i_{N},j_{N})\}$. Since $\mathbf{M}_{ij}=0$ for all $(i,j)\notin\mathcal{I}_{\mathbf{M}}$, these entries are independent of the model parameters. The chain rule
\[
  \frac{\partial \log\pi}{\partial\theta_{l}} = \sum_{i,j}\overline{\mathbf{M}}_{ij}\frac{\partial\mathbf{M}_{ij}}{\partial\theta_{l}}
  = \sum_{(i,j) \in \mathcal{I}_{\mathbf{M}}}\overline{\mathbf{M}}_{ij}\frac{\partial\mathbf{M}_{ij}}{\partial\theta_{l}}
\]
therefore depends only on adjoint entries for $(i,j)\in\mathcal{I}_{\mathbf{M}}$. If the $N$ structurally nonzero entries of $\mathbf{M}$ are represented by the vector $\boldsymbol{v}\in\mathbb{R}^{N}$, differentiation $\overline{v}_{k}=\partial\log\pi/\partial v_{k}$ can thus be carried out directly with respect to these stored values. Importantly, an entry remains part of this representation even if its numerical value happens to be zero, so the stored adjoint follows the same fixed structural pattern as the original matrix.

\subsection{Implementation and computational cost}\label{sub:SparseADImplementation}

To utilize the sparsity during backpropagation, we construct formulas for obtaining the sparse adjoints without forming dense matrices. Let $\mathbf{M}\in\mathbb{R}^{d\times d}$ be a sparse and symmetric positive definite with pattern $\mathcal{I}_{\mathbf{M}}$. Starting with a linear solve $\mathbf{M}\mathbf{U}=\mathbf{B}$ for a dense $\mathbf{B}\in\mathbb{R}^{d\times q}$, the differential $\mathrm{d}\mathbf{U}=\mathbf{M}^{-1}(\mathrm{d}\mathbf{B}-\mathrm{d}\mathbf{M}\,\mathbf{U})$ gives adjoints $\overline{\mathbf{B}}=\mathbf{M}^{-\mathrm{T}}\overline{\mathbf{U}}$ and $\overline{\mathbf{M}}=-\overline{\mathbf{B}}\mathbf{U}^{\mathrm{T}}$ \citep{Minka2000}. Computing $\overline{\mathbf{B}}$ requires one solve, and can due to symmetry reuse the Cholesky factorization of $\mathbf{M}$ from the forward pass. The matrix $-\overline{\mathbf{B}}\mathbf{U}^{\mathrm{T}}$ need not be formed, as the values on the stored pattern are
\[
  \overline{v}_{k} = -\sum_{r=1}^{q}\overline{\mathbf{B}}_{i_{k}r}\mathbf{U}_{j_{k}r},
  \qquad k=1,\ldots,N.
\]
Thus, once $\overline{\mathbf{B}}$ has been computed, the sparse adjoint can be assembled directly on the stored pattern without forming a dense matrix. The required sparse matrix products and solves are available for both CPU and GPU backends through sparsejax.

For the log-determinant $t=\log|\mathbf{M}|$, \citet{giles_collected_2008} provides differential of the determinant
\[
  \mathrm{d}|\mathbf{M}|
  = |\mathbf{M}|\operatorname{tr}\!\left(\mathbf{M}^{-1}\mathrm{d}\mathbf{M}\right).
\]
Since $|\mathbf{M}|>0$, applying the scalar chain rule to the logarithm gives
\[
  \mathrm{d}t
  = \frac{1}{|\mathbf{M}|}\mathrm{d}|\mathbf{M}|
  = \operatorname{tr}\!\left(\mathbf{M}^{-1}\mathrm{d}\mathbf{M}\right),
\]
and the reverse-mode adjoint is $\overline{\mathbf{M}}=\bar{t}\,\mathbf{M}^{-\mathrm{T}}$. Although $\mathbf{M}^{-1}$ is generally dense, we see that the fixed sparsity pattern gives
\[
  \frac{\partial t}{\partial\theta_l}
  = \sum_{(i,j)\in\mathcal{I}_{\mathbf{M}}}
    (\mathbf{M}^{-1})_{ji}
    \frac{\partial \mathbf{M}_{ij}}{\partial\theta_l}.
\]
Hence, the log-determinant gradient only requires inverse entries corresponding to structural nonzeros of $\mathbf{M}$. These entries can be obtained from the Cholesky factorization of $\mathbf{M}$ using the partial inverse recursions of \citet{takahashi1973formation,rue_approximate_2007}, thereby avoiding formation of the dense inverse. From an implementational perspective, high-performance partial-inverse implementations such as PSelInv and sTiles \citep{lin_selinv_2011,lin_fast_2011, fattah_gpu_accelerated_2025} use supernodal formulations to exploit parallelism and dense block operations. In the sparsejax implementation, the CPU backend relies on a sequential implementation of the Takahashi recurrence, which is less efficient than the supernodal approach. The GPU backend does not currently implement the partial inverse and instead obtains the required inverse elements through multiple dense solves. Consequently, both the CPU and GPU implementations can be optimized further to better exploit sparsity and parallelism and reduce memory overhead.

\section{Additional results from the simulation study}\label{seq:Additional_simulation_study}

This section presents additional results from the simulation study. Unless stated otherwise, we fit the same four candidate models as in Section 5.5 of the main paper, and generate data using the fractional and non-stationary model with smoothness $\nu = 0.5$.

\subsection{Parameter estimates from the main simulation study}

For the analysis in Section 5.5, we report the parameter estimates obtained when each candidate model is fitted to a single realization from the data-generating process. Only the stationary parts of the non-stationary parameters are reported in the tables. Note that a non-stationary model can compensate for a misspecified stationary part by adjusting its non-stationary parameters, meaning that the stationary parameters of the non-stationary models may not be directly comparable to those of the stationary models.

Table \ref{tab:ParamEstimsFracStat} reports the parameter estimates obtained from data generated by the stationary model. Both fractional models recover the smoothness parameter with a relatively low bias, and they estimate the correlation range closer to the true value than the non-fractional models. The non-fractional also overestimate the observation noise, which may be a consequence of the misspecified smoothness. Also note that the estimated stationary correlation range for F-NS is too small, indicating that the non-stationary parameters could have some influence. 

The estimates from the experiment with the non-stationary data-generating process are reported in Table \ref{tab:ParamEstimsFracNonStat}. As for the stationary case, the non-fractional models underestimate the stationary correlation range, and the NF-S overestimates the observation noise. The smoothness estimate of the F-NS also has a slightly larger bias than in the stationary case..

\begin{table}[pos=H]
  \centering
  \caption{Average and empirical standard deviation of parameter estimates over 25 repeated simulations. The models are fitted using 500 observations from a single realization, generated by a stationary model with smoothness $\nu = 0.5$.}
  \label{tab:ParamEstimsFracStat}
    \begin{tabular}{lcccccc}
      \toprule
      \multirow{2}{*}{\textbf{Model}} & \multicolumn{6}{c}{\textbf{Parameters}} \\
      \cmidrule(lr){2-7}
      & $\nu$ & $\rho_{0}$ & $a_{0}$ & $\psi_{0} \; [^{\circ}]$ & $\sigma_{0}$ & $\sigma_{\mathrm{N}} \left[10^{-1}\right]$ \\
      \midrule
      True  & 0.5 & 5.0 & 1.0 & $-$ & 1.0 & 1.0 \\
      \midrule
      NF-S  & 1.0 & 2.5 (0.6) & 1.2 (0.1) & 5 (58) & 0.9 (0.1) & 1.6 (0.2) \\
      NF-NS & 1.0 & 2.3 (0.6) & 1.2 (0.1) & 0 (54) & 0.8 (0.1) & 1.5 (0.2) \\
      F-S   & 0.5 (0.1) & 4.6 (1.7) & 1.2 (0.1) & -20 (57) & 1.0 (0.1) & 1.0 (0.2) \\
      F-NS  & 0.6 (0.1) & 3.5 (1.3) & 1.2 (0.1) & 3 (58) & 0.9 (0.1) & 1.0 (0.2) \\
      \bottomrule
    \end{tabular}
\end{table}

\begin{table}[pos=H]
  \centering
  \caption{Average and empirical standard deviation of parameter estimates over 25 repeated simulations. The models are fitted using 500 observations from a single realization, generated by a non-stationary model with smoothness $\nu = 0.5$.}
  \label{tab:ParamEstimsFracNonStat}
    \begin{tabular}{lcccccc}
      \toprule
      \multirow{2}{*}{\textbf{Model}} & \multicolumn{6}{c}{\textbf{Parameters}} \\
      \cmidrule(lr){2-7}
      & $\nu$ & $\rho_{0}$ & $a_{0}$ & $\psi_{0} \; [^{\circ}]$ & $\sigma_{0}$ & $\sigma_{\mathrm{N}} \left[10^{-1}\right]$ \\
      \midrule
      True  & 0.5 & 2.1 & 1.0 & $-$ & 0.9 & 1.0 \\
      \midrule
      NF-S  & 1.0 & 1.2 (0.3) & 1.3 (0.2) & 26 (56) & 0.9 (0.1) & 1.6 (0.6) \\
      NF-NS & 1.0 & 1.3 (0.2) & 1.1 (0.1) & -11 (54) & 0.8 (0.1) & 1.1 (0.2) \\
      F-S   & 0.5 (0.2) & 1.9 (0.9) & 1.3 (0.2) & 28 (51) & 1.0 (0.1) & 1.0 (0.4) \\
      F-NS  & 0.7 (0.1) & 1.7 (0.4) & 1.1 (0.1) & -9 (48) & 0.8 (0.1) & 0.9 (0.1) \\
      \bottomrule
    \end{tabular}
\end{table}

\subsection{Effect of the number of realizations}

We next examine how the predictive performance depends on the amount of data used during estimation. Each candidate model is fitted using $1$, $10$ or $25$ realizations from a fractional and non-stationary data-generating model, and with $500$ observations per realization. The resulting RMSE and CRPS scores are shown in Figure \ref{fig:viz_frac_non_stat_rmse_crps_varying_realizations}. The predictive performance of the stationary models remains relatively stable as the number of realizations grows, while the non-stationary models show a clear improvement when moving from one to ten realizations. The parameter estimates obtained with $25$ realizations are reported in Table \ref{tab:ParamEstimsVaryingReal}. Here the empirical standard deviations are noticeably smaller than in Table \ref{tab:ParamEstimsFracNonStat}, and the F-NS estimates the smoothness with a lower bias than in the single-realization scenario.

\begin{figure}[pos=H]
  \centering
  \includegraphics[width=0.7\textwidth]{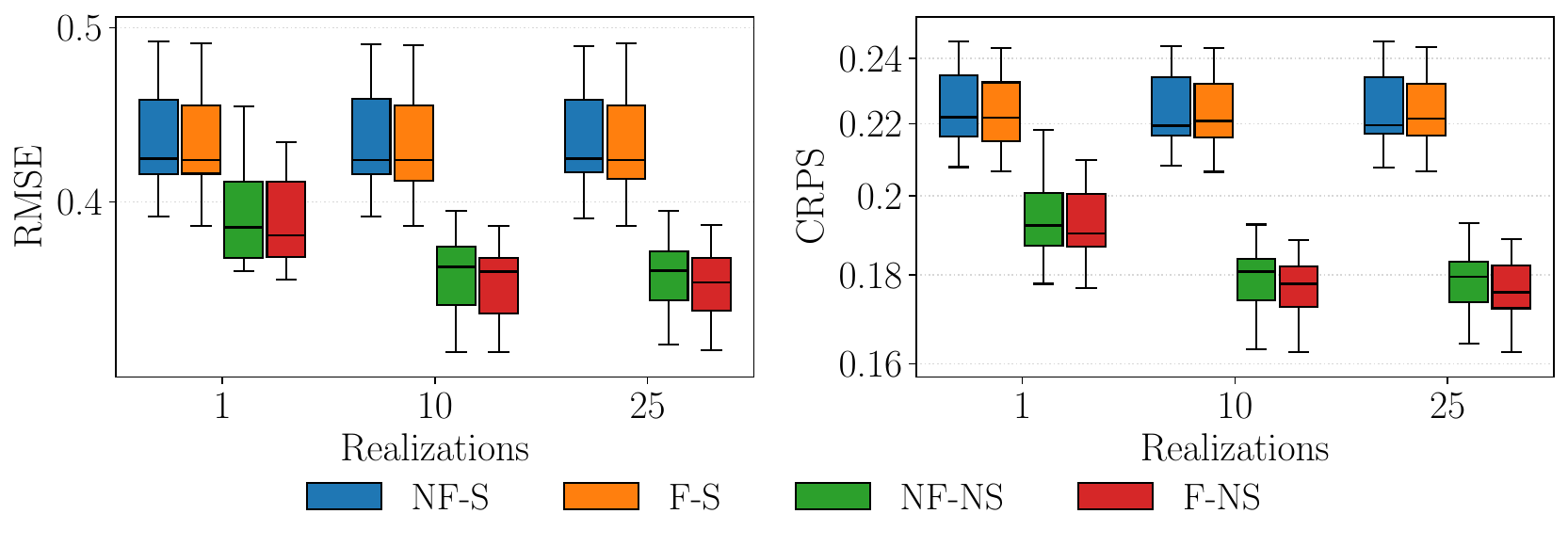}
  \caption{Log-scaled RMSE and CRPS scores for data generated by a non-stationary model with smoothness $\nu = 0.5$, plotted as functions of the number of realizations used during estimation.}
  \label{fig:viz_frac_non_stat_rmse_crps_varying_realizations}
\end{figure}

\begin{table}[pos=H]
  \centering
  \caption{Average and empirical standard deviation of parameter estimates over 25 repeated simulations using $25$ realizations of $500$ observations each, generated by a non-stationary model with smoothness $\nu = 0.5$.}
  \label{tab:ParamEstimsVaryingReal}
    \begin{tabular}{lcccccc}
      \toprule
      \multirow{2}{*}{\textbf{Model}} & \multicolumn{6}{c}{\textbf{Parameters}} \\
      \cmidrule(lr){2-7}
      & $\nu$ & $\rho_{0}$ & $a_{0}$ & $\psi_{0} \; [^{\circ}]$ & $\sigma_{0}$ & $\sigma_{\mathrm{N}} \left[10^{-1}\right]$ \\
      \midrule
      True  & 0.5 & 2.1 & 1.0 & $-$ & 0.9 & 1.0 \\
      \midrule
      NF-S  & 1.0 & 1.2 (0.1) & 1.2 (0.1) & 49 (20) & 0.9 (0.0) & 1.6 (0.2) \\
      NF-NS & 1.0 & 1.4 (0.0) & 1.0 (0.0) & 5 (41) & 0.8 (0.0) & 1.3 (0.0) \\
      F-S   & 0.4 (0.1) & 1.9 (0.2) & 1.2 (0.1) & 35 (38) & 0.9 (0.0) & 0.9 (0.2) \\
      F-NS  & 0.5 (0.0) & 2.0 (0.1) & 1.0 (0.0) & 3 (48) & 0.9 (0.0) & 1.0 (0.0) \\
      \bottomrule
    \end{tabular}
\end{table}

\subsection{Sensitivity to the rational approximation order}

Figure \ref{fig:viz_frac_non_stat_rmse_crps_varying_m} compares the impact of the rational approximation order $k$ used when fitting the F-NS model. Data are generated by the same non-stationary model as in the main simulation study, but where the order of the rational approximation is increased to $k = 4$. The F-NS model is then fitted with orders $k = 1, \ldots, 4$. Note that the mesh is kept at the same resolution as in the main simulation study. It could be that a finer mesh would amplify differences between the approximation orders. However, this is not tested as part of the simulation study.

Figure \ref{fig:viz_frac_non_stat_rmse_crps_varying_m} shows that the predictive performance is similar across the different approximation orders. For $k=4$ appears to be slightly better than the other for $500$ observations, while the $k=1$ is slightly worse at $1000$ observations. The parameter estimates for $500$ observations are reported in Table \ref{tab:ParamEstimsCompareK}. Interestingly, see that the bias of the parameter estimates decreases as the approximation order increases. The $k=1$ estimates $\nu$ to around $0.9$, while we estimate $0.6$ using order $k=4$ This indicates that the approximation order has some influence on the estimation of the parameters. However, as increasing the order also increases the computational cost, it may be that $k=2$ or $k=3$ is a good compromise between accuracy and efficiency.

\begin{figure}[pos=H]
  \centering
  \includegraphics[width=0.7\textwidth]{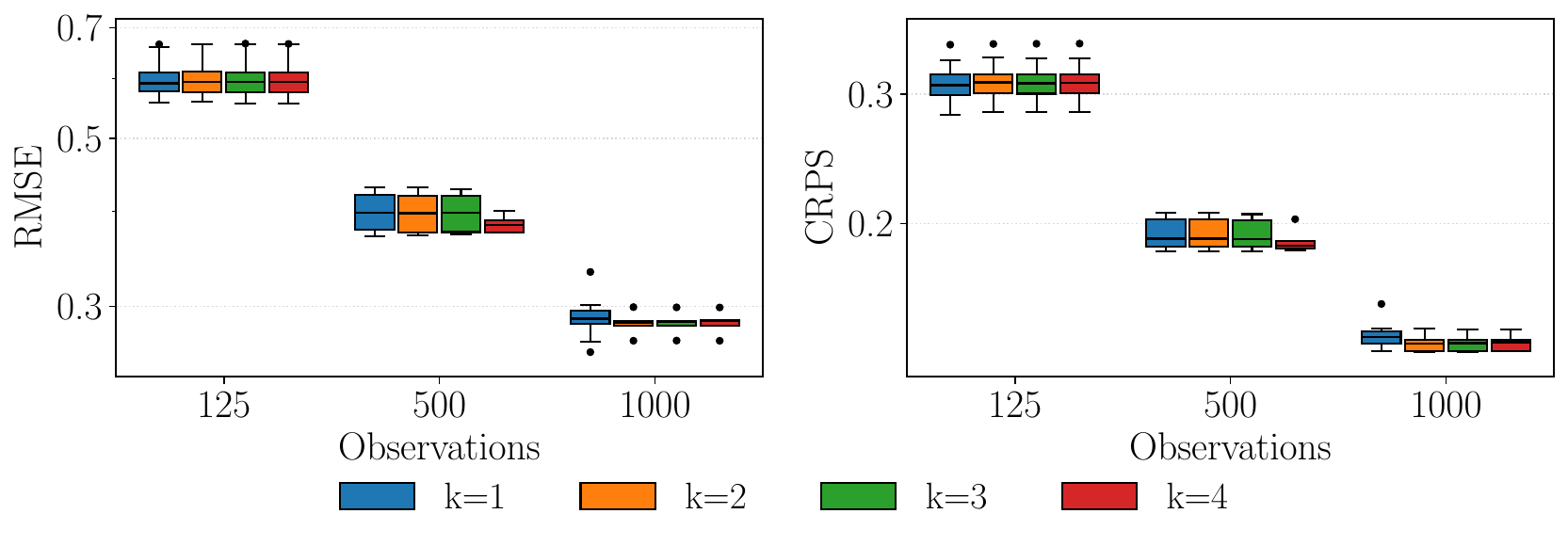}
  \caption{Log-scaled RMSE and CRPS scores for the F-NS model fitted with different rational approximation orders $k$. Data are generated by a non-stationary model with smoothness $\nu = 0.5$ using $k=4$, and the results are plotted as functions of the number of realizations.}
  \label{fig:viz_frac_non_stat_rmse_crps_varying_m}
\end{figure}

\begin{table}[pos=H]
  \centering
  \caption{Average and empirical standard deviation of parameter estimates over 25 repeated simulations. Data are generated by a non-stationary model with smoothness $\nu = 0.5$ using rational approximation order $k = 4$.}
  \label{tab:ParamEstimsCompareK}
    \begin{tabular}{lcccccc}
      \toprule
      \multirow{2}{*}{\textbf{Model}} & \multicolumn{6}{c}{\textbf{Parameters}} \\
      \cmidrule(lr){2-7}
      & $\nu$ & $\rho_{0}$ & $a_{0}$ & $\psi_{0} \; [^{\circ}]$ & $\sigma_{0}$ & $\sigma_{\mathrm{N}} \left[10^{-1}\right]$ \\
      \midrule
      True & 0.5 & 2.1 & 1.0 & $-$ & 0.9 & 1.0 \\
      \midrule
      $k=1$  & 0.9 (0.1) & 1.7 (0.3) & 1.1 (0.1) & -14 (51) & 0.8 (0.1) & 0.8 (0.2) \\
      $k=2$   & 0.7 (0.1) & 1.8 (0.4) & 1.1 (0.1) & -16 (52) & 0.8 (0.1) & 0.9 (0.2) \\
      $k=3$ & 0.7 (0.1) & 1.8 (0.4) & 1.1 (0.1) & -18 (53) & 0.8 (0.1) & 0.9 (0.2) \\
      $k=4$  & 0.6 (0.1) & 1.9 (0.3) & 1.1 (0.1) & -37 (22) & 0.8 (0.1) & 1.0 (0.3) \\
      \bottomrule
    \end{tabular}
\end{table}

\subsection{Different true smoothness}

We also check how the results change when the data-generating process has a different smoothness. We set the true smoothness to $\nu = 0.75$, while keeping the target correlation range, marginal standard deviation and anisotropy the same as in the main simulation study. Figure \ref{fig:viz_frac_non_stat_rmse_crps_nu075} shows that the results are very similar to the $\nu = 0.5$ case. The non-stationary models again outperform the stationary models, while there is little difference between the fractional and non-fractional models. The parameter estimates in Table \ref{tab:ParamEstimsNu075} shows that the F-NS model estimates the smoothness with a similar bias as in the $\nu = 0.5$ case. The F-S estimates a slightly lower smoothness than the truth, while the F-NS estimates a smoothness slightly above the correct value.

\begin{figure}[pos=H]
  \centering
  \includegraphics[width=0.7\textwidth]{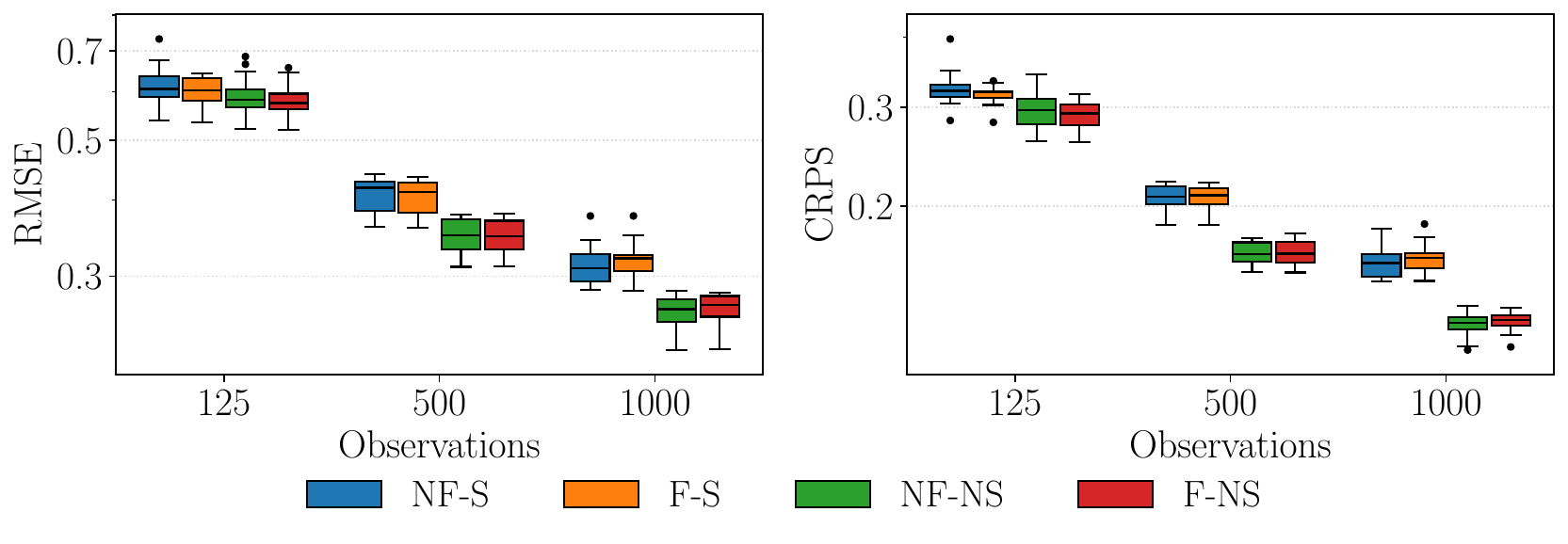}
  \caption{Log-scaled RMSE and CRPS scores for data generated by a non-stationary model with smoothness $\nu = 0.75$, plotted as functions of the number of observations.}
  \label{fig:viz_frac_non_stat_rmse_crps_nu075}
\end{figure}

\begin{table}[pos=H]
  \centering
  \caption{Average and empirical standard deviation of parameter estimates over 25 repeated simulations. Data are generated by a non-stationary model with smoothness $\nu = 0.75$ and 500 observations per realization.}
  \label{tab:ParamEstimsNu075}
    \begin{tabular}{lcccccc}
      \toprule
      \multirow{2}{*}{\textbf{Model}} & \multicolumn{6}{c}{\textbf{Parameters}} \\
      \cmidrule(lr){2-7}
      & $\nu$ & $\rho_{0}$ & $a_{0}$ & $\psi_{0} \; [^{\circ}]$ & $\sigma_{0}$ & $\sigma_{\mathrm{N}} \left[10^{-1}\right]$ \\
      \midrule
      True  & 0.75 & 2.1 & 1.0 & $-$ & 0.9 & 1.0 \\
      \midrule
      NF-S  & 1.0 & 1.3 (0.2) & 1.4 (0.2) & 48 (42) & 0.9 (0.1) & 1.4 (0.5) \\
      NF-NS & 1.0 & 1.7 (0.3) & 1.1 (0.1) & 0 (50) & 0.8 (0.1) & 1.0 (0.1) \\
      F-S   & 0.6 (0.2) & 1.9 (0.8) & 1.3 (0.2) & 34 (52) & 1.0 (0.1) & 1.0 (0.4) \\
      F-NS  & 0.8 (0.1) & 1.9 (0.3) & 1.1 (0.1) & 0 (49) & 0.9 (0.1) & 0.9 (0.1) \\
      \bottomrule
    \end{tabular}
\end{table}

\subsection{Different non-stationary data-generating model}

Finally, we test whether the conclusions change for a different non-stationary data-generating model. We keep the correlation range and marginal variance at the same constant values as in the stationary experiment, but let the anisotropy follow the same circular pattern as in the main simulation study. Fitting the four candidate models, Figure \ref{fig:viz_frac_non_stat_rmse_crps_non_stat_anisotropy} shows that the conclusions remain the same. The non-stationary models outperform the stationary models, while the fractional and non-fractional models perform similarly. Table \ref{tab:ParamEstimsAnisotropy} reports the corresponding parameter estimates.

\begin{figure}[pos=H]
  \centering
  \includegraphics[width=0.7\textwidth]{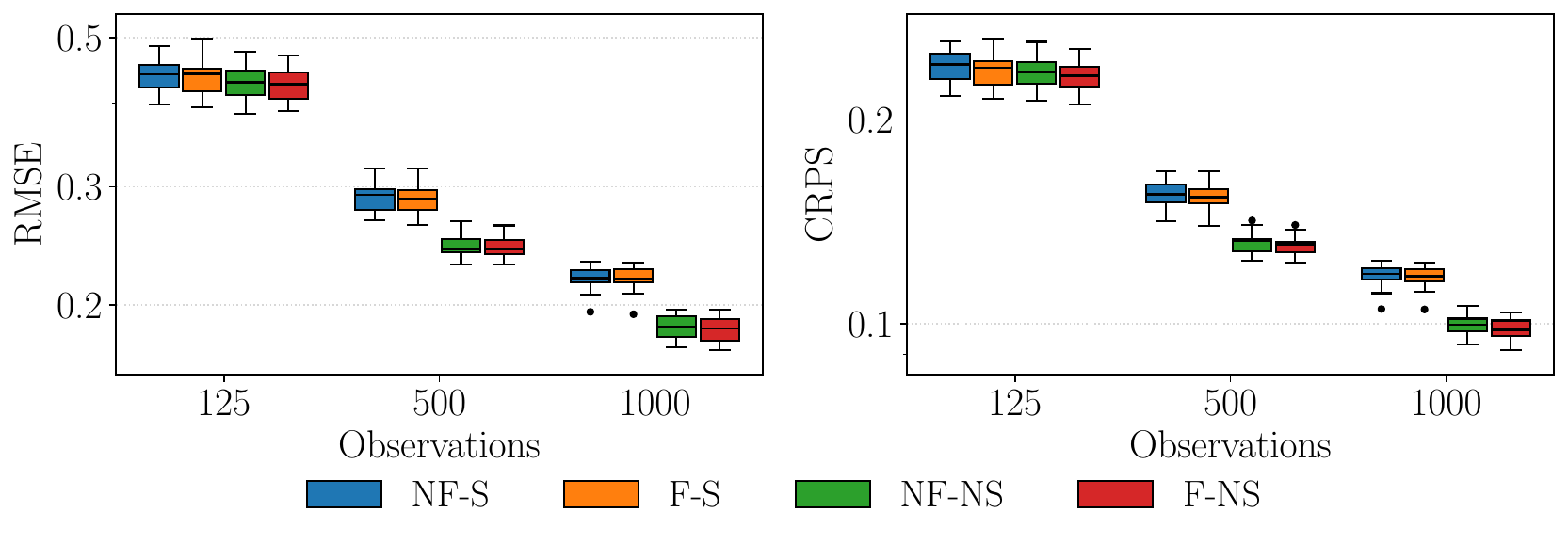}
  \caption{Log-scaled RMSE and CRPS scores for data generated by a model with smoothness $\nu = 0.5$ and non-stationary anisotropy, plotted as functions of the number of observations.}
  \label{fig:viz_frac_non_stat_rmse_crps_non_stat_anisotropy}
\end{figure}

\begin{table}[pos=H]
  \centering
  \caption{Average and empirical standard deviation of parameter estimates over 25 repeated simulations. Data are generated by a model with smoothness $\nu = 0.5$ in which only the anisotropy is non-stationary.}
  \label{tab:ParamEstimsAnisotropy}
    \begin{tabular}{lcccccc}
      \toprule
      \multirow{2}{*}{\textbf{Model}} & \multicolumn{6}{c}{\textbf{Parameters}} \\
      \cmidrule(lr){2-7}
      & $\nu$ & $\rho_{0}$ & $a_{0}$ & $\psi_{0} \; [^{\circ}]$ & $\sigma_{0}$ & $\sigma_{\mathrm{N}} \left[10^{-1}\right]$ \\
      \midrule
      True  & 0.5 & 5.0 & 1.0 & $-$ & 1.0 & 1.0 \\
      \midrule
      NF-S  & 1.0 & 2.3 (0.5) & 1.2 (0.1) & -14 (51) & 1.0 (0.1) & 1.4 (0.2) \\
      NF-NS & 1.0 & 2.6 (0.5) & 1.1 (0.1) & 0 (59) & 0.9 (0.1) & 1.3 (0.2) \\
      F-S   & 0.6 (0.1) & 3.7 (1.1) & 1.2 (0.1) & -12 (50) & 1.0 (0.2) & 1.0 (0.2) \\
      F-NS  & 0.6 (0.1) & 3.7 (0.8) & 1.1 (0.0) & -6 (55) & 0.9 (0.1) & 0.9 (0.2) \\
      \bottomrule
    \end{tabular}
\end{table}

\section{Parameter estimates from the case studies}\label{seq:Additional_case_study}

This section reports the parameter estimates for the models fitted in the two case studies. The empirical means and standard deviations are computed across repetitions of the hold-out procedure described in Section 6.1. As opposed to the simulation study, the true parameter values are here unknown. The only exception is the measurement noise standard deviation $\sigma_{\mathrm{N}}$ for the salinity data, which was added to the data with a known value of $0.01\,\mathrm{psu}$. 

Table \ref{tab:SINMODParameters} contains the parameter estimates obtained from the salinity dataset. The fractional smoothness parameters are estimated close to one, suggesting a near non-fractional behavior. Comparing the stationary parts of the correlation range, there is a notable difference between the NF-NS and F-NS models, which is somewhat surprising given their similar smoothness estimates. The stationary models estimate a larger stationary anisotropy than the non-stationary models, which could be a way to compensate for the lack of non-stationarity. The estimates of $\sigma_{\mathrm{N}}$ match the true value, indicating that the models capture the measurement noise.

\begin{table}[pos=H]
  \centering
  \caption{Average and empirical standard deviation of parameter estimates from the salinity case study, computed across repetitions of the hold-out procedure described in Section 6.1.}
  \label{tab:SINMODParameters}
    \begin{tabular}{lcccccc}
      \toprule
      \multirow{2}{*}{\textbf{Model}} & \multicolumn{6}{c}{\textbf{Parameters}} \\
      \cmidrule(lr){2-7}
      & $\nu$ & $\rho_{0} \; [\mathrm{km}]$ & $a_{0}$ & $\psi_{0} \; [^{\circ}]$ & $\sigma_{0} \; [\mathrm{psu}]$ & $\sigma_{\mathrm{N}} \left[10^{-1}\,\mathrm{psu}\right]$ \\
      \midrule
      NF-S  & 1.0 & 1.2 (0.3) & 1.9 (0.2) & -85 (15) & 1.1 (0.3) & 0.1 (0.0) \\
      NF-NS & 1.0 & 1.8 (0.4) & 1.2 (0.2) & 73 (32) & 1.1 (0.1) & 0.1 (0.0) \\
      F-S   & 0.9 (0.1) & 1.4 (0.4) & 1.9 (0.2) & -85 (27) & 1.2 (0.3) & 0.1 (0.0) \\
      F-NS  & 0.9 (0.1) & 1.2 (0.6) & 1.2 (0.2) & 64 (39) & 1.0 (0.1) & 0.1 (0.2) \\
      \bottomrule
    \end{tabular}
\end{table}

Table \ref{tab:PercipParameters} reports the parameter estimates obtained from the precipitation dataset. Both fractional models estimate the smoothness parameter close to $0.4$. The fractional models also estimate a larger correlation range than the non-fractional models, while the marginal variance estimates are similar across all four models. The stationary anisotropy angle $\psi_{0}$ differs between the stationary and non-stationary models, but this is likely a consequence of the estimated non-stationary correlation structure shown in Figure 10.

\begin{table}[pos=H]
  \centering
  \caption{Average and empirical standard deviation of parameter estimates from the precipitation case study, computed across repetitions of the hold-out procedure described in Section 6.1.}
  \label{tab:PercipParameters}
    \begin{tabular}{lcccccc}
      \toprule
      \multirow{2}{*}{\textbf{Model}} & \multicolumn{6}{c}{\textbf{Parameters}} \\
      \cmidrule(lr){2-7}
      & $\nu$ & $\rho_{0} \; [^{\circ}]$ & $a_{0}$ & $\psi_{0} \; [^{\circ}]$ & $\sigma_{0} \; [10^{-1}\,\mathrm{cm}]$ & $\sigma_{\mathrm{N}} \left[10^{-2}\,\mathrm{cm}\right]$ \\
      \midrule
      NF-S  & 0.5 & 2.9 (0.1) & 1.2 (0.1) & 5 (4) & 0.8 (0.2) & 1.9 (0.4) \\
      NF-NS & 0.5 & 2.8 (0.3) & 1.3 (0.2) & 11 (7) & 0.8 (0.1) & 1.8 (0.3) \\
      F-S   & 0.4 (0.1) & 4.4 (0.3) & 1.3 (0.2) & 5 (3) & 0.8 (0.2) & 1.9 (0.3) \\
      F-NS  & 0.4 (0.2) & 3.4 (0.4) & 1.3 (0.2) & 11 (9) & 0.8 (0.2) & 1.7 (0.2) \\
      \bottomrule
    \end{tabular}
\end{table}

\fi

\printcredits

\bibliographystyle{cas-model2-names}

% \bibliography{cas-refs}
\bibliography{bibliography}

@article{lindgren2011explicit,
  author = {Lindgren, Finn and Rue, Håvard and Lindstrøm, Johan},
  title = {An explicit link between {G}aussian fields and {G}aussian {M}arkov
           random fields: the stochastic partial differential equation approach},
  note = {With discussion and a reply by the authors},
  journal = {J. R. Stat. Soc. Ser. B Stat. Methodol.},
  fjournal = {Journal of the Royal Statistical Society. Series B. Statistical
              Methodology},
  volume = {73},
  year = {2011},
  number = {4},
  pages = {423--498},
  issn = {1369-7412,1467-9868},
  mrclass = {62M30 (60H15 62F15 62H99 62M40 62P12)},
  mrnumber = {2853727},
  mrreviewer = {C\'ecile\ Hardouin},
  doi = {10.1111/j.1468-9868.2011.00777.x},
  url = {https://doi.org/10.1111/j.1467-9868.2011.00777.x},
}

@article{bolin2020rational,
  author = {Bolin, David and Kirchner, Kristin},
  title = {The rational {SPDE} approach for {G}aussian random fields with
           general smoothness},
  journal = {J. Comput. Graph. Statist.},
  fjournal = {Journal of Computational and Graphical Statistics},
  volume = {29},
  year = {2020},
  number = {2},
  pages = {274--285},
  issn = {1061-8600,1537-2715},
  mrclass = {62M30 (60G60 60H15 60H35 62-08)},
  mrnumber = {4116041},
  doi = {10.1080/10618600.2019.1665537},
  url = {https://doi.org/10.1080/10618600.2019.1665537},
}

@article{bolin2024covariance,
  author = {Bolin, David and Simas, Alexandre B. and Xiong, Zhen},
  title = {Covariance-based rational approximations of fractional {SPDE}s for
           computationally efficient {B}ayesian inference},
  journal = {J. Comput. Graph. Statist.},
  fjournal = {Journal of Computational and Graphical Statistics},
  volume = {33},
  year = {2024},
  number = {1},
  pages = {64--74},
  issn = {1061-8600,1537-2715},
  mrclass = {99-01},
  mrnumber = {4713943},
  doi = {10.1080/10618600.2023.2231051},
  url = {https://doi.org/10.1080/10618600.2023.2231051},
}

@book{baker1996pade,
  author = {Baker, Jr., George A. and Graves-Morris, Peter},
  title = {Pad\'e{} approximants},
  series = {Encyclopedia of Mathematics and its Applications},
  volume = {59},
  edition = {Second},
  publisher = {Cambridge University Press, Cambridge},
  year = {1996},
  pages = {xiv+746},
  isbn = {0-521-45007-1},
  mrclass = {41-02 (41A21 65D15)},
  mrnumber = {1383091},
  mrreviewer = {Annie\ A. M. Cuyt},
  doi = {10.1017/CBO9780511530074},
  url = {https://doi.org/10.1017/CBO9780511530074},
}

@article{llamazareselias2024parameterization,
  title = {A parameterization of anisotropic Gaussian fields with penalized
           complexity priors},
  author = {Liam Llamazares-Elias and Jonas Latz and Finn Lindgren},
  year = {2026},
  journal = {Journal of the American Statistical Association},
  note = {In press},
}

@article{2022_SPDE_10_years,
  author = {Lindgren, Finn and Bolin, David and Rue, Håvard},
  title = {The {SPDE} approach for {G}aussian and non-{G}aussian fields: 10
           years and still running},
  journal = {Spat. Stat.},
  fjournal = {Spatial Statistics},
  volume = {50},
  year = {2022},
  pages = {Paper No. 100599, 29},
  issn = {2211-6753},
  mrclass = {99-01},
  mrnumber = {4439328},
  doi = {10.1016/j.spasta.2022.100599},
  url = {https://doi.org/10.1016/j.spasta.2022.100599},
}

@article{fuglstad2015doesnonstationaryspatialdata,
  author = {Fuglstad, Geir-Arne and Simpson, Daniel and Lindgren, Finn and Rue,
            Håvard},
  title = {Does non-stationary spatial data always require non-stationary random
           fields?},
  journal = {Spat. Stat.},
  fjournal = {Spatial Statistics},
  volume = {14},
  year = {2015},
  pages = {505--531},
  issn = {2211-6753},
  mrclass = {62M40 (60H15 62M30)},
  mrnumber = {3431054},
  doi = {10.1016/j.spasta.2015.10.001},
  url = {https://doi.org/10.1016/j.spasta.2015.10.001},
}

@article{fuglstad2019constructing,
  author = {Fuglstad, Geir-Arne and Simpson, Daniel and Lindgren, Finn and Rue,
            Håvard},
  title = {Constructing priors that penalize the complexity of {G}aussian random
           fields},
  journal = {J. Amer. Statist. Assoc.},
  fjournal = {Journal of the American Statistical Association},
  volume = {114},
  year = {2019},
  number = {525},
  pages = {445--452},
  issn = {0162-1459,1537-274X},
  mrclass = {62M40 (62F15)},
  mrnumber = {3941267},
  mrreviewer = {Onur\ Dikmen},
  doi = {10.1080/01621459.2017.1415907},
  url = {https://doi.org/10.1080/01621459.2017.1415907},
}

@article{fuglstad2015exploring,
  author = {Fuglstad, Geir-Arne and Lindgren, Finn and Simpson, Daniel and Rue,
            Håvard},
  title = {Exploring a new class of non-stationary spatial {G}aussian random
           fields with varying local anisotropy},
  journal = {Statist. Sinica},
  fjournal = {Statistica Sinica},
  volume = {25},
  year = {2015},
  number = {1},
  pages = {115--133},
  issn = {1017-0405,1996-8507},
  mrclass = {62H10 (60G60)},
  mrnumber = {3328806},
}

@article{simpson2015penalisingmodelcomponentcomplexity,
  author = {Simpson, Daniel and Rue, Håvard and Riebler, Andrea and Martins,
            Thiago G. and Sørbye, Sigrunn H.},
  title = {Penalising model component complexity: a principled, practical
           approach to constructing priors},
  journal = {Statist. Sci.},
  fjournal = {Statistical Science. A Review Journal of the Institute of
              Mathematical Statistics},
  volume = {32},
  year = {2017},
  number = {1},
  pages = {1--28},
  issn = {0883-4237,2168-8745},
  mrclass = {62F15},
  mrnumber = {3634300},
  doi = {10.1214/16-STS576},
  url = {https://doi.org/10.1214/16-STS576},
}

@misc{paszke2019pytorchimperativestylehighperformance,
  title = {PyTorch: An Imperative Style, High-Performance Deep Learning Library},
  author = {Adam Paszke and Sam Gross and Francisco Massa and Adam Lerer and
            James Bradbury and Gregory Chanan and Trevor Killeen and Zeming Lin
            and Natalia Gimelshein and Luca Antiga and Alban Desmaison and
            Andreas Köpf and Edward Yang and Zach DeVito and Martin Raison and
            Alykhan Tejani and Sasank Chilamkurthy and Benoit Steiner and Lu Fang
            and Junjie Bai and Soumith Chintala},
  year = {2019},
  eprint = {1912.01703},
  archiveprefix = {arXiv},
  primaryclass = {cs.LG},
  url = {https://arxiv.org/abs/1912.01703},
}

@misc{kingma2017adammethodstochasticoptimization,
  title = {Adam: A Method for Stochastic Optimization},
  author = {Diederik P. Kingma and Jimmy Ba},
  year = {2017},
  eprint = {1412.6980},
  archiveprefix = {arXiv},
  primaryclass = {cs.LG},
  url = {https://arxiv.org/abs/1412.6980},
}

@manual{rSPDEpackage,
  title = {rSPDE: Rational Approximations of Fractional Stochastic Partial
           Differential Equations},
  author = {David Bolin and Alexandre B. Simas},
  year = {2023},
  note = {R package version 2.3.3},
  url = {https://CRAN.R-project.org/package=rSPDE},
}

@book{MR2848400,
  AUTHOR = {Cressie, Noel and Wikle, Christopher K.},
  TITLE = {Statistics for spatio-temporal data},
  SERIES = {Wiley Series in Probability and Statistics},
  PUBLISHER = {John Wiley \& Sons, Inc., Hoboken, NJ},
  YEAR = {2011},
  PAGES = {xxii+588},
  ISBN = {978-0-471-69274-4},
  MRCLASS = {62M30 (62-01)},
  MRNUMBER = {2848400},
  MRREVIEWER = {Thomas\ R.\ Boucher},
}

@article{hu2015spatialmodellingtemperaturehumidity,
  author = {Ingebrigtsen, Rikke and Lindgren, Finn and Steinsland, Ingelin and
            Martino, Sara},
  title = {Estimation of a non-stationary model for annual precipitation in
           southern {N}orway using replicates of the spatial field},
  journal = {Spat. Stat.},
  fjournal = {Spatial Statistics},
  volume = {14},
  year = {2015},
  pages = {338--364},
  issn = {2211-6753},
  mrclass = {62M30 (60H15 62F15)},
  mrnumber = {3431045},
  doi = {10.1016/j.spasta.2015.07.003},
  url = {https://doi.org/10.1016/j.spasta.2015.07.003},
}

@book{banerjee2003hierarchical,
  author = {Banerjee, Sudipto and Carlin, Bradley P. and Gelfand, Alan E.},
  title = {Hierarchical modeling and analysis for spatial data},
  series = {Monographs on Statistics and Applied Probability},
  volume = {135},
  edition = {Second},
  publisher = {CRC Press, Boca Raton, FL},
  year = {2015},
  pages = {xxii+562},
  isbn = {978-1-4398-1917-3},
  mrclass = {62-01 (62-07 62H11)},
  mrnumber = {3362184},
}

@misc{nytko2023optimizedsparsematrixoperations,
  title = {Optimized Sparse Matrix Operations for Reverse Mode Automatic
           Differentiation},
  author = {Nicolas Nytko and Ali Taghibakhshi and Tareq Uz Zaman and Scott
            MacLachlan and Luke N. Olson and Matt West},
  year = {2023},
  eprint = {2212.05159},
  archiveprefix = {arXiv},
  primaryclass = {cs.LG},
  url = {https://arxiv.org/abs/2212.05159},
}

@misc{sintef1987,
  author = {SINTEF},
  year = {1987},
  title = {SINMOD},
  url = {https://www.sintef.no/programvare/sinmod/},
  note = {Accessed: 2025-03-07},
}

@article{SLAGSTAD20051,
  title = {Modeling the ecosystem dynamics of the Barents sea including the
           marginal ice zone: I. Physical and chemical oceanography},
  journal = {Journal of Marine Systems},
  volume = {58},
  number = {1},
  pages = {1-18},
  year = {2005},
  issn = {0924-7963},
  url = {https://www.sciencedirect.com/science/article/pii/S0924796305001296},
  author = {Dag Slagstad and Thomas A. McClimans},
}

@article{berild2023spatiallyvaryinganisotropygaussian,
  author = {Berild, Martin Outzen and Fuglstad, Geir-Arne},
  title = {Spatially varying anisotropy for {G}aussian random fields in
           three-dimensional space},
  journal = {Spat. Stat.},
  fjournal = {Spatial Statistics},
  volume = {55},
  year = {2023},
  pages = {Paper No. 100750, 32},
  issn = {2211-6753},
  mrclass = {99-01},
  mrnumber = {4583070},
  doi = {10.1016/j.spasta.2023.100750},
  url = {https://doi.org/10.1016/j.spasta.2023.100750},
}

@book{webb_introduction_nodate,
  title = {Descriptive Physical Oceanography: An Introduction},
  author = {Talley, Lynne D.},
  isbn = {9780080939117},
  year = {2011},
  publisher = {Academic Press},
}

@article{berild2024nonstationaryspatiotemporalmodelingusing,
  author = {Berild, Martin Outzen and Fuglstad, Geir-Arne},
  title = {Non-stationary spatio-temporal modeling using the stochastic
           advection-diffusion equation},
  journal = {Spat. Stat.},
  fjournal = {Spatial Statistics},
  volume = {64},
  year = {2024},
  pages = {Paper No. 100867, 22},
  issn = {2211-6753},
  mrclass = {99-01},
  mrnumber = {4823150},
  doi = {10.1016/j.spasta.2024.100867},
  url = {https://doi.org/10.1016/j.spasta.2024.100867},
}

@mastersthesis{lilleborge2021bivariate,
  author = {Karina Lilleborge},
  title = {Bivariate Spatial Models using Stochastic Partial Differential
           Equations},
  school = {Norwegian University of Science and Technology (NTNU)},
  year = {2021},
  type = {Project thesis},
  advisor = {Geir-Arne Fuglstad},
  language = {English},
}

@manual{lindgrenFmesher,
  title = {fmesher: Triangle Meshes and Related Geometry Tools},
  author = {Finn Lindgren},
  year = {2024},
  note = {R package version 0.1.7.9008, https://github.com/inlabru-org/fmesher},
  url = {https://inlabru-org.github.io/fmesher/},
}

@book{stein1999interpolation,
  title = {Interpolation of spatial data: some theory for kriging},
  author = {Stein, Michael L},
  year = {1999},
  publisher = {Springer Science \& Business Media},
}

@article{heaton2019case,
  title = {A case study competition among methods for analyzing large spatial
           data},
  author = {Heaton, Matthew J and Datta, Abhirup and Finley, Andrew O and Furrer
            , Reinhard and Guinness, Joseph and Guhaniyogi, Rajarshi and Gerber,
            Florian and Gramacy, Robert B and Hammerling, Dorit and Katzfuss,
            Matthias and others},
  journal = {Journal of agricultural, biological and environmental Statistics},
  volume = {24},
  number = {3},
  pages = {398--425},
  year = {2019},
  publisher = {Springer},
}

@article{schmidt2020flexible,
  title = {Flexible spatial covariance functions},
  author = {Schmidt, Alexandra M and Guttorp, Peter},
  journal = {Spatial Statistics},
  volume = {37},
  pages = {100416},
  year = {2020},
  publisher = {Elsevier},
}

@incollection{sampson2010constructions,
  title = {Constructions for nonstationary spatial processes},
  author = {Sampson, Paul D},
  booktitle = {Handbook of spatial statistics},
  pages = {119--130},
  year = {2010},
  publisher = {CRC/Press},
}

@online{CopernicusClimateReanalysis,
  author = {Copernicus Climate Change Service C3S},
  title = {Climate reanalysis},
  year = 2025,
  url = {https://climate.copernicus.eu/climate-reanalysis},
  urldate = {2025-05-28},
}

@article{genton2015cross,
  author = {Genton, Marc G. and Kleiber, William},
  title = {Cross-covariance functions for multivariate geostatistics},
  journal = {Statist. Sci.},
  fjournal = {Statistical Science. A Review Journal of the Institute of
              Mathematical Statistics},
  volume = {30},
  year = {2015},
  number = {2},
  pages = {147--163},
  issn = {0883-4237,2168-8745},
  mrclass = {62P12},
  mrnumber = {3353096},
  doi = {10.1214/14-STS487},
  url = {https://doi.org/10.1214/14-STS487},
}

@article{sampson1992nonparametric,
  title = {Nonparametric estimation of nonstationary spatial covariance
           structure},
  author = {Sampson, Paul D and Guttorp, Peter},
  journal = {Journal of the American Statistical Association},
  volume = {87},
  number = {417},
  pages = {108--119},
  year = {1992},
  publisher = {Taylor \& Francis},
}

@article{paciorek2006spatial,
  title = {Spatial modelling using a new class of nonstationary covariance
           functions},
  author = {Paciorek, Christopher J and Schervish, Mark J},
  journal = {Environmetrics: The official journal of the International
             Environmetrics Society},
  volume = {17},
  number = {5},
  pages = {483--506},
  year = {2006},
  publisher = {Wiley Online Library},
}

@article{schmidt2003bayesian,
  title = {Bayesian inference for non-stationary spatial covariance structure
           via spatial deformations},
  author = {Schmidt, Alexandra M and O'Hagan, Anthony},
  journal = {Journal of the Royal Statistical Society Series B: Statistical
             Methodology},
  volume = {65},
  number = {3},
  pages = {743--758},
  year = {2003},
  publisher = {Oxford University Press},
}

@article{damian2001bayesian,
  title = {Bayesian estimation of semi-parametric non-stationary spatial
           covariance structures},
  author = {Damian, Doris and Sampson, Paul D and Guttorp, Peter},
  journal = {Environmetrics: The official journal of the International
             Environmetrics Society},
  volume = {12},
  number = {2},
  pages = {161--178},
  year = {2001},
  publisher = {Wiley Online Library},
}

@article{neto2014accounting,
  title = {Accounting for spatially varying directional effects in spatial
           covariance structures},
  author = {Neto, Joaquim Henriques Vianna and Schmidt, Alexandra M and Guttorp,
            Peter},
  journal = {Journal of the Royal Statistical Society Series C: Applied
             Statistics},
  volume = {63},
  number = {1},
  pages = {103--122},
  year = {2014},
  publisher = {Oxford University Press},
}

@article{risser2020bayesian,
  title = {Bayesian inference for high-dimensional nonstationary Gaussian
           processes},
  author = {Risser, Mark D and Turek, Daniel},
  journal = {Journal of Statistical Computation and Simulation},
  volume = {90},
  number = {16},
  pages = {2902--2928},
  year = {2020},
  publisher = {Taylor \& Francis},
}

@article{risser2017local,
  title = {Local likelihood estimation for covariance functions with
           spatially-varying parameters: the convoSPAT package for R},
  author = {Risser, Mark D and Calder, Catherine A},
  journal = {Journal of Statistical Software},
  volume = {81},
  pages = {1--32},
  year = {2017},
}

@article{higdon1998process,
  title = {A process-convolution approach to modelling temperatures in the North
           Atlantic Ocean},
  author = {Higdon, David},
  journal = {Environmental and Ecological Statistics},
  volume = {5},
  number = {2},
  pages = {173--190},
  year = {1998},
  publisher = {Springer},
}

@article{risser2015regression,
  title = {Regression-based covariance functions for nonstationary spatial
           modeling},
  author = {Risser, Mark D and Calder, Catherine A},
  journal = {Environmetrics},
  volume = {26},
  number = {4},
  pages = {284--297},
  year = {2015},
  publisher = {Wiley Online Library},
}

@article{Minka2000,
  author = {Thomas. P. Minka},
  title = {Old and new matrix algebra useful for statistics},
  year = {2000},
  url = {https://tminka.github.io/papers/matrix/},
}

@incollection{giles_collected_2008,
  author = {Giles, Mike B.},
  title = {Collected matrix derivative results for forward and reverse mode
           algorithmic differentiation},
  booktitle = {Advances in automatic differentiation},
  series = {Lect. Notes Comput. Sci. Eng.},
  volume = {64},
  pages = {35--44},
  publisher = {Springer, Berlin},
  year = {2008},
  isbn = {978-3-540-68935-5},
  mrclass = {15A99},
  mrnumber = {2531677},
}

@inproceedings{takahashi1973formation,
  title = {Formation of sparse bus impedance matrix and its application to short
           circuit study},
  author = {Takahashi, Kazuhiro},
  booktitle = {Proc. PICA Conference, June, 1973},
  year = {1973},
}

@article{rue_approximate_2007,
  author = {Rue, Håvard and Martino, Sara},
  title = {Approximate {B}ayesian inference for hierarchical {G}aussian {M}arkov
           random field models},
  journal = {J. Statist. Plann. Inference},
  fjournal = {Journal of Statistical Planning and Inference},
  volume = {137},
  year = {2007},
  number = {10},
  pages = {3177--3192},
  issn = {0378-3758,1873-1171},
  mrclass = {62M05 (62F15 62M40)},
  mrnumber = {2365120},
  doi = {10.1016/j.jspi.2006.07.016},
  url = {https://doi.org/10.1016/j.jspi.2006.07.016},
}

@misc{pereira_geostatistics_2023,
  title = {Geostatistics for large datasets on {Riemannian} manifolds: a
           matrix-free approach},
  shorttitle = {Geostatistics for large datasets on {Riemannian} manifolds},
  url = {http://arxiv.org/abs/2208.12501},
  doi = {10.48550/arXiv.2208.12501},
  urldate = {2026-04-22},
  publisher = {arXiv},
  author = {Pereira, Mike and Desassis, Nicolas and Allard, Denis},
  month = jan,
  year = {2023},
  note = {arXiv:2208.12501 [math]},
}

@misc{antil_efficient_2022,
  title = {Efficient algorithms for {Bayesian} {Inverse} {Problems} with {
           Whittle}--{Matérn} {Priors}},
  url = {http://arxiv.org/abs/2205.04417},
  doi = {10.48550/arXiv.2205.04417},
  urldate = {2026-04-22},
  publisher = {arXiv},
  author = {Antil, Harbir and Saibaba, Arvind K.},
  month = may,
  year = {2022},
  note = {arXiv:2205.04417 [math]},
}

@software{jax2018github,
  author = {James Bradbury and Roy Frostig and Peter Hawkins and Matthew James
            Johnson and Yash Katariya and Chris Leary and Dougal Maclaurin and
            George Necula and Adam Paszke and Jake Vander{P}las and Skye
            Wanderman-{M}ilne and Qiao Zhang},
  title = {{JAX}: composable transformations of {P}ython+{N}um{P}y programs},
  url = {http://github.com/jax-ml/jax},
  version = {0.3.13},
  year = {2018},
}

@article{cholmod2008,
  author = {Chen, Yanqing and Davis, Timothy A. and Hager, William W. and
            Rajamanickam, Sivasankaran},
  title = {Algorithm 887: CHOLMOD, Supernodal Sparse Cholesky Factorization and
           Update/Downdate},
  year = {2008},
  issue_date = {October 2008},
  publisher = {Association for Computing Machinery},
  address = {New York, NY, USA},
  volume = {35},
  number = {3},
  issn = {0098-3500},
  url = {https://doi.org/10.1145/1391989.1391995},
  doi = {10.1145/1391989.1391995},
  journal = {ACM Trans. Math. Softw.},
  month = oct,
  articleno = {22},
  numpages = {14},
}

@article{scipy2020,
  author = {Virtanen, Pauli and Gommers, Ralf and Oliphant, Travis E. and
            Haberland, Matt and Reddy, Tyler and Cournapeau, David and Burovski,
            Evgeni and Peterson, Pearu and Weckesser, Warren and Bright, Jonathan
            and {van der Walt}, St{\'e}fan J. and Brett, Matthew and Wilson,
            Joshua and Millman, K. Jarrod and Mayorov, Nikolay and Nelson, Andrew
            R. J. and Jones, Eric and Kern, Robert and Larson, Eric and Carey, C
            J and Polat, {\.I}lhan and Feng, Yu and Moore, Eric W. and {
            VanderPlas}, Jake and Laxalde, Denis and Perktold, Josef and Cimrman,
            Robert and Henriksen, Ian and Quintero, E. A. and Harris, Charles R.
            and Archibald, Anne M. and Ribeiro, Ant{\^o}nio H. and Pedregosa,
            Fabian and {van Mulbregt}, Paul and {SciPy 1.0 Contributors}},
  title = {{{SciPy} 1.0: Fundamental Algorithms for Scientific Computing in
           Python}},
  journal = {Nature Methods},
  year = {2020},
  volume = {17},
  pages = {261--272},
  adsurl = {https://rdcu.be/b08Wh},
  doi = {10.1038/s41592-019-0686-2},
}

@inproceedings{cusparse2009implementing,
  author = {Bell, Nathan and Garland, Michael},
  title = {Implementing sparse matrix-vector multiplication on
           throughput-oriented processors},
  year = {2009},
  isbn = {9781605587448},
  publisher = {Association for Computing Machinery},
  address = {New York, NY, USA},
  url = {https://doi.org/10.1145/1654059.1654078},
  doi = {10.1145/1654059.1654078},
  booktitle = {Proceedings of the Conference on High Performance Computing
               Networking, Storage and Analysis},
  articleno = {18},
  numpages = {11},
  location = {Portland, Oregon},
  series = {SC '09},
}

@misc{nvidia_cudss,
  author = {{NVIDIA Corporation}},
  title = {{NVIDIA cuDSS: GPU-Accelerated Direct Sparse Solver Library}},
  year = {2025},
  howpublished = {\url{https://developer.nvidia.com/cudss}},
  note = {Accessed: 2026-04-28},
}

@misc{svee_sparsejax,
  author = {Elling Svee},
  title = {sparsejax: Efficient sparse adjoints in JAX},
  year = {2026},
  howpublished = {\url{https://github.com/ellingsvee/sparsejax}},
  note = {Accessed: 2026-04-29},
}

@misc{hill_sparser_2025,
  title = {Sparser, {Better}, {Faster}, {Stronger}: {Sparsity} {Detection} for {
           Efficient} {Automatic} {Differentiation}},
  url = {http://arxiv.org/abs/2501.17737},
  urldate = {2026-05-10},
  publisher = {arXiv},
  author = {Hill, Adrian and Dalle, Guillaume},
  year = {2025},
}

@misc{juliadiffcontributorsJuliaDiffSparseDiffToolsjl2024,
  author = {{JuliaDiff contributors}},
  title = {{JuliaDiff/SparseDiffTools.jl}},
  year = {2024},
  month = oct,
  url = {https://github.com/JuliaDiff/SparseDiffTools.jl},
  note = {Accessed: 2026-05-10},
}

@article{geoga2023fitting,
  title = {Fitting Mat{\'e}rn smoothness parameters using automatic
           differentiation},
  author = {Geoga, Christopher J and Marin, Oana and Schanen, Michel and Stein,
            Michael L},
  journal = {Statistics and Computing},
  volume = {33},
  number = {2},
  pages = {48},
  year = {2023},
  publisher = {Springer},
}

@article{zhang2004inconsistent,
  title = {Inconsistent estimation and asymptotically equal interpolations in
           model-based geostatistics},
  author = {Zhang, Hao},
  journal = {Journal of the American Statistical Association},
  volume = {99},
  number = {465},
  pages = {250--261},
  year = {2004},
  publisher = {Taylor \& Francis},
}

@article{hildeman2021deformed,
  title = {Deformed SPDE models with an application to spatial modeling of
           significant wave height},
  author = {Hildeman, Anders and Bolin, David and Rychlik, Igor},
  journal = {Spatial Statistics},
  volume = {42},
  pages = {100449},
  year = {2021},
  publisher = {Elsevier},
}

@misc{baydin_automatic_2015,
  title = {Automatic differentiation in machine learning: a survey},
  shorttitle = {Automatic differentiation in machine learning},
  url = {https://arxiv.org/abs/1502.05767v4},
  language = {en},
  urldate = {2026-05-21},
  journal = {arXiv.org},
  author = {Baydin, Atilim Gunes and Pearlmutter, Barak A. and Radul, Alexey
            Andreyevich and Siskind, Jeffrey Mark},
  month = feb,
  year = {2015},
}

@article{LBFGSB,
  author = {Zhu, Ciyou and Byrd, Richard H. and Lu, Peihuang and Nocedal, Jorge},
  journal = {ACM Trans. Math. Softw.},
  number = 4,
  pages = {550-560},
  title = {Algorithm 778: L-BFGS-B: Fortran Subroutines for Large-Scale
           Bound-Constrained Optimization.},
  url = {http://dblp.uni-trier.de/db/journals/toms/toms23.html#ZhuBLN97},
  volume = 23,
  year = 1997,
}

@software{Optax2020,
  title = {The {D}eep{M}ind {JAX} {E}cosystem},
  author = {DeepMind and Babuschkin, Igor and Baumli, Kate and Bell, Alison and
            Bhupatiraju, Surya and Bruce, Jake and Buchlovsky, Peter and Budden,
            David and Cai, Trevor and Clark, Aidan and Danihelka, Ivo and Dedieu,
            Antoine and Fantacci, Claudio and Godwin, Jonathan and Jones, Chris
            and Hemsley, Ross and Hennigan, Tom and Hessel, Matteo and Hou,
            Shaobo and Kapturowski, Steven and Keck, Thomas and Kemaev, Iurii and
            King, Michael and Kunesch, Markus and Martens, Lena and Merzic, Hamza
            and Mikulik, Vladimir and Norman, Tamara and Papamakarios, George and
            Quan, John and Ring, Roman and Ruiz, Francisco and Sanchez, Alvaro
            and Sartran, Laurent and Schneider, Rosalia and Sezener, Eren and
            Spencer, Stephen and Srinivasan, Srivatsan and Stanojevi\'{c}, Milo\v
            {s} and Stokowiec, Wojciech and Wang, Luyu and Zhou, Guangyao and
            Viola, Fabio},
  url = {http://github.com/google-deepmind},
  year = {2020},
}

@misc{noack_gp2scale_2026,
  title = {{gp2Scale}: {A} {Class} of {Compactly} {Supported} {Non}-{Stationary}
           {Kernels} and {Distributed} {Computing} for {Exact} {Gaussian} {
           Processes} on 10 {Million} {Data} {Points}},
  publisher = {arXiv},
  author = {Noack, Marcus M. and Risser, Mark D. and Luo, Hengrui and Tekriwal,
            Vardaan and Pandolfi, Ronald J.},
  year = {2025},
}

@article{risser_compactly-supported_2025,
  title = {Compactly-{Supported} {Nonstationary} {Kernels} for {Computing} {
           Exact} {Gaussian} {Processes} on {Big} {Data}},
  volume = {36},
  issn = {1099-095X},
  url = {https://onlinelibrary.wiley.com/doi/abs/10.1002/env.70054},
  number = {8},
  journal = {Environmetrics},
  author = {Risser, Mark D. and Noack, Marcus M. and Luo, Hengrui and Pandolfi,
            Ronald J.},
  year = {2025},
  pages = {e70054},
}

@article{bolin2020numerical,
  title = {Numerical solution of fractional elliptic stochastic PDEs with
           spatial white noise},
  author = {Bolin, David and Kirchner, Kristin and Kov{\'a}cs, Mih{\'a}ly},
  journal = {IMA Journal of Numerical Analysis},
  volume = {40},
  number = {2},
  pages = {1051--1073},
  year = {2020},
  publisher = {Oxford University Press},
}

@article{bolin2018weak,
  title = {Weak convergence of Galerkin approximations for fractional elliptic
           stochastic PDEs with spatial white noise},
  author = {Bolin, David and Kirchner, Kristin and Kov{\'a}cs, Mih{\'a}ly},
  journal = {BIT Numerical Mathematics},
  volume = {58},
  number = {4},
  pages = {881--906},
  year = {2018},
  publisher = {Springer},
}

@article{lin_fast_2011,
  title = {A {Fast} {Parallel} {Algorithm} for {Selected} {Inversion} of {
           Structured} {Sparse} {Matrices} with {Application} to {2D} {Electronic
           } {Structure} {Calculations}},
  volume = {33},
  issn = {1064-8275},
  url = {https://epubs.siam.org/doi/10.1137/09077432X},
  number = {3},
  urldate = {2026-09-07},
  journal = {SIAM Journal on Scientific Computing},
  publisher = {Society for Industrial and Applied Mathematics},
  author = {Lin, Lin and Yang, Chao and Lu, Jianfeng and Ying, Lexing and E,
            Weinan},
  year = {2011},
  pages = {1329--1351},
}

@article{lin_selinv_2011,
  title = {{SelInv} - {An} algorithm for selected inversion of a sparse
           symmetric matrix},
  volume = {37},
  issn = {0098-3500},
  url = {
         https://collaborate.princeton.edu/en/publications/selinv-an-algorithm-for-selected-inversion-of-a-sparse-symmetric-/
         },
  language = {English (US)},
  number = {4},
  journal = {ACM Transactions on Mathematical Software},
  publisher = {Association for Computing Machinery},
  author = {Lin, Lin and Yang, Chao and Meza, Juan C. and Lu, Jianfeng and Ying,
            Lexing and Weinan, E.},
  year = {2011},
  pages = {40},
}

@misc{fattah_gpu_accelerated_2025,
  title = {{GPU}-{Accelerated} {Parallel} {Selected} {Inversion} for {Structured
           } {Matrices} {Using} {sTiles}},
  url = {http://arxiv.org/abs/2504.19171},
  publisher = {arXiv},
  author = {Fattah, Esmail Abdul and Ltaief, Hatem and Rue, Havard and Keyes,
            David},
  month = sep,
  year = {2025},
}



\end{document}